\documentclass[useAMS,usenatbib]{mn2e}
\usepackage{graphicx}
\usepackage{latexsym}
\usepackage{epsfig}
\usepackage{amsmath,amssymb}
\usepackage{color}
\usepackage{url}
\usepackage{ulem}
\usepackage{longtable,lscape}

\title{Rapid Evolution of the Innermost Dust Disk of Protoplanetary
Disks Surrounding Intermediate-mass Stars}

\author[Chikako Yasui, Naoto Kobayashi, Alan T. Tokunaga, and Masao Saito]
{Chikako Yasui,$^{1}$\thanks{E-mail: ck.yasui@astron.s.u-tokyo.ac.jp
 (CY)}
Naoto Kobayashi,$^{2}$ Alan T. Tokunaga$^{3}$ and Masao Saito$^{4}$\\
$^1$Department of Astronomy, Graduate School of Science,
 University of Tokyo, Hongo 7-3-1, Bunkyo-ku, 113-0033, Tokyo, Japan \\
$^2$Institute of Astronomy, School of Science, University
of Tokyo, 2-21-1 Osawa, Mitaka, Tokyo 181-0015, Japan\\
$^3$Institute for Astronomy, University of Hawaii, 2680 Woodlawn
Drive, Honolulu, HI 96822, USA \\
$^{4}$National Astronomical Observatory of Japan 2-21-1
Osawa, Mitaka, Tokyo, 181-8588, Japan}

\begin{document}

\pagerange{\pageref{firstpage}--\pageref{lastpage}} \pubyear{2002}

\maketitle

\label{firstpage}

\def\mnras{MNRAS}
\def\apj{ApJ}
\def\aap{A\&A}
\def\apjl{ApJL}
\def\apjs{ApJS}
\def\bain{BAIN}
\def\araa{ARA\&A}
\def\pasp{PASP}
\def\aj{AJ}
\def\pasj{PASJ}
\def\ga{\sim}
\voffset=-0.5in

\begin{abstract}

We derived the intermediate-mass {($\simeq$1.5--7\,$M_\odot$)} disk
fraction (IMDF) in the near-infrared {\it JHK} photometric bands as well
as in the mid-infrared (MIR) bands for young clusters in the age range
of 0 to $\sim$10\,Myr.
From the {\it JHK} IMDF, the lifetime of the innermost dust disk
($\sim$0.3\,AU; hereafter the $K$ disk) is estimated to be $\sim$3\,Myr,
 suggesting a stellar mass ($M_*$) dependence of $K$-disk lifetime
 $\propto M^{-0.7}_*$.
However, from the MIR IMDF, the lifetime of the inner disk ($\sim$5\,AU;
hereafter the MIR disk) is estimated to be $\sim$6.5\,Myr, 
suggesting a very weak stellar mass dependence ($\propto M^{-0.2}_*$).
The much shorter $K$-disk lifetime compared to the MIR-disk lifetime
for intermediate-mass (IM) stars suggests that IM stars with {\it
transition disks}, which have only MIR excess emission but no $K$-band
excess emission, are more common than classical Herbig Ae/Be stars,
which exhibit both.  We suggest that this prominent early disappearance
of the $K$ disk for IM stars is due to dust settling/growth in the
protoplanetary disk, and it could be one of the major reasons for the
paucity of close-in planets around IM stars.

\end{abstract}


\begin{keywords}
circumstellar matter --
planetary systems: formation --
planetary systems: proto-planetary discs --
stars: pre-main-sequence --
infrared: stars --
stars: variables: T Tauri, Herbig Ae/Be.
\end{keywords}

\section{Introduction}
\label{sec:intro}

Understanding protoplanetary disks is not only essential for understanding the
star formation process, 
it is also critical for understanding planet formation
\citep[e.g.][]{LadaLada2003}.  The lifetime of protoplanetary disks is
one of the most fundamental parameters of a protoplanetary disk because
it directly restricts the time for planet formation
\citep[e.g.][]{Williams2011}.  Many studies that derive the lifetime of
protoplanetary disks are now available.  In a pioneering work,
\citet{Strom1989} studied the frequency of disk-harbouring stars with
known ages in the Taurus molecular cloud that have a $K$-band excess and
suggested that the disk lifetime is in the range from $\ll$3\,Myr to
$\sim$10\,Myr.  Subsequently, a more direct method using the `disk
fraction', which is the frequency of near-infrared (NIR) or mid-infrared
(MIR) excess stars within a young cluster {\it with an assumed age}, has
been widely adopted to study the disk lifetime following the work by
\citet{Lada1999} and \citet*{Haisch2001ApJL}.  Using the disk fraction
that monotonically decreases as a function of cluster age, the disk
lifetime is estimated at about 5--10\,Myr in the solar neighborhood
(\citealt{{Lada1999}, {Haisch2001ApJL},{Hernandez2008}}; see also
\citealt{My2010ApJ}).  \citet{Mamajek2009} compiled disk fractions for
about 20 clusters and derived the characteristic disk decay time-scale
($\tau$) of 2.5\,Myr, assuming the disk ${\rm fraction} [{\rm \%}]
\propto \exp ( - t[{\rm Myr}] / \tau)$.

These estimated disk lifetimes were mainly derived from the disk
fraction with all detected cluster members and thus the estimated
lifetime has been primarily for low-mass stars ($<$2\,$M_\odot$),
considering the characteristic mass of the initial mass function (IMF)
\citep[$\sim$0.3\,$M_\odot$,][]{Elmegreen2009} and the typical stellar
mass detection limit ($\sim$0.1\,$M_\odot$).  However, a number of
assessments of the effect of the stellar mass dependence of the disk
lifetime have recently suggested a shorter disk lifetime for the
higher-mass stars
\citep[e.g.][]{{Hernandez2005},{Carpenter2006},{Kennedy2009}}.  Although
the existence of disks of high-mass stars ($\gtrsim$8\,$M_\odot$) is
still under debate \citep[e.g.][]{{Mann2009},{Fuente2002}}, the disks of
IM stars have been extensively studied and are well characterized
\citep{Hernandez2005,Hernandez2007_662,Hernandez2008}.  IM stars with
optically thick disks are known as Herbig Ae/Be (HAeBe) stars.  They
were originally discovered with strong emission-lines by
\citet{Herbig1960}.  After \citet{Hernandez2004} established the method
for selecting HAeBe stars using the spectral energy distribution (SED)
slope from the $V$-band to the IRAS 12-$\mu$m band,
\citet{Hernandez2005} derived the HAeBe-star disk fraction for six
clusters in the age range of 3--10\,Myr.  They showed that the disk
fraction is lower compared to the previously derived disk fraction for
low-mass (LM) stars, in particular by a factor of $\sim$10 lower at
$\sim$3\,Myr.

Recently, the disk fraction studies have shifted to longer wavelengths
(MIR or submm) mainly because of the interest in tracing the outer disk,
where most of the disk mass resides \citep[][]{Williams2011}.  However,
the disk fractions can also be estimated using only {\it JHK} data, in
particular, for the IM stars.  Originally, the disk lifetime was
estimated with disk fractions derived by using the color--color diagram
based on imaging in the {\it JHK} and {\it JHKL} photometric bands
\citep[e.g.][]{{Haisch2001ApJL}, {Lada1999}}.  After the advent of the
{\it Spitzer Space Telescope}, the SED slope ($\alpha = d \ln \lambda
F_\lambda / d \ln \lambda$) in the MIR wavelength range (3.6, 4.5, 5.8,
and 8.0\,$\mu$m) is used for selecting disk-harbouring stars (e.g.
$\alpha \ge -2.0$; \citealt{Lada2006}).  However, the derived disk
fractions and disk lifetime with {\it Spitzer} data were found to be
almost the same as those with {\it JHKL} data (see
\citealt{{Sicilia-Aguilar2006}, {Hernandez2008}}), and even with {\it
JHK} data \citep{Lada1999,My2010ApJ}.
Although the NIR disk fractions are known to show values
systematically smaller and with a larger uncertainty due to
contamination of the non--disk-harbouring stars on the colour--colour
diagram \citep{{Haisch2001a},{My2010ApJ}}, \citet{Hernandez2005} showed
that the {\it JHK} colour--colour diagram can clearly distinguish
disk-harbouring IM stars from non-disk IM stars.  The {\it JHK} disk
fraction value is robust both because of the large infrared excess and
the higher stellar effective temperature of the IM stars, compared to LM
stars.

In this paper, we derived the {\it JHK} IMDF using the Two Micron All
Sky Survey (2MASS) Point Source Catalog of a large number ($\sim$20) of
well-established nearby ($D \lesssim 1.5$\,kpc) young clusters with an
age span of 0 to $\sim$10\,Myr in order to quantitatively and
comprehensively study the lifetime of protoplanetary disks surrounding
IM stars ($\simeq$1.5--7\,$M_\odot$).  In particular, we included as
many clusters as possible with ages $<$5\,Myr.  To securely identify IM
cluster members, we made use of the spectral types of each cluster
member from the literature, assuming a single age for each cluster.
With the derived {\it JHK} IMDFs for a large number of younger clusters
($<$5\,Myr), we estimated the disk lifetime of the IM stars.  We then
estimated the stellar mass dependence of the disk lifetime by comparing
the lifetime of IM stars to that of LM stars.  We also derived the MIR
IMDFs with {\it Spitzer} data in the literature to compare them to the
{\it JHK} IMDFs.  We found that the derived {\it JHK} IMDFs are
significantly lower than the MIR IMDFs, in particular at younger ages
($<$3\,Myr), which results in shorter lifetime of the $K$ disk than the
MIR disk.  This suggests a potentially larger fraction of `transition
disks' for IM stars compared to those for LM stars.  We discuss the
implications of these results for dust growth and planet formation.

Because the sample clusters and the selection of the IM stars is
critical for this paper, we discuss this in detail in
Section~\ref{sec:sample}.  The definition and derivation of the {\it
JHK} IMDF and the MIR IMDF are described in Sections~\ref{sec:JHK IMDF}
and $\S$~\ref{sec:MIR IMDF}, respectively.  Before interpreting the
results of the IMDFs, the definition of the disk lifetime is discussed
in Section~\ref{sec:def_time}. Section~\ref{sec:kdisk} then discusses
the results for {\it JHK} IMDFs.  The mid-IR disk fraction is discussed
in Section~7.  Subsequently, Section~\ref{sec:DF_NIR_MIR} discusses the
large difference between {\it JHK} and MIR IMDFs found in this study and
potential disk evolution consequences for IM stars.  Finally,
Section~\ref{sec:physical} discusses the possible physical mechanisms of
this rapid evolution of the $K$ disk.  At the end, in
Section~\ref{sec:planet}, we briefly discuss possible implications for
planet formation.  Section~\ref{sec:conclusion} summarizes this paper.

\section{Target Clusters and  Selection of Intermediate-mass Star Samples} 
\label{sec:sample}

\subsection{Target clusters} 

We selected our target clusters from previous studies of the disk
fraction/disk evolution (\citealt{Haisch2001ApJL};
\citealt{Hernandez2005}, \citealt{Hernandez2008}; \citealt{Kennedy2009};
\citealt{Mamajek2009}; \citealt{Gaspar2009}; \citealt{Fedele2010};
\citealt{Roccatagliata2011}).  For estimating the disk lifetime with
acceptable accuracy, it is necessary to derive the IMDFs for as many as
young clusters as possible, ideally more than ten.  We thus selected our
target young clusters from the above papers, but with the following
criteria:
(1) Cluster ages are spaced from 0 to $\sim$10\,Myrs, to cover the time
period of disk dispersal.
(2) The cluster membership is well defined from a variety of
observations (astrometry, radial velocity, variability, H$\alpha$,
X-ray, NIR excess, MIR excess, optical spectroscopy, NIR spectroscopy,
etc.). This criterion naturally leads to clusters in the solar
neighborhood (distance $<$1.5\,kpc).
(3) The spectral types of a large number of cluster members are
available by spectroscopy.
(4) Well-defined NIR and MIR photometry of the cluster members with
$M_{\rm limit} \sim 1$\,$M_\odot$ is published or in a widely available
catalogue, such as 2MASS.
(5) At least three IM stars are available per cluster for IMDF derivation.

The resultant 19 target clusters are summarized in Table
\ref{tab:clusters} along with the age, distance, and references for the
disk fraction study.  Almost all young ($<$5\,Myr) clusters in the
references (Table \ref{tab:clusters}) are included, though three young
clusters (MBM 12, NGC 6231, and NGC 7129) are excluded.  This is because
it appears that no IM stars are present in MBM 12 \citep{Luhman2001} and
the spectral types of stars in NGC 6231 and NGC 7129 are limited only to
brightest members (OB, A stars), and is not adequate to cover the entire
IM star mass range down to 1.5\,$M_\odot$.  Some older clusters
($>$5\,Myr), mostly those from \cite{Fedele2010}, are excluded because
they do not satisfy the above criterion.  For several well-known
clusters within the target clusters (Trapezium, Ori OB1a, Ori OB1bc, Per
OB2), we could derive only {\it JHK} disk fractions because we could not
find published {\it Spitzer} MIR data for the IM stars, probably because
of saturation.  As a result, we obtained the {\it JHK} IMDF for 19
clusters and the MIR IMDF for 13 clusters.

\subsection{Selection of intermediate-mass stars} 
\label{subsec:selection_IMstars}

Although the original definition of mass for HAeBe is
$\sim$2--10\,$M_\odot$ with spectral types of B and A (and in a few cases F)
\citep{Herbig1960}, the presence of disks around stars earlier than B5
($\simeq$6--7\,$M_\odot$ in the main-sequence phase) is not well
established since the disk lifetime of high-mass stars is very rapid,
e.g. $\sim$1\,Myr \citep{{Zinnecker2007},{Fuente2002}}.
Also, the number of high-mass stars ($>$6--7\,$M_\odot$) is
very small because of the IMF, and the number stochastically fluctuates from
cluster to cluster.  Therefore, we set the upper-mass limit as
7\,$M_\odot$ in this paper.  This is also a good match with the mass
range of the isochrone model by \citet*{Siess2000} ($M_{\rm
max}=7\,M_\odot$), which is used throughout this paper.  As for the
lower-mass limit, we employed 1.5\,$M_\odot$, which corresponds to
spectral type `F1' for main sequences, following past comprehensive
works of disks for IM stars by \citet{Hernandez2005} and
\citet{Kennedy2009}. The latter defined a mass range bin of
1.5$-$7\,$M_\odot$, which can be directly compared to our results.

The IM star selection is a critical item for this study. Ideally,
stellar mass and the age of each cluster member are determined from the HR
diagram with the extinction-corrected luminosity and
spectroscopically determined effective temperature 
through an isochrone model. However, this requires a time-consuming
observational program, and thus the number of target clusters is limited
as in the previous studies. Even if we had the complete observational
data, the value of mass and age depends on the isochrone model, and
could strongly depend on the extinction correction with different R$_V$
(e.g. \citealt{Hernandez2004}, \citeyear{Hernandez2005}).  Another
approach is to use a limited number of parameters, such as only the
spectral type, to pick up cluster members in a broad mass range, such as
IM or LM stars, and to use a larger number of clusters.  Although
sacrificing the accuracy of the mass estimate, a study including a
larger number of clusters is possible.  Although some past studies, in
fact, focus on targets of certain spectral types (e.g. earlier than F1)
to pick up IM stars (\citealt{Hernandez2005}; \citealt*{Uzpen2009}),
{\it the true mass for a star of a certain spectral type varies with the
age of the star}, and such spectral-type--limited samples should be
viewed with caution (\citealt{Kennedy2009}).  We assume the cluster age
is the age of the members in the selection of the IM stars.

The choice of the isochrone model is critical for the mass
estimate. Although a number of recent isochrone models are available
\citep[e.g., such as][]{{Yi2003}, {Tognelli2011}}, we choose
\citet{Siess2000} because it is the most used isochrone track in the
target mass range with the reliability through various tests and
application to many observational data.
Using dynamically and kinematically determined stellar masses,
\citet{Hillenbrand2004} confirmed that virtually any isochrone model
provides similar mass estimate for masses more than 1.2\,$M_\odot$.
\citet{Hillenbrand2004} also noted that the introduction of new
isochrone models tend to bring new systematic uncertainty and should be
used with caution. 
For this study, using Siess's isochrone is also critical for comparison
with the previous studies which used Siess's isochrone in most cases
(e.g., \citealt{Kennedy2009}).

 We take particular note of the fact that the age spread of young
clusters in the solar neighborhood is in many cases small enough so that
a single age can be assumed for each cluster (see Table
\ref{tab:clusters}).  Therefore, the boundary masses of the IM stars (7
and 1.5\,$M_\odot$) theoretically correspond to a unique spectral type
for each cluster, which enables IM star selection only with the spectral
type of the members without considering differential extinction.  This
method should be effective, in particular, for IM stars because most of
the time they evolve along the Henyey track, which is roughly horizontal
on the HR diagram, and even when the IM stars are on Hayashi track
before switching to the Henyey track, the spectral type does not change
because the track is almost vertical on the HR diagram. Table
\ref{tab:age} shows the unique spectral types corresponding to the
boundary masses for each cluster age based on the isochrone model by
\citet{Siess2000}.

However, there are several points to take note for using the above
method in selecting the IM stars. 
First, each spectral type, in particular the later spectral type
(G7--K5), corresponds to a slightly broader mass range as shown in the
third column of Table \ref{tab:age}. For example, the boundary spectral
type K5 corresponds to 1.2--1.5\,$M_\odot$ for the age of 2\,Myr.
Therefore, the sampling by spectral type naturally leads to the
inclusion of stars with a mass of slightly lower than the nominal
1.5\,$M_\odot$.
Next, note that the age spread of each cluster may cause contamination
of lower-mass stars in our IM-star samples in the case where the age of
the star is older than the cluster age. Table \ref{tab:age} also shows
the possible mass range for an age spread of $\Delta t = 2$\,Myr, which
is the maximum possible age spread in most cases (typically $\Delta t =
1$\,Myr: see Table \ref{tab:clusters}).
Although the age of most stars are within the age spread of 2\,Myr,
there are $\sim$15\,\% stars at most which are older than the age spread
and are actually lower-mass stars (see Figs.~1--5 in
\citealt{Palla2000}).
Lastly, the distance uncertainties of target clusters may also influence
on the selection of the IM stars. The typical uncertainties of distance
is about 10\,\% for the clusters in the solar neighborhood
\citep{{Reipurth2008hsf1},{Reipurth2008hsf2}}.
For the clusters studied by \citet{Hernandez2005}, the uncertainties are
even smaller (less than 5\,\%) with {\it Hipparcos} data.  For deriving
the mass and age of a star on H-R diagram, the effective temperature is
independent of the distance because it is derived from spectroscopy,
while luminosity is directly affected.
However, the luminosity can differ by only 0.2\,mag with the assumed
distance uncertainties, which then cause a mass difference of
$\lesssim$0.1\,$M_\odot$ around lower mass limit of this study,
1.5\,$M_\odot$\footnote{Note that the age differences are $\le$1\,Myr
for stars $\le$3\,Myr and $\lesssim$2\,Myr for older stars.  These
differences are within the age spread (2\,Myr) we are considering.},
from the isochrone models by \citet{Siess2000} in the target age range
of this paper ($\le$11\,Myr).  Because this mass uncertainty is very
small, the distance uncertainties for the selection of IM stars does not
affect our results.  The above three points (or any other unconsidered
uncertainties) might mask the possible lifetime difference between IM
and LM stars. However, if we find any significant difference, it is
likely to be real and should be clearly seen with better selected
IM-star samples in the future. Note that contamination of higher-mass
stars can occur, but that should not affect the lifetime differences
between the disks of IM and LM stars.  We discuss the effect of IM star
selection on the derived IMDF in section 8.1.2.

\subsection{Selected samples}
\label{subsec:sample IM-stars}

We searched the literature to gather all of the available spectral type
information for the stars in the sample clusters. We then made a list of
all the IM stars by selecting cluster members 
by spectral type earlier than that of the lower-mass boundary and also
later than that of the higher-mass boundary.  The clusters chosen are
shown in column 1 and the references to the papers from which the IM
stars were selected is shown in column 2 of Table 3.
Following the fifth criterion in Section~2.1, we removed any target
clusters for which less than three IM stars can be identified.  Also,
because IMDFs for clusters with age of $>$5\,Myr are found to be
$\simeq$0\,per cent as discussed in the following sections, we obtained
IMDFs for only about 10 clusters.
Disk fractions for clusters with age of $\le$5\,Myr are the most useful
for studying stellar mass dependence of disk dispersal (cf.
\citealt{Kennedy2009}).

As a result, the total number of stars used for deriving JHK and MIR
IMDF become 799 and 365, respectively.
In Appendix A, the IM star samples for all clusters are summarized in
tables as well as in colour-colour diagrams.
For the following five clusters, the spectral type information for
lower-mass stars in the literatures is incomplete, and we could not
reach to the mass-limit of 1.5\,$M_\odot$: $\gamma$ Vel
(F5:~2\,$M_\odot$), $\lambda$ Ori (G0:~2\,$M_\odot$), Per OB2
(G8:1.8\,$M_\odot$), OB1bc (G3:~2.2\,$M_\odot$), and OB1a
(G6:1.7\,$M_\odot$).  Although it is desirable to set exactly the same
mass limit, such as 2\,$M_\odot$, we used 1.5\,$M_\odot$ as the lowest
mass for the other clusters in order to obtain as many IM stars as
possible.\footnote{We checked how much the IMDF changes with a mass
limit of 2\,$M_\odot$ and confirmed that the resultant IMDFs do not
change within the uncertainty.}

\section{{\it JHK} IMDF}
\label{sec:JHK IMDF}

The optical-NIR SED difference between stars with and without disks is
more prominent for IM stars than LM stars
\citep{{Lada1992},{Carpenter2006}}.  This is mainly because the stellar
SED for stars with higher masses peaks at the shorter wavelength side of
the $U$ and $B$ bands (e.g. $\sim$0.3\,$\mu$m for A0V stars, with mass
of $\sim$3\,$M_\odot$ and $T_{\rm eff}$ of 9790\,K; \citealt{Allen's
AQ}) compared to the disk SED that peaks near the $K$ band
($\gtrsim$2\,$\mu$m).  HAeBe stars also have a large infrared excess
from the optically thick disk `wall', which arises from the inner edge
of the disks and where dust disk is so hot as to evaporate
(\citealt{Natta2001}; \citealt*{Dullemond2001}).
Therefore, even in the case of using only {\it JHK} photometry, IM stars
with disks can be much more easily and more accurately selected than LM
stars with disks. Indeed, \citet{Hernandez2005} found an intrinsic
region for HAeBe stars on a {\it JHK} colour--colour diagram (see their
fig.~2).  Also, the photometric uncertainties for IM stars in nearby
clusters are very small, typically $\lesssim$0.02\,mag for all {\it
JHK}-bands.  We make use of these characteristics to derive the {\it
JHK} IMDF for each target cluster with the selected sample of IM stars.

\subsection{Identification of IM stars in the colour--colour diagram}
\label{subsec:IMDFdef_JHK}

On the {\it JHK} colour--colour diagram, the stars with disks are known
to be lying in the infrared-excess region that is separated from the
region of stars without disks \citep[e.g.][]{Lada1992}.  For low-mass
stars, disk fractions of various young clusters have been derived using
{\it JHK} colour--colour diagram
(\citealt{Lada1999}; \citealt{My2010ApJ}; see also
\citealt{Hillenbrand2005}).  For IM stars, \citet{Lada1992} showed that
HAeBe stars occupy completely separated regions even from those for
classical Be (CBe) stars based on the modelling of disk emission. CBe
stars show near-infrared excess from gaseous free-free emission and are
often confused with HAeBe stars, but the disk excess from HAeBe stars is
much larger.  After \citet{Hernandez2005} defined the locus of HAeBe
stars on the intrinsic {\it JHK} colour--colour diagram,
\citet*{Wolff2011} identified HAeBe stars of IC 1805 using their
definition.  \citet{Comeron2008a} defined the disk excess region for
HAeBe stars on the {\it JHK} colour--colour diagram (non-intrinsic) by
using a line that passes through $(H-K, J-H) = ( 0.11, 0)$ and is
parallel to the reddening vector as a border-line between the HAeBe
stars and CBe stars.

\begin{figure}
\begin{center}
\includegraphics[scale=0.5]{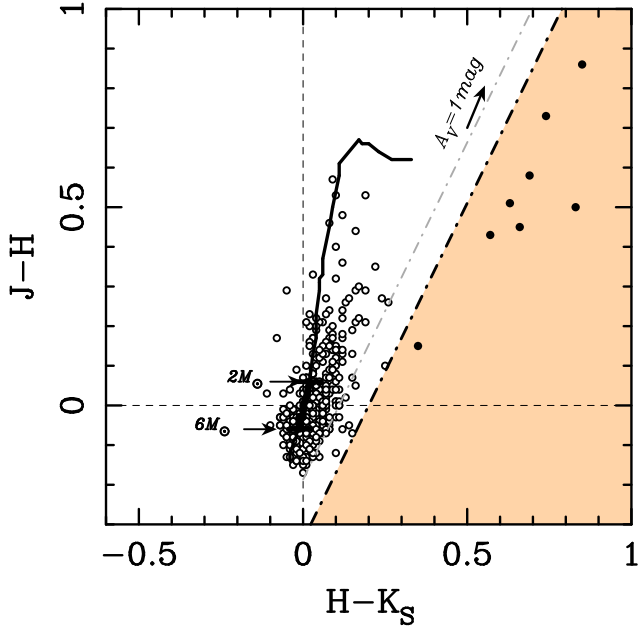}
\caption{{\it JHK} colour--colour diagram for IM stars.  The observed
colours of HAeBe stars (filled circles) and CBe stars (open circles) in
nearby clusters from \citet[][Upper Scorpius, Per OB2, Lac OB1, Ori
OB1a, and Ori OB1bc]{Hernandez2005} are shown with the dwarf track
\citep{Bessell1988} (black line).  The estimated border-line, which is
parallel to the reddening vector and distinguishes HAeBe stars from
other objects, is shown with a dot-dashed line: the gray dot-dashed line
shows the definition by \citet{Comeron2008a} while the black dot-dashed
line shows our definition.  The region to the right of the border-line
(orange color) is defined as the `IM disk excess region'.}
\label{fig:Her_all}.
\end{center}
\end{figure}

In Fig.~\ref{fig:Her_all}, we plot the {\it observed} colors of the
HAeBe stars (filled circles) and the CBe stars (open circles) 
for all the samples in \citet[][Upper Sco, Lac OB1, Ori OB 1a, Ori OB
1bc, and Tr 37]{Hernandez2005} on the {\it JHK} colour--colour diagram.
This shows that the HAeBe stars are spatially separated from the
main-sequence track \citep[][black line in the colour--colour
diagram]{Bessell1988}.  Note that the data points shown are not
corrected for reddening.  There is a clear division between HAeBe stars
and other objects, and a border-line can be set as the dot-dashed line,
which passes through the point of $(H-K_S, J-H) = ( 0.2, 0)$ and is
parallel to reddening vector (black arrow).  This border-line is
slightly shifted to the right compared to Comer{\'o}n et al.'s
border-line (Fig.~\ref{fig:Her_all}) to completely avoid contamination
from CBe stars.  Hereafter, we call the right-side region of the
border-line the `IM disk excess region' (orange shaded region).  This
border-line is more precise for dividing HAeBe stars and CBe stars than
that of \citet{Comeron2008a} (gray dot-dashed line in
Fig.~\ref{fig:Her_all}) because some CBe stars are included in the IM
disk excess region when using Comeron's border-line.  Therefore, we use
the line passing through $(H-K_S, J-H) = (0.2, 0)$ as the border-line,
and the {\it JHK} IMDF is defined to be the ratio of the stars located
in the IM disk excess region to the total number of stars in a cluster
that are selected with the criteria in
Section~\ref{subsec:selection_IMstars}.

Note that there is a well-known classification for HAeBe stars by
\citet{Meeus2001}, Group I for younger flared disk phase and Group II
for older flat disk phase. We confirmed that all stars in Group I and
II, except for one star (HD 135344 in Group I) out of 14 stars, are
recognized as HAeBe stars in our method.

\subsection{Determination of the IMDF}
\label{sec:IMDFderivation_JHK}

We used the 2MASS Point Source Catalog\footnote{
\url{http://www.ipac.caltech.edu/2mass/releases/allsky/doc/sec6_2.html}}
to obtain the {\it JHK} magnitudes of all the sample IM stars.  We
rejected all IM stars that do not have an `A' photometric quality flag
(signal-to-noise $\geq$10 for all {\it JHK} bands) in the 2MASS
catalogue.  We then obtained the IMDF of the IM stars from the {\it JHK}
colour--colour diagrams of each target cluster.

From previous studies of the disk fractions for low-mass stars, the
systematic errors of the disk fraction are known to be less than the
statistical errors when using data with small photometric uncertainties
(\citealt*{Liu2003}; \citealt{My2009ApJ}). The present data should be in
the same situation in view of the small uncertainties in {\it JHK}
photometry of the IM star samples.  For estimating the statistical
errors of the disk fraction, we assumed that the errors are dominated by
Poisson errors ($\sqrt{N_{\rm disk}}$), and we used $\sqrt{N_{\rm disk}}
/ N_{\rm all}$ for the one-sigma uncertainty of the disk fraction, where
$N_{\rm disk}$ is the number of stars with optically thick disks ($=$
HAeBe stars) and $N_{\rm all}$ is the number of all cluster members,
respectively.  However, if the number of HAeBe stars is zero, the
statistical error was calculated assuming one HAeBe star in the examined
target cluster to give a one-sigma uncertainty of $1 / N_{\rm all}$
\citep[e.g.][]{Hernandez2005}.  Table~\ref{tab:CL_list} summarizes the
derived {\it JHK} IMDFs for all the target clusters.

\section{MIR IMDF} \label{sec:MIR IMDF}

In the previous studies utilizing the data from the {\it Spitzer Space
Telescope}, the SED slope ($\alpha = d \ln \lambda F_\lambda / d \ln
\lambda$; \citealt*{Adams1987}) in the MIR wavelength range (3.6, 4.5,
5.8, and 8.0\,$\mu$m) is used for selecting disk-harbouring stars (e.g.
$\alpha \ge -2.0$; \citealt{Lada2006}; \citealt{Hernandez2007_671}).
The number of such IM stars should be precisely determined with this
method since disks show a large flux excess compared to the central star
continuum in the MIR.  For the derivation of the MIR IMDF, we made use
of the published {\it Spitzer} photometric results in the literature
because of the signal-to-noise and uniformity across target clusters.

For the definition of the MIR disk fraction, we followed the procedure
by \cite{Kennedy2009}, who derived $\alpha$ using the SED slope of {\it
Spitzer}'s Infrared Array Camera (IRAC) [3.6] to [8] and regarded those
with $\alpha > -2.2$ as cluster members with MIR dust disks.  We
estimated $\alpha$ of the IM stars only in the cases where reliable
photometry in all four IRAC bands is available.  However, for the
derivation of $\alpha$, we used only [3.6] and [8.0] because those two
bands determine $\alpha$ for almost all cases. For several clusters
(e.g. IC 348, which shows moderate extinction), we cross-checked our
$\alpha$ values with those in the literature \citep{Hernandez2008} and
confirmed that they are almost the same.  Following \cite{Kennedy2009},
we set the boundary at $\alpha=-2.2$ to separate all the IM stars into
the categories of `with disk' and `without disk'.

For the target clusters with published {\it Spitzer} data (13 clusters
out of 19 target clusters; see Table~\ref{tab:CL_list}), we estimated
$\alpha$ for the IM stars. Unfortunately, the MIR {\it Spitzer}
photometry of some IM stars in the nearby star-forming regions could not
be obtained because they are too bright for {\it Spitzer}.  Therefore,
the number of IM stars for the MIR IMDF is, in most cases, less than
those for the {\it JHK} IMDF (e.g. Tr 37).  In some cases, we have more
sample stars for the MIR than those for {\it JHK} (e.g. $\lambda$ Ori)
because some of the MIR stars do not have good {\it JHK} photometry with
2MASS.  In this case, we calculated the MIR IMDF by rationing the number
of stars with disks by the total number of stars in each MIR sample. The
results are summarized in Table~\ref{tab:CL_list}. The treatment of
uncertainty is similar to that for the {\it JHK} IMDF, as described in
Section\,\ref{sec:JHK IMDF}).

\section{Definition of the disk lifetime} 
\label{sec:def_time}

Different terms have been used for the disk dispersal time-scale,:
e.g. disk lifetime \citep{{Lada1999}, {Haisch2001ApJL},{Hernandez2008}},
disk decay time-scale \citep{Mamajek2009}, and disk dissipation
time-scale \citep{Fedele2010}.  These terms are based on the observed
cluster age--disk fraction plot with age on the horizontal axis and disk
fractions on the vertical axis.  However, these terms are not
consistently used.  Moreover, the value of disk fraction at zero age has
not been considered with care because these studies are performed mainly
for low-mass stars and all low-mass stars are thought to initially have
disks in the standard picture of low-mass star formation
\citep*{Shu1987}.  Since this may not be the case for IM stars, we
define these terms explicitly in this section.

\begin{figure}
\begin{center}
\includegraphics[scale=0.4]{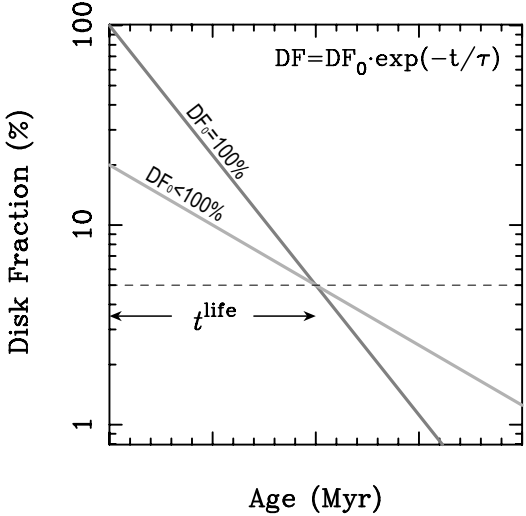}
\caption{Definitions of disk decay time-scale ($\tau$), disk
 lifetime ($t^{\rm life}$), and initial disk fraction (DF$_0$).}
\label{fig:define}
\end{center}
\end{figure}

To fit with a single function to the disk fraction evolution curve, an
exponential function is appropriate.  The {\it `disk decay time-scale'}
($\tau$) is defined as: ${\rm DF} [\%] \propto \exp (- t [{\rm Myr}]/
\tau)$ \citep[e.g. ][]{Mamajek2009}.  The decay time-scale is
proportional to the slope of the curve on a semilog plot, log
(IMDF)--age plot (Fig.~\ref{fig:define}).  On the other hand, the most
often used term {\it `disk lifetime'} ($t^{\rm life}$) is originally
defined as the $x$-intercept of the cluster age-disk fraction plotted as
a linear function.  However, fitting with a linear function does not
appear appropriate to describe the shape of disk fraction evolution,
which appears to decrease and level out at about 5--10\,per cent
\citep{Hernandez2008}.  Therefore, we define the disk lifetime to be the
time when the disk fraction is 5\,per cent ($t^{\rm life}$) and use this
for the discussion throughout this paper.

We define the {\it `initial disk fraction'} (DF$_0$) as the disk
fraction at $t=0$.  Fig.~\ref{fig:define} shows two possible cases:
DF$_0 = 100$\, per cent and DF$_0 < 100$\, per cent for the same $t^{\rm
life}$.  The value of DF$_0 = 100$\, per cent (the dark gray line in
Fig.~\ref{fig:define}) means that all stars initially have disks, while
that of DF$_0 < 100$\, per cent (the light gray line in
Fig.~\ref{fig:define}) means that all stars do not necessarily have
disks from the beginning or that some disks disappear quickly within a
very short time-scale that is not recognized within the accuracy of the
age determination.  Note that if DF$_0$ is constant, then the disk
lifetime is proportional to disk decay time-scale.

\section{Evolution of the $K$ disk} 
\label{sec:kdisk} 

The stellocentric distance of the $K$ disk ($r_K$) for HAeBe stars has a
wide range, $\sim$0.1--1.0\,AU for HAe stars to $\sim$1--10\,AU
for HBe stars \citep{Millan-Gabet2007}. However, because a large part of
the IM stars in this paper are HAe stars, $r_K$ of $\sim$0.3\,AU is
taken to be the nominal radius in this paper.  The {\it JHK} IMDF derived in
this paper is the fraction of the HAeBe stars whose disks at a
stellocentric distance ($r_K$) of $\sim$0.3\,AU are optically thick with a
temperature of $\sim$1500\,K (see e.g. fig.~2 in
\citealt{Millan-Gabet2007}).

In this section, we discuss the evolution of the $K$ disk of IM stars
traced by $K$ band excess emission and on the {\it JHK} IMDF change with
cluster age.

\subsection{Disk lifetime} 
\label{subsec:kdisk-lifetime} 

\begin{figure}
\begin{center}
\includegraphics[scale=0.5]{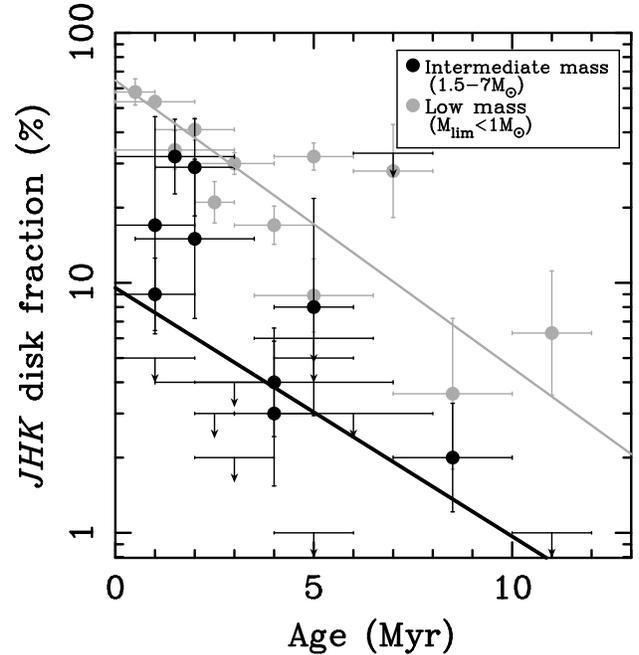}
\caption{{\it JHK} IMDF (black) and {\it JHK} LMDF (gray) of young
 clusters in the solar neighbourhood as a function of cluster age. For
 the IMDF, the data for clusters in Table~\ref{tab:CL_list} are shown
 with black filled circles, while upper-limits are shown with downward
 arrows.  The fitted curve with the survival analysis for all the
 clusters, including the upper limits is shown with the black line.  For
 comparison, the LMDF from \citet{{My2009ApJ},{My2010ApJ}} is shown with
 the gray line.}
\label{fig:IM_lowDFvsage}
\end{center}
\end{figure}

By making use of the method described in Section~\ref{sec:JHK IMDF}, the
{\it JHK} IMDFs of $\sim$20 clusters are derived for the first time, in
particular for clusters at ages $<$3\,Myr.  Fig.~\ref{fig:IM_lowDFvsage}
shows the derived IMDF as a function of ages (black filled circles).
The {\it JHK} IMDF is found to show an exponentially decreasing trend
with increasing cluster age as seen in previous studies.  There is a
large scatter with many upper-limits at 1--3\,Myr.  In view of the
upper-limit points, we used the astronomical survival analysis methods
(\citealt{Isobe86}; \citealt*{Lavalley92}) as a primary analysis tool.
We used the {\tt schmidttbin} task in the {\tt iraf/stsdas} package.
The resultant disk decay time-scale is 
 $\tau = 4.4 \pm 2.2$\,Myr with DF$_0 = 10\pm 4$ per cent. The resulting
 disk lifetime is $t^{\rm life}_{{\rm IM,} {\it JHK}} = 2.8\pm
 2.4$\,Myr.
The fitted curve is shown in Fig.~\ref{fig:IM_lowDFvsage} as a thick
black line.  We refer to this fitting as `survival fitting'.

The IMDF data points at ages $<$3\,Myr show a rather large scatter with
upper-limit points.  The clusters with zero disk fraction ($\rho$ Oph,
IC 348, $\sigma$ Ori, NGC 2264) do not appear to have obvious common
features.  As for the initial disk fraction (DF$_0$), the fitting
results show a low value of $<$20\,per cent.  Although some non-zero
data points at 1--2\,Myr are apparently above the fitted line, all of
them show relatively low values that are not more than $\sim$40\,per
cent.  Whether all high-mass stars initially have disks or not is still
under debate \citep[e.g.][]{Zinnecker2007}. On the other hand, all
low-mass stars, are thought to initially have disks in the standard
picture of low-mass star formation \citep{Shu1987}.  Our results suggest
a possible low initial disk fraction for IM stars, and further study of
the IMDF is important, in particular, those of the younger clusters.

\subsection{Comparison with low-mass stars} 
\label{subsec:kdisk-comparison}

In Fig.~\ref{fig:IM_lowDFvsage}, we show the {\it JHK} LMDF for
comparison. The LMDF data points are for 12 clusters in the solar
neighbourhood from \citet{{My2009ApJ}, {My2010ApJ}}.  The fitting result
of LMDFs is shown with a gray line (Fig.~\ref{fig:IM_lowDFvsage}), which
has {$\tau_{{\rm LM,} {\it JHK}} = 3.8 \pm 0.4$\,Myr and ${\rm LMDF}_{0,
{\it JHK}} = 64 \pm 6$\,per cent, leading to a disk lifetime of $t^{\rm
life}_{{\rm LM,} {\it JHK}} = 9.7 \pm 1.1$\,Myr}.
Fig. 3 clearly shows the difference in the disk fraction value, as well
as the disk lifetime difference, between the IM and LM stars.  The IMDFs
for older clusters (${\rm age} > 3$\,Myr) show systematically lower
values compared to the LMDF, as \cite{Hernandez2005} initially found for
5 clusters (Tr 37, Ori OB1bc, Upper Sco, Per OB2, and Ori OB1a).  We
increased the number of target clusters of age $>$3\,Myr to 10 and
confirmed this tendency.  However, while fig.~10 in \cite{Hernandez2005}
shows an IMDF curve with only a single zero disk fraction point (Per
OB2), our results in Table 3 show about half of the target clusters with
a zero disk fraction (Fig.~\ref{fig:IM_lowDFvsage}).  Because the number
of IM stars for each cluster is typically more than 20, a simple
stochastic effect due to a small number of stars is not likely to be the
reason for the many zero disk fraction points.

As a result of the fitting, the lifetime for the IM stars is found to be
significantly shorter than that for the LM stars.  The above results are
summarized in Table~\ref{tab:JHK_MIR_timescale}.  In the case of
fitting, including the upper limits (survival fitting), the estimated
lifetime ($t^{\rm life}_{{\rm IM,} {\it JHK}} = 2.8 \pm 2.4$\,Myr), is
much shorter than that of the LM stars ($t^{\rm life}_{{\rm LM} {\it
JHK}} = 9.7 \pm 1.1$\,Myr) by about 7\,Myr.
These results clearly show the existence of a stellar mass dependence for the 
lifetime of the innermost disk.

\subsection{Stellar mass dependence of the disk lifetime} 
\label{subsec:kdisk-massdependence}

The stellar mass dependence of the disk lifetime can be a strong
constraint on the disk dispersal mechanism and the theory of planet
formation, as discussed by Kennedy \& Kenyon (2009).  They compared the
disk fraction for different mass bins, $\sim$1\,$M_\odot$
(0.6--1.5\,$M_\odot$) and $\sim$3\,$M_\odot$ (1.5--7\,$M_\odot$), in
seven clusters and suggested that their data are more consistent with
$\tau_{\rm KK09} \propto M^{-1/2}_*$ than with $\propto M^{-1/4}_*$.
$\tau_{\rm KK09}$ is the disk decay time-scale defined by their model,
in which the disks are dispersed when the accretion rate drops below the
wind-loss rate.  However, only four clusters appear to be the main
contributors to the resultant mass dependency (see fig.~9 in
\citealt{Kennedy2009}) -- the H$\alpha$ disk fraction for three clusters
and the MIR disk fraction for one cluster.  Although
\cite{Hernandez2005} and \cite{Carpenter2006} found similar stellar mass
dependence for clusters with ages $>$3\,Myr, the dependence is uncertain
because of an insufficient number of clusters, in particular those with
ages $\leq$3\,Myr.  Obviously, it is necessary to increase the number of
data points to clarify the mass dependence.  Also, the large uncertainty
in previous studies might be the result of differences in the evolution
of the $K$ and MIR disks.  Thus studying of only the {\it JHK} disk (or
only the MIR disk) might show a clearer mass dependence.

The stellar mass dependence of the disk lifetime can be quantitatively
estimated by combining the time-scales for the two mass ranges.  For the
IMDF, stars with mass of 1.5--7\,$M_\odot$ are used in this paper.
Considering the larger number of lower-mass stars with the typical
universal IMF \citep[e.g.][]{Kroupa2002}, the characteristic mass is set
as 2--3\,$M_\odot$, or 2.5\,$M_\odot$.  We estimated the stellar mass
dependence with a characteristic mass from 2--3\,$M_\odot$, but no
significant difference was found within the uncertainties.  The
characteristic mass of 0.5$\pm$0.5\,$M_\odot$ (0.1--1\,$M_\odot$) for
the LMDF is set by considering the IMF and mass detection limit
($\sim$0.1\,$M_\odot$) for clusters used to derive disk fractions.
Assuming the stellar mass dependence of the disk lifetime as a power-law
function of stellar mass, we find $t^{\rm life}_{\it JHK} \propto
M^{-0.8 \pm 0.7}_*$ using the survival fitting.
These results are tabulated in Table~\ref{tab:JHK_MIR_timescale}.
Our result is consistent with the results by \citet{Kennedy2009}, who
found $\tau_{\rm KK09}$ is proportional to about $M^{-0.5}$.  However,
note again that our results are derived only from the $K$-disk data, while
\citet{Kennedy2009} used mostly data from the MIR disk or the H$\alpha$
gas disk.  We discuss the difference of disk lifetimes of the $K$ disk,
MIR disk, and gas accretion disk in \S~\ref{sec:DF_NIR_MIR}.

\section{Evolution of the MIR disk} 
\label{sec:mirdisk} 

In this section, we discuss the evolution of the inner disk of IM stars
traced by the MIR excess emission and using the results on the MIR IMDF
derived in Section~\ref{sec:MIR IMDF}.

\subsection{Disk lifetime} 
\label{subsec:mirdisk-lifetime}

\begin{figure}
\begin{center}
\includegraphics[scale=0.5]{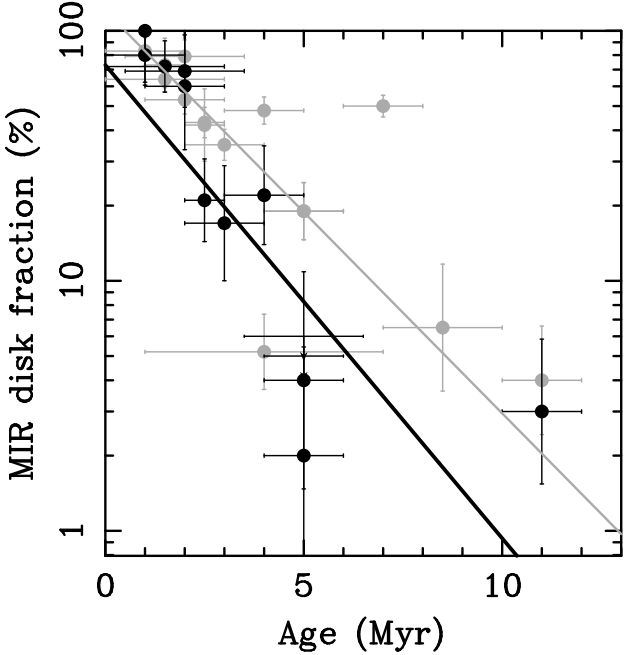}
\caption{MIR IMDF (black) and MIR LMDF (gray) of young clusters in the
 solar neighborhood as a function of cluster age. For the IMDF, the data
 for clusters in Table~\ref{tab:CL_list} are shown with black filled
 circles, while upper-limits are shown with downward arrows (only two
 clusters at 5\,Myr, NGC 2362 and $\gamma$ Vel).  The fitted curve using
 survival analysis for all the data including the upper limits is shown
 with the thick black line.  The fitting for the LMDF is shown with the
 gray line.}
\label{fig:IMDF+LMDFvsage}
\end{center}
\end{figure}

In Fig.~\ref{fig:IMDF+LMDFvsage}, we plot the MIR IMDFs, showing a
relatively clear exponential decay curve from cluster age zero to
$\sim$10\,Myr.  We performed fitting in the same way as for the $K$
disks (Section~\ref{sec:kdisk}).  By including the upper limits (two
clusters: NGC 2362 and $\gamma$ Vel), a survival analysis was performed
to obtain $\tau = 2.3 \pm 1.4$\,Myr and DF$_{\rm 0 (MIR)}$ of $73 \pm
57$\,per cent, which leads to a disk lifetime of $t^{\rm life}_{\rm IM,
MIR} = 6.1 \pm 4.2$\,Myr.
The resultant fitted line is shown with the black line in
Fig.~\ref{fig:IMDF+LMDFvsage}.  The fitted results are consistent with a
100 per cent initial disk fraction for the MIR IMDF.

\subsection{Comparison with low-mass stars} 
\label{subsec:mirdisk-comparison}

We also plot the MIR LMDF for comparison in
Fig.~\ref{fig:IMDF+LMDFvsage}.  The data points for MIR LMDFs are for 18
clusters from {\it Spitzer} observations: 16 clusters in
\citet{Roccatagliata2011}\footnote{For the LMDF, we excluded two
clusters from their 18 target clusters. First, NGC 2244 is excluded
because the detection limit is not given. Second, $\gamma$ Vel is
excluded because the completeness limit for this cluster does not reach
the low mass ($<$1\,$M_\odot$) limit.}  and Orion OB1bc and Orion
OB1a/25 Ori in \cite{Kennedy2009}.  We performed fitting in the same way
as in Section~\ref{sec:kdisk} to obtain $t^{\rm life}_{\rm LM, MIR} =
8.6 \pm 0.7$\,Myr and DF$_{0 {\rm (MIR)}} = 120 \pm 12$\,per cent with a
reduced $\chi^2$ value of 1.0 with a degree of freedom of 16. This
result is consistent with 100 per cent initial disk fraction for the MIR
LMDF.  There is no significant difference in the disk fraction lifetime
between the IMDF and LMDF disk, unlike for the $K$ disk in the previous
section.

\subsection{Stellar mass dependence of the disk lifetime} 
\label{subsec:mirdisk-massdependence}

The results of the lifetimes for the MIR disks of both IM and LM stars
are summarized in Table~\ref{tab:JHK_MIR_timescale}.  We derived the
stellar mass dependence of the MIR disk lifetime as {$t^{\rm life}
\propto M^{-0.2 \pm 0.3}_*$}, assuming a power-law function and using
the characteristic masses for the two mass ranges as for the {\it JHK}
disk lifetime (Section~\ref{subsec:kdisk-comparison}) and the results
for the survival fitting.  These results are tabulated in
Table~\ref{tab:JHK_MIR_timescale}.

Our results show no significant stellar mass dependence of the disk
lifetime, which is apparently inconsistent with \cite{Kennedy2009}, who
derived a steeper stellar mass dependence of $\tau_{\rm KK09} \propto
M_*^{-0.5}$.  However, note that their results are based on the lifetime
of both dust and gas disks.  The strong dependency appears to be mainly
contributed from the inclusion of H$\alpha$ gas disk.  They suggested a
$M_*^{-0.5}$ dependence rather than $M_*^{-0.25}$ dependence mostly
based on the data for three clusters (Taurus (H$\alpha$), Tr 37
(H$\alpha$ \& MIR), and OB1bc (H$\alpha$); see their fig.~9), but the
existence of the disks is based mostly on the H$\alpha$ gas disk for
those three clusters.
Using their data, we attempted to estimate the mass dependence and
confirmed that $\tau$ $\propto M_*^{-0.5}$ is obtained in the case of
using only the H$\alpha$ disk fraction for the eight clusters in their
list except for OB1a/25Ori, while $\tau$ $\propto M_*^{-0.2}$ is
obtained in the case of using only the MIR disk fraction for the same
eight clusters.  Therefore, we conclude that there is no mass dependence
of the lifetime of an MIR disk within the uncertainties.

\section{DIFFERENCE IN THE EVOLUTION OF $K$ AND MIR DISKS}
\label{sec:DF_NIR_MIR}

In the previous sections, we discussed the disk lifetime of the $K$
disk, which traces the innermost dust disk, and the MIR disk, which
traces the inner disk outside of the $K$ disk.
In this section, we compare the {\it K} and MIR disk fractions and
discuss the evolution of the $K$ disk and the MIR disk.  We also discuss
the relation of the MIR disk to the inner {\it gas} disk, which is
traced by accretion signatures, such as the H$\alpha$ emission line.

\begin{figure}
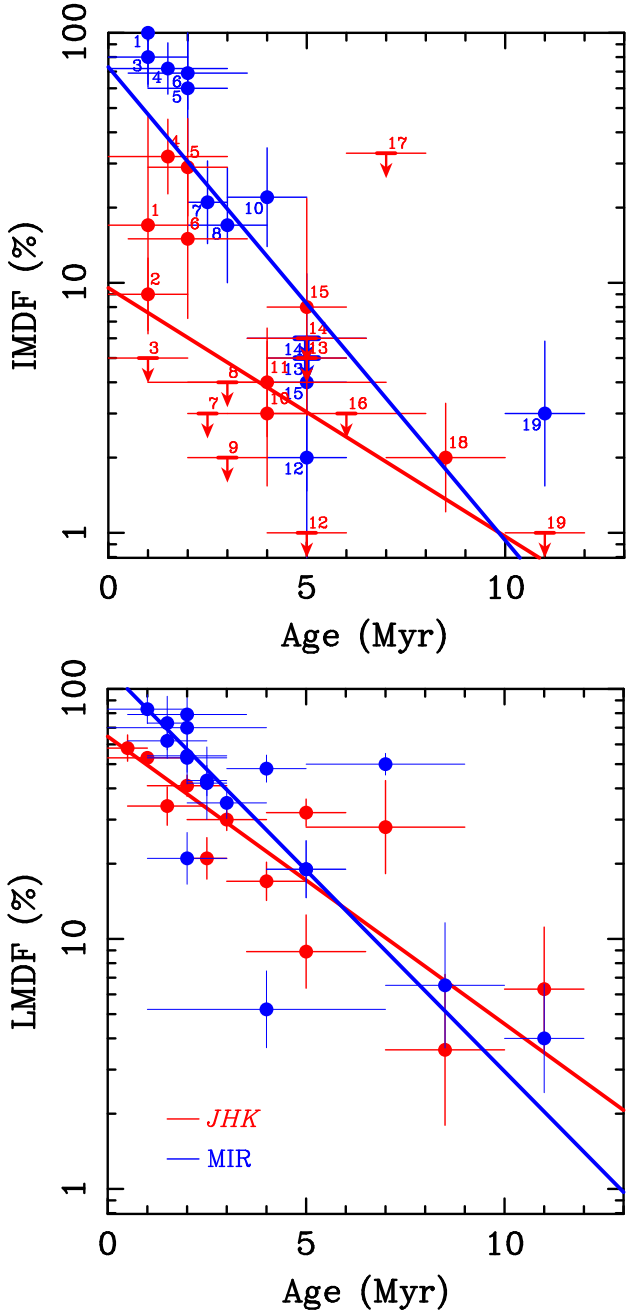

\begin{center}
\includegraphics[scale=0.5]{fig5L.eps}
\hspace{3em}
\includegraphics[scale=0.5]{fig5R.eps}
\caption{Comparison of {\it JHK} disk fraction (red) to MIR disk
fraction (blue) as a function of cluster age.  The left figure is for
intermediate-mass stars (IMDF), while the right figure is for low-mass
stars (LMDF).  For the IMDF, the red filled circles show the {\it JHK}
IMDF from Section~\ref{sec:kdisk} (Table~\ref{tab:CL_list}), while the
blue filled circles show the MIR IMDFs from the same table.  The arrows
show the upper limits.  
Both circles and arrows are labeled with the cluster numbers in Table
 1. The lines show the fits with survival analysis including the upper
 limits.  For the LMDF, red filled circles are from
 \citet{{My2009ApJ},{My2010ApJ}}, while blue filled circles are mainly
 from \citet{Roccatagliata2011} (see the text for the details).}
\label{fig:DF_NIR_MIR}
\end{center}
\end{figure}

\subsection{Comparison of the $K$ disk and the MIR disk} 
\label{subsec:JHK_MIR_IMDF}

\subsubsection{Low-mass stars}

Before discussing the case for the IM stars, we take a look at the case
for the LM stars as a reference.  Fig.~\ref{fig:DF_NIR_MIR} (right)
shows the comparison of the {\it JHK} LMDF (red) and the MIR LMDF
(blue).  The derived lifetime for the $K$ disk (9.7$\pm$1.1\,Myr) and
the MIR disk (8.6$\pm$0.7\,Myr) are identical within the uncertainties
(see Table~\ref{tab:JHK_MIR_timescale}), which suggests that the $K$
disk and the MIR disk disperse almost simultaneously in the disks of LM
stars.  This is consistent with the recent view of disk dispersal that
the {\it entire} disk disperses almost simultaneously for low-mass stars
($\Delta t \lesssim 0.5$\,Myr; \citealt{Andrews2005}).

\subsubsection{Intermediate-mass stars}

\begin{figure}
\begin{center}
\includegraphics[scale=0.5]{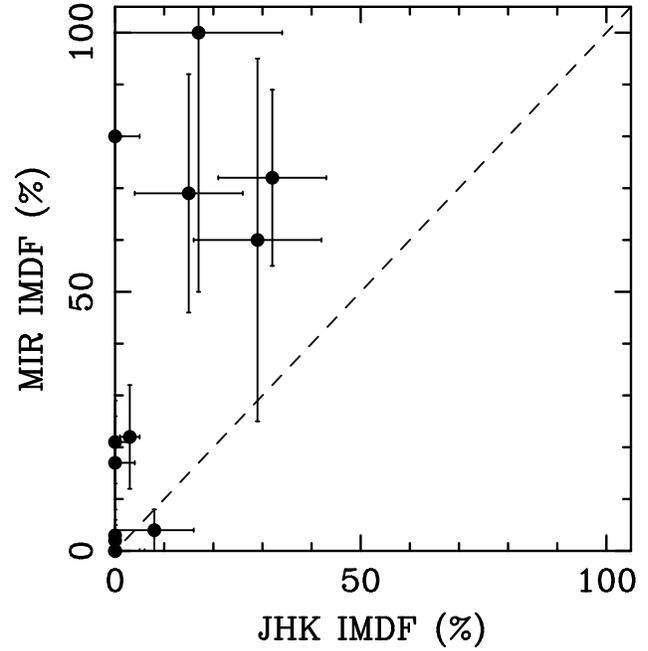}
\caption{Comparison of {\it JHK} IMDF to MIR IMDF
for 13 target clusters for which both {\it JHK} and MIR IMDFs are available
 (see Table \ref{tab:CL_list}).}
\label{fig:JHK_MIR_IMDF} 
\end{center}
\end{figure}

We compared {\it JHK} and the MIR IMDFs in Fig.~\ref{fig:DF_NIR_MIR}
(left).  The filled circles showing IMDFs and the arrows showing upper
limits are labeled with the cluster numbers in Table 1.
This figure immediately suggests that the MIR IMDFs are systematically
 larger than the {\it JHK} IMDFs. The MIR IMDF appears to be almost as
 high as 100\,per cent at $t\sim0$ and exponentially declines, while the
 {\it JHK} IMDF is less than 50\,per cent at the beginning and keeps
 smaller values than the MIR IMDF throughout the age span.  Because this
 offset might be due to an incomplete cluster sampling that favors only
 the higher MIR or the lower {\it JHK} IMDFs, we directly compared the
 MIR and {\it JHK} IMDFs for the clusters that have both fractions
 estimated.  The results (see Figure~\ref{fig:JHK_MIR_IMDF}) show that
 the MIR IMDFs are systematically larger than the {\it JHK} IMDFs for
 all the 13 clusters that have both.  We thus conclude that the large
 offset of the IMDFs is real, and that the smaller {\it JHK} IMDF is a
 unique property of the IM stars disk lifetimes compared to those of the
 LM stars.
The significantly lower disk fraction of the $K$ disks means they
disappear much {\it earlier} than the MIR disks.  The lifetime
difference is about 3\,Myr ($6.1 - 2.8 = 3.3$\,Myr; see
Table~\ref{tab:JHK_MIR_timescale}).

As suggested in section 2.2, possible contamination of LM stars in
selecting IM stars may affect the above discussion.  Therefore, it is
safer to set the lower limit mass for IM-stars as 2\,$M_\odot$, which is
slightly larger than the nominal mass limit in this paper
(1.5\,$M_\odot$).
With this lower limit mass, we derived the JHK/MIR IMDFs in the same way
as in section 2, 3, and 4.  As a result, the derived IMDFs do not
largely differ, and the estimated lifetimes of $K$- and MIR-disk are
$2.7 \pm 3.6$\,Myr and $6.0 \pm 6.1$\,Myr, respectively, which are very
close to the results for the lower mass limit of 1.5\,$M_\odot$ although
the uncertainties for both disk fractions and disk lifetimes become
larger.
Therefore, we conclude that the effect of possible contamination of LM
stars to our IM-star samples is very small, and that it does not change
the conclusion.

\begin{figure}
\begin{center}
\includegraphics[scale=0.5]{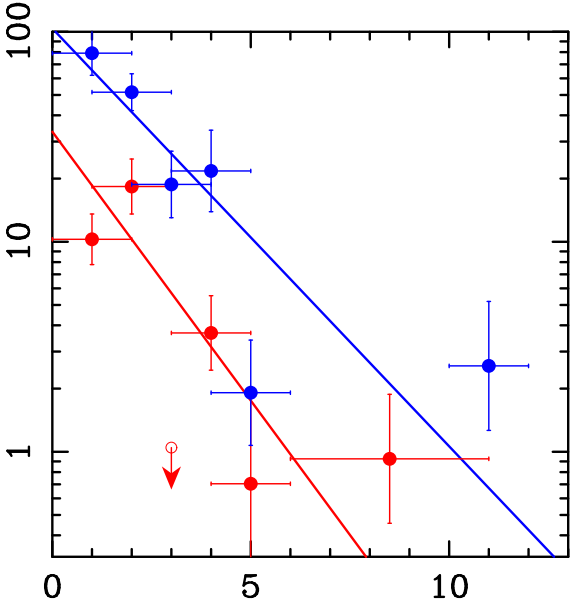}
\caption{Comparison of the {\it JHK} IMDF (red) to the MIR IMDF (blue)
as a function of age.  Same as the left figure of
Figure~\ref{fig:DF_NIR_MIR}, but the data are binned in the age axis
direction (see Section~\ref{subsec:JHK_MIR_IMDF} for details).  The
straight lines show the fits to the data points with the upper limit at
3\,Myr excluded from the fitting of {\it JHK} IMDF.}
\label{fig:DF_NIR_MIR_bin}.
\end{center}
\end{figure}

An alternative approach was tried to confirm these results. To increase
the statistical significance, we binned all of the disk-harbouring stars
and cluster members in the cluster age range from 1 to 5\,Myr with
1\,Myr bins and 1\,Myr steps and computed both the {\it JHK} and MIR
IMDFs.  To have enough clusters, all the members of the four clusters in
the 6 to 11\,Myr range are accumulated to estimate a binned IMDF at
8.5$\pm$2.5\,Myr for the {\it JHK} IMDF.  Because we have only one data
point at $t = 11$\,Myr for the MIR IMDF, we simply used it without
binning.  This `binning' process effectively reduces the number of
upper-limit points, and the disk fraction curve becomes clearer with
less scatter.

The results are plotted in Figure~\ref{fig:DF_NIR_MIR_bin}, which
 suggests the disk fraction offset between the {\it JHK} and the MIR
 IMDFs as well as the lifetime difference.  The fitting results for the
 {\it JHK} IMDFs are as follows: disk lifetime ($t^{\rm life}_{{\rm IM,}
 {\it JHK}}$) of 3.3$\pm$0.9\,Myr with DF$_{ 0 ({\it JHK})}$ of
 35$\pm$13 per cent with $\chi^2_\nu$ of 1.0 with degree of freedom of
 3.
The derived lifetime is, in fact, very close to the results without
binning (2.8\,Myr).  Note that the data point at $t=3$\,Myr was not used
because it remains an upper limit due to many upper limits in this age
bin. As for MIR IMDF, a disk lifetime ($t^{\rm life}_{\rm IM, MIR}$) of
6.7$\pm$1.1\,Myr with DF$_{\rm 0 (MIR)}$ of 104$\pm$26\,per cent with
$\chi^2_\nu$ of 2.2 with degree of freedom of 4 were obtained.
In summary, these `binning' fitting results (see
Table~\ref{tab:JHK_MIR_timescale}) confirm the survival analysis results
without binning although there are very few data points in the binning
fitting and we should be cautious of any unknown biases.
The stellar mass dependence for this binning analysis is also listed in
Table~\ref{tab:JHK_MIR_timescale} and is consistent with previous
results.  Therefore, we conclude that there is a lifetime difference of
$\sim$3--4\,Myr between {\it K} and MIR disks.

\subsection{Comparison with the submm disk}

To investigate further the dependence on stellocentric distance, we also
compared the disk fraction and lifetime of the K and MIR disks with that
of the outer cold ($\sim$10\,K) dust disk traced by the submm and mm
continuum. There are a number studies of submm observations of IM stars
in Taurus (1.5\,Myr), $\rho$ Oph (2\,Myr), and Upper Sco (5\,Myr)
\citep{{Andrews2005}, {Andrews2007}, {Mathews2012}}.  We confirmed that
the MIR disk is well correlated with the submm disk for the IM stars.
Out of the observed 20 B-, A-, F, and G-type stars in the above papers,
 19 stars are detected with MIR and submm disks, and only one star, HIP
 76310 in Upper Sco, lacks a MIR disk but has a submm disk. The strong
 correlation clearly suggests that the MIR inner disk and submm outer
 disk disperse almost simultaneously for the IM stars.
This behaviour is similar to that for LM stars \citep{{Andrews2005},
{Andrews2007}, {Mathews2012}}.  Thus the early disappearance of the
innermost $K$ disk again appears to be the only unique property of the
IM stars compared to the LM stars.

\subsection{Comparison with the H$\alpha$ gas disk}

\begin{figure}
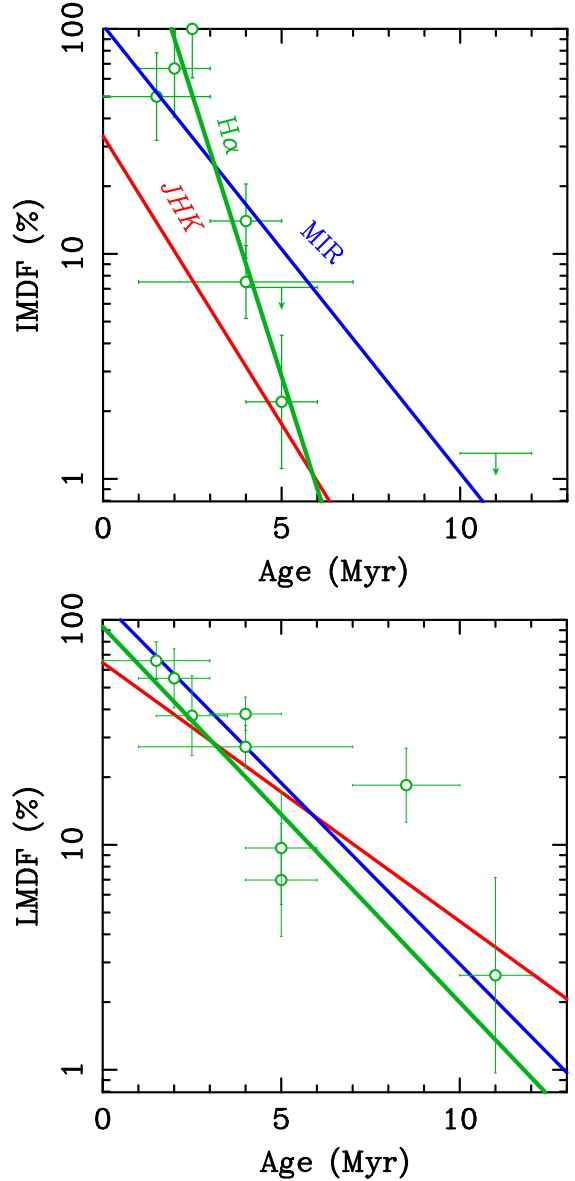

\begin{center}
\includegraphics[scale=0.45]{fig8L.eps}
\hspace{3em}
\includegraphics[scale=0.45]{fig8R.eps}
\caption{Comparison of disk fraction curves of the $K$ disk (red line),
MIR disk (blue line), and H$\alpha$ disk (green line).  For the {\it
JHK} and the MIR IMDFs, the binned fitting results are shown (see
Section~8.1.2). The H$\alpha$ disk fraction from \citet{Kennedy2009} is
plotted with open circles. The left figure is for the IMDF, while the
right figure is for the LMDF.}
\label{fig:JHK_Ha_MIRdf_age}
\end{center}
\end{figure}

Another question we investigated is how the dust disk evolution is
synchronized with the {\it gas} disk evolution. We compared the disk
fraction and lifetime of the {\it K} and MIR disks with those of the
innermost gas disk traced by the H$\alpha$ emission as has been
comprehensively studied by \citet{Fedele2010}. They used spectroscopy of
the H$\alpha$ emission for the clusters in the solar neighbourhood.
Because the H$\alpha$ emission was not observed for many IM stars, we
used the H$\alpha$ disk fractions from \citet{Kennedy2009} (their
1.5--7\,$M_\odot$ samples for IMDF and all mass range samples for LMDF)
and directly compared them with those of {\it K} and MIR disks.  This is
shown in Fig.~\ref{fig:JHK_Ha_MIRdf_age}.
The left panel includes eight clusters (Taurus, Cha I, IC 348, Tr 37, Ori
OB1bc, Upper Sco, NGC 2362, and NGC 7160), and the right panel includes
an additional cluster (OB1a/25Ori).

In the right panel of Fig.~\ref{fig:JHK_Ha_MIRdf_age}, the H$\alpha$
LMDF closely traces {\it JHK} and MIR LMDFs, and this shows the
co-evolution of the dust and gas disks for LM stars.  This is consistent
with the results of \citet{Fedele2010}, who found that the time-scale of
H$\alpha$ mass accretion is almost the same as that of the dust disk.
In the left figure, however, the H$\alpha$ IMDF shows a different
cluster age dependence compared to the IMDF of the dust disk.  We note
that (1) it overlaps the MIR IMDF at younger ages ($<$5\,Myr), and (2)
it is systematically larger than the {\it JHK} IMDF at younger ages with
a longer lifetime than the $K$ disk.  While the first point suggests the
co-evolution of the gas and dust disk, which is suggested by
\citet{Fedele2010}, the second point has not been noted before.  Only
the $K$ disk appears to have a unique cluster age dependence among the
different disk components of the IM stars.

\subsection{Long transition disk phase for IM stars}
\label{subsec:inner_to_outer}

In summary, for the IM stars there appears to be the following lifetime
sequence for the various stellocentric radii: $t^{\rm life}_K < t^{\rm
life}_{{\rm H}\alpha} \lesssim t^{\rm life}_{\rm MIR} \sim t^{\rm
life}_{\rm submm}$.
On the other hand, all these time-scales are nearly the same for the LM
stars \citep{{Andrews2005}, {Andrews2007}, {Mathews2012}}.  The above
result suggests that for the IM stars the $K$ disk has a shorter
time-scale and an evolutionary history that is different from that of
the LM stars.  The observed longer lifetime with larger stellocentric
distance is qualitatively consistent with the recent view of the disk
dispersal sequence for protoplanetary disks of LM stars
\citep{Williams2011}.  However, the lifetime difference between the {\it
K} and MIR disks for the IM stars ($\sim$3--4\,Myr) is significantly
longer than that suggested previously for low-mass stars ($\Delta t
\lesssim 0.5$\,Myr; e.g.  \citealt{Williams2011}).

That a time lag is clearly seen {\it only for the IM stars} gives us a
clue to the mechanism of disk evolution. The time-lag between $K$- and
MIR-disk lifetimes can be interpreted as a {\it transition disk phase},
in which the innermost $K$ disk disappears while the outer MIR disk
remains. Disks with no {\it JHK} excess emission and with MIR excess are
called `classical' transition disks \citep{Muzerolle2010}, while the
original definition of `transition disk` is a disk that has no or little
excess emission at $\lambda < 10$\,$\mu$m and a significant excess at
$\lambda \ge 10$\,$\mu$m \citep{{Strom1989},{Wolk1996}}.  The two
significant processes, disk dispersal (e.g. \citealt{Muzerolle2010}) and
planet formation (e.g.  \citealt{Calvet2002}), are thought to happen
during this phase.  Therefore, our finding suggests that such critical
evolutionary events can be clearly recognized in the transition disk
phase for IM stars as a time lag between the dispersal of the $K$ disk
and the dispersal of the MIR disk, while both events happen nearly
simultaneously for the LM stars ($\Delta t \lesssim 0.5$\,Myr).

\section{PHYSICAL MECHANISM OF THE DISK EVOLUTION OF INTERMEDIATE-MASS STARS}
\label{sec:physical}

In this section, we discuss the implications of the observed time-scales
of the gas/dust disks on the mechanism of disk evolution of the IM
stars.  Although there are many detailed processes related to disk
evolution, we focus on discussing the following two categories, which
are not intended to be comprehensive but broadly cover basic processes
related to disk evolution:
(1) the disk dispersal processes, such as mass accretion and dissipation
by photoevaporation, and  
(2) the dust settling to the disk midplane and dust growth, which could
be connected to planetesimal formation and planet formation.  We suggest
that the latter process is more likely for the early disappearance of
the $K$ disks.

Before discussing the detailed evolution mechanisms, we first remark on
the radial configuration of the dust disk in the steady state. 
If we consider an optically thick disk for disks with IR excess, the
radius ($R$) with a temperature ($T$) is given by $R = (L_* / 4 \pi T^4
\sigma)^{1/2}$, where $L_*$ is the stellar luminosity and $\sigma$ is
the Stefan-Boltzmann constant.
The dust temperature is about 1500\,K for the $K$ disk and $\sim$500\,K
for the MIR disk as inferred from the peak wavelength of the black body
emission. From those typical temperatures, the stellocentric distances
to those disks regions are estimated to be $r_K \sim 0.3$\,AU, $r_{\rm
MIR} \simeq 5$\,AU for IM stars (with the characteristic mass $M_* \sim
2.5$\,$M_\odot$) and $r_K \simeq 0.1$\,AU, $r_{\rm MIR} \simeq 1$\,AU
for LM stars (with the characteristic mass $M_* \sim 0.5$\,$M_\odot$)
(see Fig.~9), considering the effective temperatures of the central star
(see \citealt{Millan-Gabet2007}).
Because the radius $R$ of an optically thick disk with IR excess is
expressed with $R = (L_* / 4 \pi T^4 \sigma)^{1/2}$, $R$ is proportional
to $M_*^2$ with the mass-luminosity relation of $L_* \propto M_*^4$
\citep{Siess2000}.  Therefore, $r_K$ and $r_{\rm MIR}$ should be roughly
proportional to $M_*^2$.

\begin{figure*}
\begin{center}
 \includegraphics[scale=0.6]{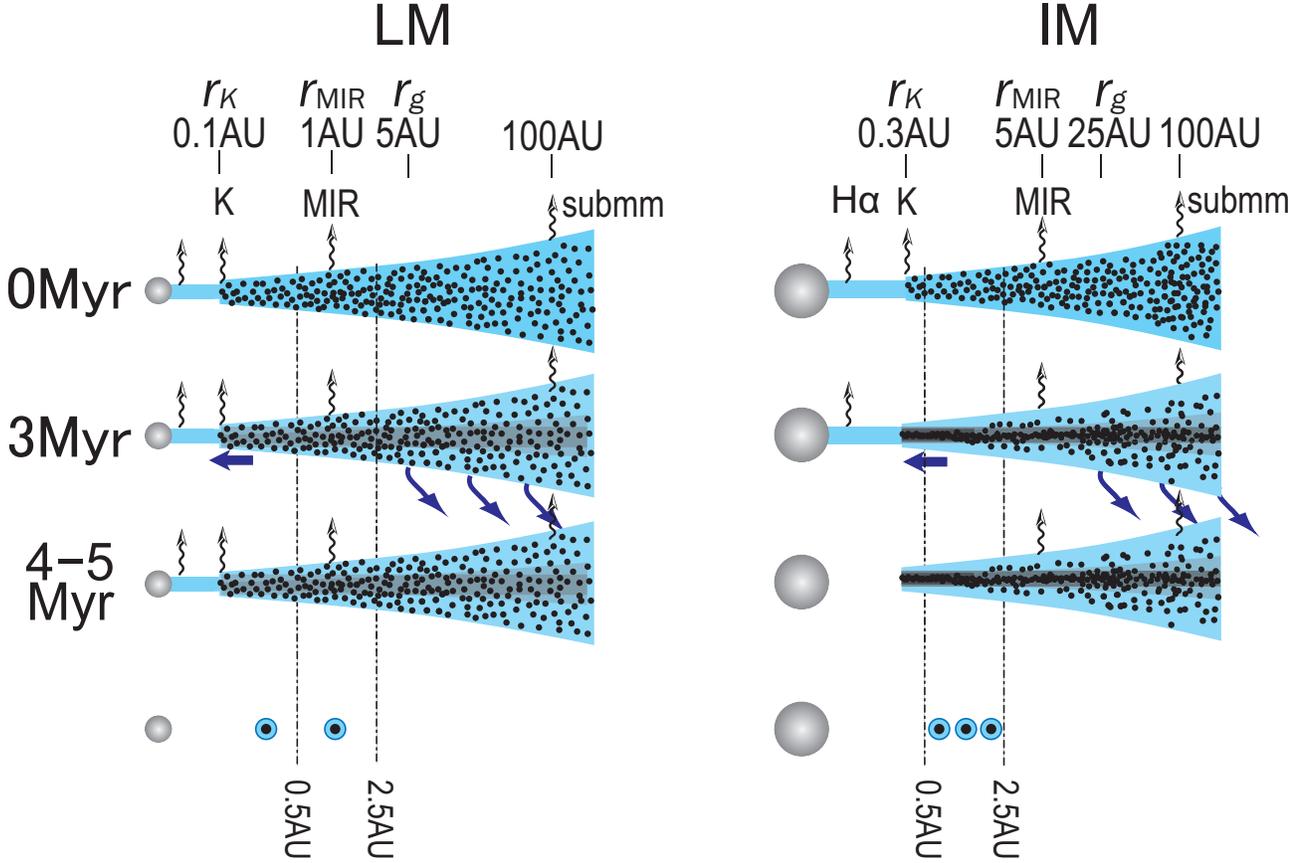}
\caption{Proposed disk evolution sequences for LM stars (left) and IM
 stars (right) as discussed in Section~\ref{sec:physical}.
 The radius in the horizontal direction is roughly shown with a
 logarithmic scale.  $r_K$, $r_{\rm MIR}$, and $r_g$ denote the K-disk
 radii, MIR-disk radii, and photoevaporation radii, respectively.
Blue arrows show the dispersal of gas/dust.  Black dots and the cyan
region show the dust and gas distribution, respectively, while
Jupiter-mass planets are shown by circles with cyan and black.  $K$,
MIR, submm, and H$\alpha$ emissions are shown by arrows with wavy lines.
After 4--5\,Myr, the entire gas/dust disk disperses before the dust
settling is completed for LM stars (left), while the entire gas/dust
disk disperses {\it after dust settling is completed in the $K$ disk}
for IM stars (right).  After the dispersal, the Jupiter-mass planets are
left (bottom).  It is known that a larger number of Jupiter-mass planets
are distributed at $<$2.5\,AU for IM stars than for LM stars).  The thin
vertical lines at $r= 0.5$\,AU show the inner region where close-in
planets are rarely found for IM stars (see Section~9.3). Note that this
is intended to describe the typical case and is not applicable to all
stars.}  \label{fig:disk_evolv}
\end{center}
\end{figure*}

\subsection{Disk dispersal?} 

The disk dispersal process consists of two kinds of processes: {\it mass
accretion} onto the central star and {\it dissipation} into interstellar
space \citep*{Hollenbach2000PPIV}.  The combination of mass accretion
and dissipation due to photoevaporation \citep[e.g.  the so called
`UV-switch model';][]{Alexander2008} is thought to be one of the major
mechanisms of overall disk dispersal \citep{Williams2011}, because this
can explain the almost simultaneous dispersal of the entire disk
($\Delta t \lesssim 0.5$\,Myr), and thus the short transition disk phase
as implied in Fig.~\ref{fig:JHK_Ha_MIRdf_age} (right). Although there
are a number of other proposed dispersal mechanisms (e.g. stellar
encounter, disk wind), our discussion here focuses on the dispersal due
to photoevaporation.

\subsubsection{Accretion onto the central star?} 

The first possible mechanism for the short $K$-disk lifetime of the IM
stars is the faster mass accretion onto the central star for higher-mass
stars. \citet{Mendigutia2011} suggested a very strong mass dependence of
the mass accretion rate from observations of UV Balmer excess. However,
our results suggest that the gas accretion disk has a longer lifetime
than the $K$ disk (Fig.~\ref{fig:JHK_Ha_MIRdf_age}, left), about equal to
that of the MIR disk. Therefore, more rapid accretion and the resultant
deficiency of material are not likely to be the cause of the faster
destruction of the $K$ disk.

\subsubsection{Photoevaporation?} 

Photoevaporation is another strong candidate for the dispersal mechanism
that may cause the short $K$-disk lifetime. 
Photoevaporation is known to be effective for outside of the
gravitational radius, $r_g$, where the thermal energy balances with
gravitational potential.
This radius scales with the stellar mass as $r_g \sim GM_* /c^2$
\citep{Alexander2008}, and $r_g$ for IM stars (2.5\,$M_\odot$) and LM
stars (0.5\,$M_\odot$) are $\sim$25\,AU and $\sim$5\,AU, respectively.
The corresponding $K$ disk radii ($r_K$) for IM and LM stars are only
$\sim$0.3 and $\sim$0.1\,AU, respectively, which are less than $1/50$ of
$r_g$.
Similarly the corresponding MIR disk radii are $\sim$5 and $\sim$1\,AU,
respectively. 
Although all these radii change with the stellar mass (from 1.5 to
7\,$M_\odot$ for IM stars), the relative magnitude of the radii, {\Large
$r_K < r_{\rm MIR} < r_g$,} should not change for both IM and LM star
mass ranges, even considering the scaling with the stellar mass
mentioned above.
Therefore, photoevaporation is not likely the main cause of the fast
$K$-disk dispersal for IM stars.

\subsection{Dust settling, dust growth, and planet formation}

The transition disk phase is now interpreted as the most important phase
 in the standard planet formation scenario, and now much observational
 effort has been put into characterizing this phase
 \citep{Williams2011}.  In this interpretation, the early disappearance
 of the innermost dust disk compared to other portions of disk is due to
 dust settling to disk mid-plane \citep{{Kenyon1987},{Dullemond2005}}
 and/or dust growth \citep{{Weidenschilling1997}, {Dullemond2004}}.  We
 discuss those possibilities in the following section with a schematic
 picture shown in Fig.~9.

\subsubsection{Dust settling \& growth}

From the basic equations of protoplanetary disks in equilibrium, the
radial dependence of the dust settling time can be analytically shown to
be proportional to the Kepler rotation period, which is proportional to
$r^{3/2} M_*^{-1/2}$ \citep{Nakagawa1981}.
Although there are many new simulations incorporating more physical
processes to show the dust settling time with a different r- or
M$_*$-dependence (e.g., \citealt{Tanaka2005}), we compare our results
with this base relationship by \citet{Nakagawa1981} as an initial
consistency check.

 First, as for the radius $r$ dependence, the shorter $K$-disk lifetime
than that of MIR-disk for the IM stars (Fig.~9, right,
Table~\ref{tab:JHK_MIR_timescale}) is qualitatively consistent with the
base relationship in that the dust settling/growth is occurring more
effectively in the inner disk. Although the observed results
{($t^{\rm life} \sim 3$\,Myr at $r_K \sim 0.3$\,AU, and $t^{\rm life}
\sim 6$--7\,Myr at $r_{\rm MIR} \sim 5$\,AU)} do not quantitatively
follow the base $r^{3/2}$ relation and instead show a much weaker
$r$-dependence,
this can be interpreted as that the turbulent process or some other
processes that prevent the dust settling/growth have the opposite
$r$-dependence to reduce the $r$-dependence of dust settling/growth 
(e.g., \citealt{Dullemond2005}). 
For the LM stars, on the other hand, the lifetimes of the inner $K$,
MIR, and the outer submm disks do not show any significant difference
(see discussions in Section~8.1.1 and ~8.2).
This is even more inconsistent with the base $r^{3/2}$ relation than the
IM stars. Most likely this means that the disk disperses before dust is
totally settled in the entire disk, although dust settling/growth is
reported for some LM stars (e.g., \citealt{Pinte2008}). In this case,
the dust in the upper disk layer is dissipated in the process of mass
accretion and photoevaporation (Fig.~9, left), and the disk lifetime
($\sim$9--10\,Myr) sets the lower limit of the dust setting/growth
time-scale for most of the LM srars.
However, it should be noted that the above discussion is intended to
describe the typical case and is not applicable to all stars.
The time scale of transition disk is still under debate, short
($\sim$0.2\,Myr; e.g. Luhman et al. 2010) or long ($\sim$ a few Myr;
e.g. Sicilia- Aguilar \& Currie 2011), although the discrepancies among
these studies are largely due to differing definitions of the transition
disk and how to estimate the total disk lifetime \citep{Espaillat2014}.

Next, as for the stellar mass $M_*$ dependence, the much shorter
$K$-disk lifetime of the IM stars than that of the LM stars ({$t^{\rm
life}_{{\rm IM}, {\it JHK}}\sim 3$\,Myr, while $t^{\rm life}_{{\rm LM},
{\it JHK}} \sim 9$--10\,Myr;} see Table~\ref{tab:JHK_MIR_timescale};
Fig. 5, left) is apparently consistent with the base relation in that
dust settling/growth occurring effectively for higher-mass stars.
However, if we also consider the $r$-dependence, the characteristic IM
stars ($M_*=2.5$\,$M_\odot$, $r_K=0.3$\,AU) are expected to have longer
settling time-scale (about twice) than the characteristic LM stars
($M_*=0.5$\,$M_\odot$, $r_K=0.1$\,AU) which shows the opposite tendency
compared to the observed timescales.
This might suggest that turbulence of the innermost disk is much weaker
for the IM stars than for the LM stars so that the dust growth/settling
occurs quickly. Although the larger disk mass (surface density) for
stars with higher mass \citep{Andrews2005} might cause such a situation,
the physical process is unknown.

\subsubsection{Planet formation}
\label{subsubsec:Planet Formation?}

Planetesimal and planet formation result from the dust settling/growth
processes according to the standard core-accretion model
\citep{Lissauer_PPV}.  For IM stars, the quick dust settling/growth
processes {\it in the presence of gas} may cause effective planetesimal
formation \citep{Hubickyj2005}.  This results in effective Jupiter-mass
planet formation
\citep[e.g. ][]{{Laughlin2004},{Ida2004a},{Robinson2006}}, which could
accelerate the disappearance of the innermost disk with clearing by
migration (\citealp*{Lin1996}; \citet{Trilling1998};
\citet{Trilling2002}).  Such a scenario is consistent with the trend of
a higher probability of Jupiter-mass planets with a larger stellar mass
for stars in the mass range of 0.2--1.9\,$M_\odot$ for semimajor axes of
$<$2.5\,AU \citep{Johnson2010}.  However, the mass range for IM stars
(1.5--7\,$M_\odot$) has only a small overlap with this trend.  This
trend is generally interpreted as a result of larger disk mass (high
surface density) for larger stellar mass, which enables the rapid
formation of Jupiter-mass planets (e.g. within 1\,Myr;
\citealt{Ida2004}).

\subsubsection{Summary}

In summary, dust settling/growth (and some planet formation) can 
generally explain the shorter $K$-disk lifetime of IM stars, although the
specific physical processes are not known. This interpretation is
summarized in the schematic pictures shown in Fig.~9 (right):
(1) The $K$ disk ($r_K \sim 0.3$\,AU):
Dust settling/growth works very efficiently from the beginning of disk
evolution (cluster age $=$ 0) and is almost completed in $\sim$3\,Myr. 
Because there is no left-over IR-emitting
grains even in the upper disk layer, no NIR continuum is emitted,
(2) The MIR disk ($r \sim 5$\,AU): A significant amount of dust grains
are in the upper disk layer due to the turbulence and give rise to the
MIR continuum emission.  After $\sim$4--5\,Myr, dust settling has
occurred, or dust in the upper layer of the MIR disk is dissipated,
resulting in no emission of MIR--thermal continuum emission.  If this
picture is correct, the lifetime difference of {\it JHK} and MIR IMDFs
constrains the time-scale of this settling process in the $K$ disk to
about $\sim$4\,Myr (Section~8.1.2) (Table~4).  The low initial value of
the {\it JHK} IMDF ($\sim$50\,\%) might also be naturally explained with
the effective settling in the inner disk.  Future MIR spectroscopy of
the silicate emission lines and the SED slope \citep{Furlan2006} of
those IM stars with and without the $K$ disk will test the idea that the
disappearance of the $K$ disk is due to dust settling/growth.

\section{Implications for Planet Formation around IM
 Stars}\label{sec:planet} 

Two remarkable trends are known for the Jupiter-mass planets 
around IM stars:
(1) the lack of close-in planets with semimajor axes of
$\lesssim${0.5}\,AU orbiting stars with masses $M > 1.5$\,$M_\odot$ 
(such planets are common for stars with $M_* < 1.2$\,$M_\odot$),  and
(2) the higher probability of having planets with semimajor axes of
$<$2.5\,AU compared to low-mass stars.  In this section, we discuss
these trends in the context of our disk fraction lifetime results.

\subsection{Implications for the lack of close-in planets}

There appears to be a lack of close-in planets with semimajor axes of
$\lesssim$0.5\,AU orbiting stars with masses of 1.5--3\,$M_\odot$ in
planet-search surveys, while close-in planets are more frequent for
lower-mass stars \citep{Wright2009ApJ}.  Because planets are thought to
form in situ or migrate inward in the formation phase
\citep[e.g. ][]{Lin1996}, our suggestion of rapid planet formation in
the $K$ disk appears to be inconsistent with the paucity of close-in
planets.  However, considering the possible radial range of `$K$ disk'
(from $\sim$0.3\,AU to 1\,AU, depending on the mass of the central star,
disk mass, etc.), the higher planet occurrence for higher-mass stars
\citep{Johnson2010} may reflect the rapid planet formation at $r \gtrsim
0.5$\,AU.  The planets that formed at $r \lesssim 0.5$\,AU may have
dropped into the central stars due to migration \citep{Papaloizou2007}
because the gas disk traced by H$\alpha$ still remains for about 2\,Myr
after the disappearance of the $K$ disk.  Or, they may have disappeared
due to collisional destruction that may have effectively occurred along
with grain growth \citep{Johansen2008}.  In any case more studies are
necessary to understand the precise relation between disk lifetime and
planet formation.

Regarding the lack of close-in planets for IM stars, two major scenarios
have been proposed.  The first scenario is planet engulfment caused by
the stellar evolution of primary stars in the RGB phase
\citep{Villaver2009}.  Another scenario is that the observed differences
in orbital distribution are primordial, and they are a consequence of
the planet-formation mechanism around the more-massive stars
\citep{Currie2009}.  In this section, we focus on the latter scenario
because our results are relevant to the early stages of star formation.
\citet{Currie2009} suggests that planets around IM stars cannot migrate
to inner orbits because of the shorter gas-disk lifetime for IM stars.
In addition, \citet{Kretke2009} suggest that the inner edge of the dead
zone in protoplanetary disk, where the dead zone is the region of the
disk without magnetorotational instability \citep{Gammie1996},
effectively determines the semimajor axes of giant planets because the
dead zone traps inwardly migrating solid bodies.  Thus, they suggested
that the larger radius of the inner edge for higher mass stars explains
the lack of close-in planets.

Our results are qualitatively consistent with Currie's scenario in that
shorter disk lifetimes are expected for higher-mass stars.  We estimated
that the stellar mass dependence of gas disk lifetime $t^{\rm
life}_{{\rm H}\alpha}$ is about $M_*^{-0.5}$ ($t^{\rm life}_{{\rm IM,
H}\alpha} \sim 5$\,Myr and $t^{\rm life}_{{\rm LM, H}\alpha} \sim
10$\,Myr in Fig.~8).
However, this dependence is not as steep as assumed in
\citet{Currie2009}, $t^{\rm life}_{\rm gas} \propto M_*^{-\beta}$ with
$\beta=0.75$--1.5.  In any case, migration alone may not be able to
explain the observed sharp outward step in giant planet orbits as
pointed out in \citet{Kennedy2009}.

Our results that the innermost disks of IM stars disappear at a very
early time also seem to be consistent with Kretke's dead zone model.
\citet{Kretke2009} assumed a smooth stellar mass dependence of the inner
edge radius of the dead zone (proportional to $M_*$) from the
theoretical relationship between the radius and the mass accretion, and
they compared this to the mass accretion rate derived from observations.
Our results, showing the early disappearance of the innermost $K$ disk, 
suggest that the radius becomes even larger because of low opacity,  
which makes the formation of dead zone difficult. 
If the critical stellar mass where the time lag between the $K$ disk and
the MIR disk disk dispersal is observationally determined, then this
dead zone idea may be able to explain the lack of close-in planets even
for the sharp cut-off at 0.5 AU in the planet semimajor axes in the
distribution of Jupiter-mass planets.

In addition, the difference in the planet formation site in disks of IM
stars and LM stars may explain the lack of close-in planets for the IM
stars.  Planets are thought to form outward of the snow line,
$\sim$3\,AU for LM stars and $\sim$10\,AU for IM stars
\citep{Kennedy2008}.  This difference might explain the observed
difference in planet location even after the smearing out by the
migration processes, although this idea does not explain the sharp step
in planet semimajor axes.

\subsection{Implications for higher planet formation probability}

The probability of IM stars having Jupiter-mass planets is found to be
proportional to $M_*^{1.0}$ for semimajor axes $<$2.5\,$M_\odot$
\citep{Johnson2010}.  This observed frequency is likely to be determined
by two competing effects:
the tendency of shorter disk lifetimes for more massive stars reduces
the likelihood of giant planets forming \citep{Butler2006}, and the
tendency of higher disk masses for more massive stars increases the
probability of gas-giant planet formation \citep*{Wyatt2007}.
In \citet{Kennedy2009}, the stellar mass dependence of disk lifetime is
estimated using the H$\alpha$ disk and MIR disk fractions as $\tau_{\rm
KK09} \propto M^{-1/2}_*$, where $\tau_{\rm KK09}$ is the disk decay
time-scale defined by their model.
However, from our results, the disk lifetime at $r \gtrsim r_g$\, for IM
and LM stars is not significantly different, and the stellar mass
dependence of disk lifetime ($t^{\rm life}_{\rm MIR}$) is as small as
$M_*^{-0.2}$. 
The disk mass is known from submm observations to be roughly
proportional to the stellar mass \citep{Andrews2005}.
The stellar mass dependence of the disk lifetime is negative, while the
stellar mass dependence of the disk mass is positive.
Therefore, the higher probability of IM stars having planets compared to
LM stars seems to be due to the difference in disk mass instead of the
difference in disk lifetime.

\section{Conclusion}\label{sec:conclusion}

We derived and compiled protoplanetary disk fractions of
intermediate-mass stars (1.5--7\,$M_\odot$) for a large number of nearby
young clusters (within $3$\,kpc and $\lesssim$5\,Myr old) with the
available {\it JHK} photometric data in the literature.
From the results and by comparing them with those for other wavelengths
(H$\alpha$, MIR, and submm), we found the following results:

\begin{itemize}

\item The $K$-disk lifetime of IM stars ($t^{\rm life}_{\it JHK}$),
      which is defined as the time-scale of disk fraction to bottom out
      at 5\,per cent, is estimated to be 3.3$\pm$0.9\,Myr.

 \item The $K$-disk lifetime for the IM stars, $t^{\rm life}_{\it JHK}$,
      is about one-third of that for the LM stars. Assuming a power-law
      dependence, the stellar mass dependence of the $K$-disk lifetime
      is found to be proportional to $M_*^{-0.7\pm0.3}$.

 \item By comparing the $K$ disk ($r \sim 0.3$\,AU) evolution to that of
      the MIR disk ($r \sim 5$\,AU) for IM stars, we find that the $K$
      disk seems to disperse earlier than the MIR disk by
      $\sim$3--4\,Myr.
       Because the $K$ disk and the MIR disk disperse almost
       simultaneously in LM stars ($\Delta t \lesssim 0.5$\,Myr), the
       long time lag may be a characteristic of IM stars, suggesting that
       the transition disk is the common phase in IM stars.

\item Because the disk time-scale at $r \gtrsim r_{\rm MIR}$ for the IM
      stars does not seem to be significantly different from that of LM
      stars, the most likely cause for the time lag seems to be early
      dust growth/settling and/or Jupiter-mass planet formation in the
      innermost disk ($K$ disk) in IM stars.

 \item Our results for the $K$ disk of the IM stars suggest the possible
      reasons for the paucity of close-in planets around IM stars, but
      they are not conclusive.  Our results also suggest that the disk
      mass is a more important factor for the stellar mass dependence of
      planet occurrence than the disk lifetime.
\end{itemize}




\begin{table*}
 \centering
 \begin{minipage}{18cm}
  \caption{Summary of target clusters.\label{tab:clusters} }
  \begin{tabular}{lllll}
\hline
& Cluster & Age$^a$ & Distance$^b$ & References for the disk fraction
   study$^c$ \\ 
& & (Myr) & (pc) & \\
\hline
1 & NGC 1333 & 1$\pm$1 (He08) & 318 (LL03) & He08, Ma09, Ro11\\ 
2 & Trapezium & 1$\pm$1 (Mu02) & 450 (LL03) & Ha01, He08, Ma09 \\ 
3 & $\rho$ Oph & 1$\pm$1 (Fe10) & 125 (LL03) & Fe10 \\
4 & Taurus & 1.5$\pm$1.5 (He08) & 140 (El78) & 
 Ha01, He08, Ke09, Ma09, Fe10, Ro11 \\  
5 & Cha I & 2$\pm$1 (Ro11) & 170 (Lu08) & Ha01, He08, Ke09, Ma09, Fe10, Ro11 \\
6 & NGC 2068/71 & 2$\pm$1.5 (Ro11) & 400 (LL03) & He08, Ma09, Ro11 \\ 
7 & IC 348 & 2.5$\pm$0.5 (He08) & 320 (Ha01a) & Ha01, He08, Ke09, Ma09,
Fe10, Ro11 \\ 
8 & $\sigma$ Ori & 3$\pm$1 (Ro11) & 440 (He07a) &  He08,  Ma09, Fe10, Ro11 \\  
9 & NGC 2264 & 3$\pm$1 (He08) & 760 (Da08a) & Ha01, He08, Ma09 \\ 
10 & Tr 37 & 4$\pm$1 (He08) & 900 (Si05) & He08, Ke09, Ma09, Ro11 \\
11 & Ori OB1bc & 4$\pm$3 (He05) & 443  (He05) & He05, He08, Ke09, Ma09,
Ga09 \\ 
12 & Upper Sco & 5$\pm$1 (Pr02) & 144 (He05) & He05, He08, Ke09, Ma09, Ga09,
 Fe10, Ro11 \\ 
13 & NGC 2362 & 5$\pm$1 (He08) & 1500 (Da08b) & Ha01, He08, Ke09, Ma09, Fe10,
 Ro11 \\ 
14 & $\gamma$ Vel & 5$\pm$1.5 (He08) & 350 (He08) 
& He08, Ma09, Ro11 \\ 
15 & $\lambda$ Ori & 5$\pm$1 (He08) & 450 (He09) & He08, Ma09 \\  
16 & Per OB2 & 6$\pm$2 (He05) & 320 (He05) & He05 \\
17 & $\eta$ Cham & 7$\pm$1 (He08) & 100 (Ma99) & He08,Ma09,
  Ga09, Fe10, Ro11 \\  
18 & Ori OB1a & 8.5$\pm1.5$ (Br05) & 330 (He05) & He05, He08, Ke09,
Ma09, Ga09 \\
19 & NGC 7160 & 11$\pm$1 (He08) & 900 (Si05) & He08, Ke09, Ma09, Ro11 \\ 
\hline
\end{tabular}

\medskip
{\bf Notes:}\\
$^{\rm a}$Adopted age with reference in parenthesis.
$^{\rm b}$Distance with reference in parenthesis.
$^{\rm c}${Literatures for disk fraction study in the past. Note that some
  references show different cluster names (e.g., 25 Ori, which is named
  as Ori OB1a in our list).} \\
References: Br05: \citet{Briceno2005};
Da08a: \citet{Dahm2008hsfa};
Da08b: \citet{Dahm2008hsfb};
El78: \citet{Elias1978};
Fe10: \citet{Fedele2010};
Ga09: \citet{Gaspar2009}; 
Ha01a: \citet{Haisch2001ApJL}; 
Ha01b: \citet{Haisch2001a};
He05: \citet{Hernandez2005};
He07a: \citet{Hernandez2007_662};
He08: \citet{Hernandez2008};
He09: \citet{Hernandez2009};
Ke09: \citet{Kennedy2009};  
LL03: \citet{LadaLada2003};
Lu08: \citet{Luhman2008hsf};
Ma09: \citet{Mamajek2009};  
Ma99: \citet{Mamajek1999};
Mu02: \citet{Muench2002};
Pr02: \citet{Preibisch2002};
Ro11: \citet{Roccatagliata2011};
Si05: \citet{Sicilia-Aguilar2005ApJ130}. 
\end{minipage}
\end{table*}


\begin{table*}
 \centering
 \begin{minipage}{18cm}
\caption{Adopted spectral type for the boundary masses of
  intermediate-mass stars. 
\label{tab:age}}
  \begin{tabular}{lllllll}
\hline
 $t^a$ & \multicolumn{2}{c}{SpType$^b$} & \multicolumn{2}{c}{Boundary
 mass with $\Delta$$t^c$}\\ 
\cline{2-3} \cline{4-5}
 (Myr) & 7\,$M_\odot$ & 1.5\,$M_\odot$ & 7\,$M_\odot$ & 1.5\,$M_\odot$ \\ 
  & &  &  ($\Delta t = \pm 2$\,Myr) & ($\Delta t = \pm2$\,Myr) \\
\hline
1 & B2.5 & K5 (1.5--1.6\,$M_\odot$) & $\gtrsim$7\,$M_\odot$ &
 1.2--1.5\,$M_\odot$\\ 
1.5& B3& K5 (1.2--1.5) & $\gtrsim$7\,$M_\odot$ & 1.2--$<$2.2\,$M_\odot$\\
2 & B3 & K5 (1.2--1.5) &7 & 1.2--$<$2.2\,$M_\odot$\\
2.5 & B3& K5 (1.2--1.5) & 6--7 & 1.2--1.8\,$M_\odot$\\
3 & B3 & K4 (1.5--1.6) & $>6$--7 &1.2--1.6\,$M_\odot$\\
4 & B3 & K4 & 7 & 1.4--1.8\,$M_\odot$\\
5 & B3 & K4 (1.4--1.5) & 7 & $>$1.4--1.6\,$M_\odot$\\
6 & B3 & K3 & 7 & 1.4--1.6\,$M_\odot$\\
7 & B3 & K2 & 7 & 1.4--1.7\,$M_\odot$\\
8.5 & B3 & K1 & 7 & 1.4--1.7\,$M_\odot$\\
10 & B2 & K2 & $\sim$7 & 1.3--1.5\,$M_\odot$\\
11 & B3 & G7 & 7 & $>$1.3--$<$1.7\,$M_\odot$\\
\hline
\end{tabular}

\medskip
{\bf Notes:}\\
$^{\rm a}$Age of cluster.\\
$^{\rm b}$Spectral type for the boundary mass (7 and 1.5 M$_\odot$)
  based on the isochrone model by \citet{Siess2000}. 
The range of stellar mass corresponding to the spectral type is shown in
the parentheses when the range covers more than $\Delta M \ge
0.1$\,$M_\odot$.\\
$^{\rm c}$The possible shift of boundary mass for the age spread of
  $\pm$2\,Myr based on the isochrone model by \citet{Siess2000}. 

\end{minipage}
\end{table*}

\begin{table*}
\centering
\begin{minipage}{18.5cm}
\caption{Intermediate-mass star selection and {\it JHK}/MIR IMDF of
target clusters.\label{tab:CL_list}}
\begin{tabular}{lllllllll}
\hline
Cluster & Membership Ref$^{a}$ & Age & SpT$^{b}$ & SpT Ref$^{c}$ & {\it JHK} IMDF$^d$ & &  
MIR Ref$^{e}$ & MIR IMDF$^f$ \\  
 & & \multicolumn{1}{c}{(Myr)} &  & & \multicolumn{1}{c}{(\%)} & & & \multicolumn{1}{c}{(\%)} \\  
\hline
NGC 1333 & St76,As97,Wi04 & 1$\pm$1 & B2.5--K5 &
 Win10,Co10,SB & 17$\pm$17 $(1/6)$ & & Gu09 & 100$\pm$50 $(4 / 4)$ \\
Trapezium & Hi97 & 1{$\pm$1} & B2.5--K5 & Hi97 & 
 9$\pm$3 $(8 / 89)$ & &  --- & ---$^g$ \\ 
$\rho$ Oph & Wi08 & 1{$\pm$1} & B2.5--K5 & Wi08 & 0$\pm$5
 $(0/20)$ & & Wi08 & 80$\pm$20 $(4 / 5)$ \\   
Taurus & Fu06, Fu11 & 1.5{$\pm$1.5} & B3--K5 & 
 Fu06, Fu11 &  31$\pm$10 $(9/29)$ 
 & & Fu06,Lu06 & 72$\pm$16 $(21 / 29)$ \\
Cha I & Lu04 & 2{$\pm$1} & B3--K5 &Lu04 &  29$\pm$13 $(5/17)$ & & Lu08 &
 60$\pm$35 $(3 / 5)$ \\ 
NGC 2068/71 & Fl08 & 2{$\pm$1.5} &B3--K5 & Fl08 & 15$\pm$11 $(2/13)$ 
 & & Fl08 & 69$\pm$23 $(9 / 13)$\\   
IC 348 & Lu03 & 2.5{$\pm$0.5} & B3--K5 & Lu03 & 0$\pm$3
 $(0/34)$ & & La06 & 21$\pm$8 $(7 / 34)$\\   
 $\sigma$ Ori & He07a & 3{$\pm$1} & B3--K4 & Ca10,Re09,SB
 & 0$\pm$4 $(0/23)$ & & He07a & 17$\pm$9 $(4 / 23)$ \\   
NGC 2264 & Re02 & 3{$\pm$1} & B3--K4 & Re02 & 0$\pm$2
 $(0/55)$ & & --- & ---$^g$ \\ 
Tr 37 & Si05 & 4{$\pm$1} & B3--K4 & Si05,SB & 3$\pm$2
 $(2/69)$ & & Si05, Si06 & 22$\pm$10 $(5/23)$ \\ 
Ori OB1bc & He05 & 4{$\pm$3} & B3--K4$\dagger$ & He05 &
 4$\pm$2 $(4 / 94)$ & & --- & ---$^g$ \\  
Upper Sco & Ca06 & 5{$\pm$1} &  B3--K4 & Ca06 & 0$\pm$1
$(0/94)$ & & Ca06 &  2$\pm$2 $(2/94)^h$ \\
NGC 2362 & Da07 & 5{$\pm$1} & B3--K4 & Da07 & 0$\pm$5
 $(0/19)$ & & Da07 & 0$\pm$5 $(0/19)$ \\ 
$\gamma$ Vel & He08 & 5{$\pm$1.5} &B3--K4$\dagger$ &
 Ho78,SB & 0$\pm$6 $(0/17)$ & & He08 & 0$\pm$6 $(0/17)$\\
$\lambda$ Ori & He09 & 5{$\pm$1} & B3--K4$\dagger$ & He09
 & 8$\pm$8 $(1/13)$ & & He09 & 4$\pm$4 $(1/27)$ \\
Per OB2 & He05 & 6{$\pm$2} & B3--K3$\dagger$ & He05 &
 0$\pm$3 $(0/31)$ & & --- &---$^g$ \\ 
$\eta$ Cham & Me05 & 7{$\pm$1} &  B3--K2 & Me05 & 0$\pm$33 $(0/3)$ & &
 Me05 & ---$^i$ \\  
Ori OB1a & He05 & 8.5{$\pm$1.5} & B2--K1$\dagger$ & He05
 & 2$\pm$1 $(2/98)$ & & --- & ---$^g$ \\  
NGC 7160 & Si05 & 11{$\pm$1} & B3--G7 & Si05 & 0$\pm$1
 $(0/82)$ & & Si06 & 3$\pm$2 $(2/78)$ \\ 
\hline
\end{tabular}

\medskip
{\bf Notes:}\\ 
$^{\rm a}$References from which the members of the clusters were picked
up. The IM stars that were used for deriving the {\it JHK} IMDF were
obtained from these references. For the Trapezium Cluster, members are
selected from Hi97, but only those whose stated membership probability
is more than 50\,\% were used.\\
$^{\rm b}$The range of spectral type for the target mass range
(1.5—7\,M$_\odot$) for the cluster age listed in the third
column. $\dagger$ shows cluster for which the observed spectral
types of cluster members do not completely reach to the boundary
spectral type for the lowest mass (see the main text).\\
$^{\rm c}$References from which the spectral types in the clusters were
obtained. For some clusters for which the spectral type listing in the
published papers is incomplete, we supplemented the spectral type
information with those listed in the SIMBAD database at
http://simbad.u-strasbg.fr/simbad/ (denoted as SB).\\
$^{\rm d}$Derived {\it JHK} IMDF and uncertainties based on Poisson
errors. Numbers in parentheses show the number of disk-harbouring
members over total number of members. For the treatment of Poisson
errors for zero detection, see the main text.\\
$^{\rm e}${References for the MIR photometric data.} \\
$^{\rm f}$Derived MIR IMDF and uncertainties based on Poisson errors.
Numbers in the parentheses shows the number of disk harbouring members
over total number of members). For the treatment of Poisson errors for
zero disk harbouring members, see the main text.\\
$^{\rm g}$The clusters for which {\it Spitzer} MIR data are unavailable.\\
{$^{\rm h}$For MIR disk classification of this cluster, we use the slope
between [4.5] and [8] rather than [3.6] and [8] because Carpenter et
al. (2006) does not list photometry data in [3.6]. However, Kennedy and
Kenyon (2009) confirms that use of [4.5] instead of [3.6] does not
change the classification.} \\
$^{\rm i}$MIR IMDF was not derived because of the small number of sample IM
  stars ($<$3).\\
{\bf References:}\\
As97: \citet{Aspin1997};
Ca06: \citet{Carpenter2006}; 
Ca10: \citet{Caballero2010}; 
Co10:  \citet{Connelley2010};
Da07: \citet{Dahm2007}; 
Fl08: \citet{Flaherty2008}; 
Fu06: \citet{Furlan2006}; 
Gu09: \citet{Gutermuth2009}
He05: \citet{Hernandez2005}; 
He07a: \citet{Hernandez2007_662}; 
He08: \citet{Hernandez2008}; 
He09: \citet{Hernandez2009}; 
Hi97: \citet{Hillenbrand1997}; 
Ho78: \citet{Houk1978}; 
La06: \citet{Lada2006};
Lu03: \citet{Luhman2003}; 
Lu04: \citet{Luhman2004a};
Lu06: \citet{Luhman2006}; 
Lu08: \citet{Luhman2008}; 
Me05: \citet{Megeath2005}.
Re02: \citet{Rebull2002}; 
Re09: \citet{Renson2009}; 
Si05: \citet{Sicilia-Aguilar2005ApJ130}; 
Si06: \citet{Sicilia-Aguilar2006};
St76: \citet{Strom1976};
Wi04: \citet{Wilking2004};
Wi08: \citet{Wilking2008};
Win10: \citet{Winston2010}.
\end{minipage}
\end{table*}


\begin{table*}
 \centering
\begin{minipage}{18.5cm}
  \caption{Summary of disk lifetime.}
  \label{tab:JHK_MIR_timescale}　
\begin{tabular}{lcccccccc}
\hline
%
& $<M_*>^a$ & \multicolumn{3}{c}{$t^{\rm life}_{{\it JHK}}$ (Myr)} & &
 \multicolumn{3}{c}{$t^{\rm life}_{\rm MIR}$ (Myr)}\\ \cline{3-5}
 \cline{7-9} 
& & \citet{{My2010ApJ}} & Survival & Binning$^b$ & & \citet{{My2010ApJ}}
 & Survival & Binning$^b$\\
\hline
Intermediate-mass & 2.5\,$M_\odot$ 
& 
---  & 2.8$\pm$2.4 & 3.3$\pm$0.9 & &
---  & 6.1$\pm$4.2 & 6.7$\pm$1.1 \\
Low-mass & 0.5\,$M_\odot$ 
& 9.7$\pm$1.1 & --- & --- & 
& 8.6$\pm$0.7 & --- & ---\\
\hline
Mass dependence & 
&  & {$M_*^{-0.8 \pm 0.7}$} & {$M_*^{-0.7 \pm 0.3}$} & & & 
 {$M_*^{-0.2 \pm 0.3}$} & {$M_*^{-0.2 \pm 0.1}$} \\

\hline
\end{tabular}

\medskip
{\bf Notes:}\\ 
$^a$Characteristic mass for the mass range (see details in the main
 text).\\ 
$^b$See Section~\ref{subsec:JHK_MIR_IMDF} for the definition of this
 fitting.\\ 
\end{minipage}
\end{table*}



\appendix

\section{List of Sample Intermediate-mass Stars in Target Clusters}

In this appendix, the intermediate-mass star samples for all 19 clusters
 listed in Table~\ref{tab:CL_list} are summarized in tables
 (Tabs.~A1--A19) as well as in colour--colour diagrams (Figs.~A1--A19).

In the tables, only RA, Dec coordinates (in J2000) are shown in case
objects names are not available in the references. ``SpT'' shows the
spectral types in the literatures. The ``$K$ disk'' and ``MIR disk'' columns
show objects with a disk (o) and without a disk (X). The numbers in the
parenthesis in MIR disk column is $\alpha$ as defined in
Section~\ref{sec:intro}. The stars with $K$-disk emission are judged from the
colour--colour diagram, in which the red and black circles show  those with 
a $K$ disk and without a $K$ disk, respectively.


\clearpage

\begin{table*} 
 \centering
 \begin{minipage}{18cm}
\caption{NGC 1333.}  The $\alpha$ values for MIR disk are directly
referred from \citet{Gutermuth2009}. Though extinction is not corrected
for $\alpha$ values in this reference, it should not affect the disk
judgement because the $\alpha$ value is much larger than $-2$.
The spectral type with $*$ mark is from SIMBAD database.\\

  \begin{tabular}{llllll}
\hline
Name & RAJ2000 & DEJ2000 & SpT & {\it K}\,disk & MIR\,disk \\
& (h:m:s) & (d:m:s) &   & \\
\hline
2MASS J03291977+3124572 & 03 29 19.77605 & +31 24 57.0474 &
	       B8{*} & X & ... \\ 
2MASS J03285720+3114189 & 03 28 57.2107 & +31 14 19.056 & B* & X &...\\
2MASS J03290575+3116396 & 03 29 05.754 & +31 16 39.69 & A3 & X & o ($-$0.28) \\
2MASS J03291037+3121591	(LZK 12) & 03 29 10.379 & +31 21 59.16 & F4--G0 &
		   o & o ($-$0.40) \\
2MASS J03285930+3115485 & 03 28 59.306 & +31 15 48.52 & K2 & X & o ($-$0.25)\\ 
2MASS J03292187+3115363	(LkHA 271) & 03 29 21.873 & +31 15 36.30 & K4.0 
	       & X & o ($-$1.49) \\
\hline
\end{tabular}
\end{minipage}
\end{table*}

\begin{figure*}
\begin{center}
\includegraphics[scale=0.5]{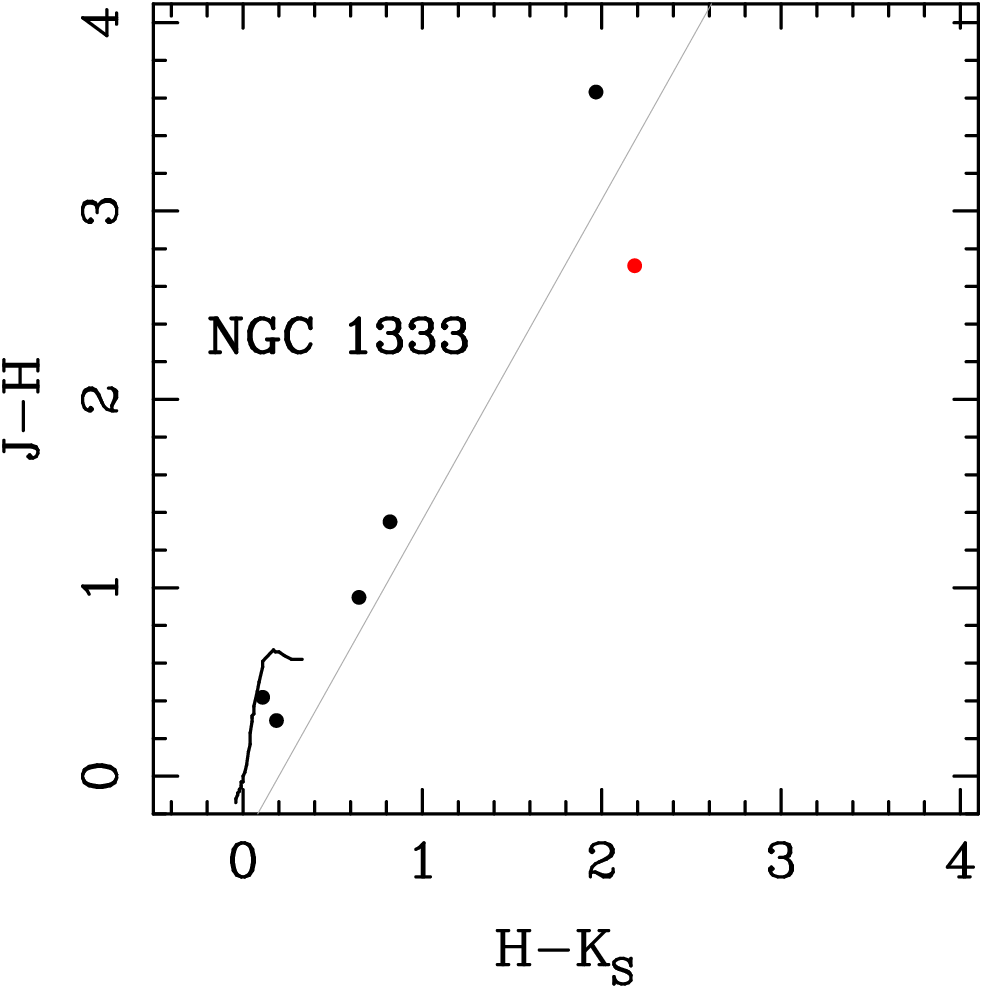}
\caption{NGC1333.}
\end{center}
\end{figure*}

\clearpage
\begin{table*} 
 \centering
 \begin{minipage}{18cm}
\caption{Trapezium.}
  \begin{tabular}{llllll}
\hline
 RAJ2000 & DEJ2000 & SpT & $K$\,disk \\
 (h:m:s) & (d:m:s) &   & \\
\hline
05 35 06.08 & $-$05 12 15.22 & B3--B5 & X \\
05 35 31.35 & $-$05 25 15.92 & B3--B6 & X \\
05 35 40.06 & $-$05 17 29.12 & B6 & X \\
05 35 54.10 & $-$05 37 42.50 & B5--B7 & X \\
05 34 39.93 & $-$05 10 06.81 & B8--A0 & X \\
05 34 55.20 & $-$05 30 21.52 & B8--B9 & X \\
05 35 00.03 & $-$05 25 15.82 & B9--A1 & X \\
05 35 58.45 & $-$05 22 30.62 & B8--B9.5 & X \\
05 35 28.32 & $-$05 26 19.82 & B9.5--A0V & X \\
05 35 16.88 & $-$05 21 45.02 & A0--A2 & X \\
05 34 46.90 & $-$05 34 14.82 & B9--A1 & X \\
05 36 27.08 & $-$05 24 30.20 & B9--A0 & X \\
05 35 55.33 & $-$05 13 55.52 & A0--A5 & X \\
05 35 35.67 & $-$05 12 20.32 & B9--A1 & X \\
05 34 49.91 & $-$05 18 44.42 & A2--A7 & o \\
05 35 50.36 & $-$05 28 34.62 & B8--A5 & o \\
05 35 19.03 & $-$05 20 38.52 & B5--A7 & X \\
05 35 18.70 & $-$05 17 28.92 & A8--F0 & o \\
05 35 15.89 & $-$05 23 52.52 & F2--F7 & X \\
05 35 31.28 & $-$05 33 08.62 & A8--F8 & o \\
05 35 54.65 & $-$05 10 55.22 & F7--G4 & X \\
05 35 05.11 & $-$05 14 50.22 & F8--G5 & X \\
05 35 18.57 & $-$05 20 33.52 & F8--K0 & X \\
05 35 11.53 & $-$05 16 57.52 & G0--K0 & X \\
05 34 24.83 & $-$05 22 05.09 & G0--G1 & X \\
05 35 21.16 & $-$05 09 15.82 & F8--K2 & o \\
05 35 26.75 & $-$05 11 07.12 & G3 & o \\
05 34 19.39 & $-$05 27 11.57 & G6 & X \\
05 34 14.39 & $-$05 28 16.30 & G6--K0 & X \\
05 35 26.10 & $-$05 27 36.22 & G3--K3 & X \\
05 35 15.15 & $-$05 22 56.42 & G6--G8 & ... \\
05 35 35.89 & $-$05 12 25.02 & G6--K2 & o \\
05 35 21.70 & $-$05 23 53.62 & G6--K3 & X \\
05 35 20.94 & $-$05 23 48.62 & G8--K5 & X \\
\hline
\end{tabular}
\end{minipage}
\end{table*}

\begin{table*}
 \centering
 \begin{minipage}{18cm}
\contcaption{}
  \begin{tabular}{llllll}
\hline
 RAJ2000 & DEJ2000 & SpT & $K$\,disk \\
 (h:m:s) & (d:m:s) &   & \\
\hline
05 35 26.19 & $-$05 08 39.72 & G8--K5.5 & ... \\
05 35 15.53 & $-$05 22 56.12 & G8--K1 & X \\
05 35 18.95 & $-$05 23 49.22 & G8--K5 & ... \\
05 35 21.21 & $-$05 12 12.42 & G8--G0 & X \\
05 35 28.49 & $-$05 31 26.12 & G & X \\
05 35 02.75 & $-$05 22 07.92 & G & X \\
05 34 53.52 & $-$05 26 36.72 & G & X \\
05 35 20.11 & $-$05 20 56.72 & F7--K3 & o \\
05 35 05.55 & $-$05 25 19.02 & A0--K4 & X \\
05 35 23.73 & $-$05 30 46.92 & G8--K3 & X \\
05 35 41.87 & $-$05 28 12.42 & K0--K1 & X \\
05 35 54.56 & $-$05 22 00.72 & K0--K2 & X \\
05 35 16.96 & $-$05 23 33.72 & K0--K2 & ... \\
05 35 11.77 & $-$05 19 26.02 & K0--K3 & ... \\
05 35 14.60 & $-$05 39 11.42 & K0--K3 & X \\
05 35 20.65 & $-$05 15 49.12 & K0--K5 & X \\
05 35 15.85 & $-$05 23 49.42 & G5--K0 & X \\
05 35 25.63 & $-$05 09 49.22 & K0--K4 & X \\
05 34 39.80 & $-$05 26 41.62 & K0--K3 & X \\
05 35 21.18 & $-$05 24 56.92 & K1--K2 & X \\
05 35 02.91 & $-$05 30 00.92 & K1--K2 & X \\
05 35 34.81 & $-$05 29 14.02 & K1--K4 & X \\
05 35 35.05 & $-$05 33 49.02 & K1--K5 & X \\
05 35 31.18 & $-$05 15 32.92 & K1 & X \\
05 35 38.74 & $-$05 12 41.72 & K2 & X \\
05 35 08.29 & $-$05 28 28.92 & K2 & X \\
05 34 51.48 & $-$05 25 12.62 & K2 & X \\
05 35 19.19 & $-$05 20 07.72 & K2 & X \\
05 34 33.87 & $-$05 28 24.22 & K2 & X \\
05 34 45.10 & $-$05 25 03.62 & K2 & X \\
05 34 55.89 & $-$05 23 12.62 & K0--K4 & X \\
05 35 53.53 & $-$05 15 41.42 & K2--K3 & X \\
05 35 17.41 & $-$05 17 39.82 & K2--K3 & X \\
05 35 11.40 & $-$05 26 01.82 & K2--K3 & X \\
05 35 13.69 & $-$05 39 10.52 & K2--K4 & X \\
05 35 24.98 & $-$05 23 46.32 & K2--K4 & X \\
05 35 15.96 & $-$05 20 36.32 & K2--K4 & X \\
05 35 18.55 & $-$05 23 13.52 & K2--K5 & ... \\
05 34 37.35 & $-$05 34 51.92 & K2--K5 & X \\
05 34 35.05 & $-$05 32 10.22 & K2 & X \\
05 35 25.30 & $-$05 10 47.92 & K3 & X \\
05 35 02.30 & $-$05 15 47.82 & K3 & X \\
05 36 10.38 & $-$05 19 44.62 & K1--K3 & X \\
05 35 29.74 & $-$05 32 53.12 & G8--K3 & X \\
05 35 22.15 & $-$05 20 29.02 & K2--K4 & X \\
05 35 17.50 & $-$05 22 56.22 & K3--K4 & X \\
05 35 18.79 & $-$05 16 13.72 & K2--K5 & X \\
05 35 14.88 & $-$05 21 59.62 & K3--K4 & X \\
05 35 31.13 & $-$05 23 39.72 & K3--K5 & X \\
05 35 05.52 & $-$05 11 50.62 & K3--K5 & X \\
05 35 35.17 & $-$05 21 26.92 & K4 & ... \\
05 35 08.75 & $-$05 31 48.52 & K4 & X \\
05 35 27.19 & $-$05 23 36.32 & K4 & X \\
05 35 26.31 & $-$05 23 01.92 & K4--K5 & ... \\
05 35 04.43 & $-$05 29 37.82 & K4--K5 & ... \\
05 35 24.14 & $-$05 25 18.32 & K0-K5 & X \\
05 35 50.66 & $-$05 16 29.02 & K5 & X \\
05 35 23.54 & $-$05 23 31.62 & K5 & ... \\
05 35 21.47 & $-$05 09 38.72 & K5 & X \\
05 35 20.90 & $-$05 31 21.22 & K5 & X \\
05 35 06.19 & $-$05 22 02.32 & K5 & X \\
05 35 04.67 & $-$05 17 42.12 & K5 & X \\
05 35 02.34 & $-$05 20 46.32 & K5 & X \\
05 34 58.71 & $-$05 21 17.52 & K5 & X \\
05 34 50.63 & $-$05 24 01.02 & K5 & ... \\
\hline
\end{tabular}
\end{minipage}
\end{table*}

\begin{figure*}
\begin{center}
\includegraphics[scale=0.5]{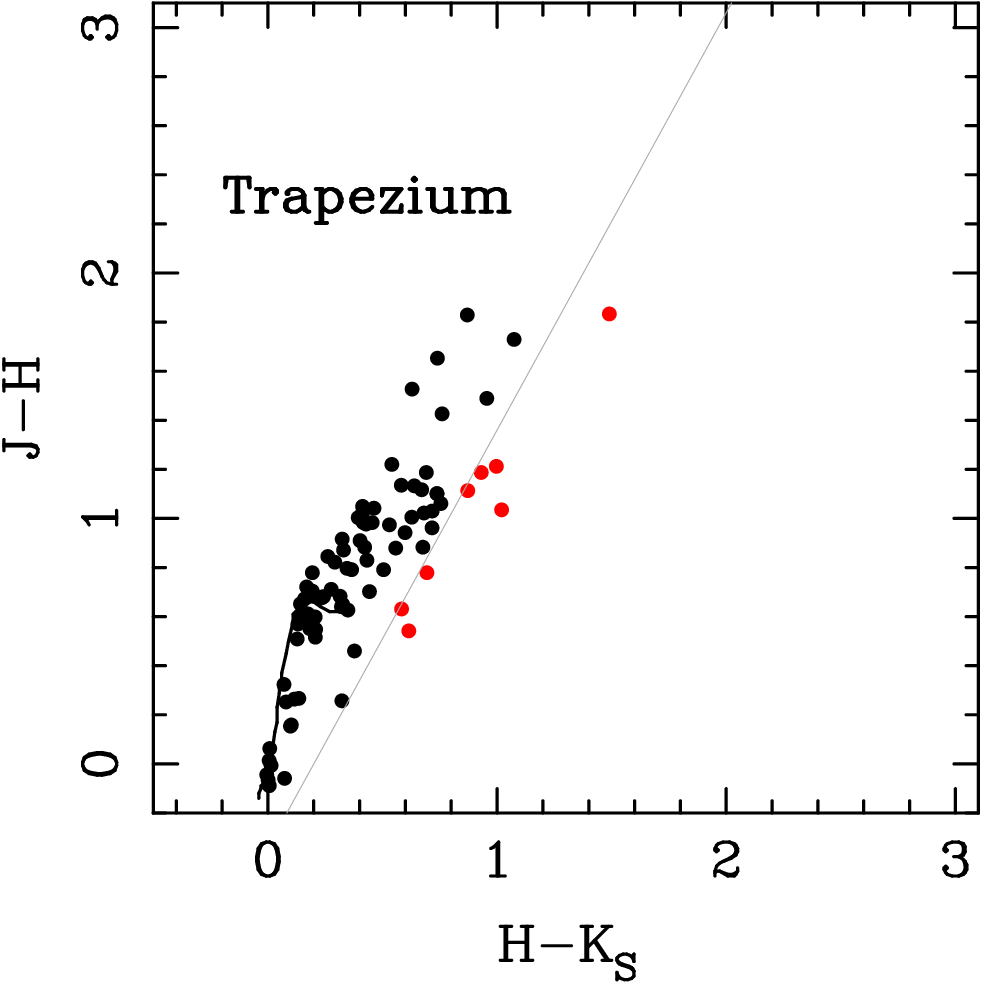}
\caption{Trapezium.}
\end{center}
\end{figure*}

\clearpage
\begin{table*}
\centering
 \begin{minipage}{18cm}
\caption{$\rho$ Oph.}
Because we could not find published IRAC photometry data, we directly
used the MIR disk classification in \citet{Wilking2008}.

  \begin{tabular}{llllll}
\hline
 RAJ2000 & DEJ2000 & SpT & $K$\,disk & MIR\,disk \\
 (h:m:s) & (d:m:s) &   & \\
\hline
16 26 9.31 & $-$24 34 12.10 & A0 V & X & ...\\
16 27 49.87 & $-$24 25 40.20 & A7 & X & X \\	
16 28 25.16 & $-$24 45 0.90 & F2 V & X & ...\\
16 25 7.93 & $-$24 31 57.20 & F5 & X & ...\\
16 27 10.28 & $-$24 19 12.70 & G1& X & o \\
16 25 19.24 & $-$24 26 52.60 & G1IV & X & ... \\
16 26 46.43 & $-$24 12 0.10 & G3.5& X & ...\\
16 26 23.36 & $-$24 20 59.80 & G6 & X & o \\
16 28 32.66 & $-$24 22 44.90 & G7 & X & ...\\
16 26 3.02 & $-$24 23 36.00 & K0 & X & ... \\
16 26 58.51 & $-$24 45 36.90 & K1 & ... & ...\\
16 27 17.08 & $-$24 47 11.20 & K1 & X & ... \\
16 25 24.35 & $-$23 55 10.30 & K3/M0: (BA92) & X & ... \\
16 25 49.64 & $-$24 51 31.90 & K3/M0 (BA92W94) & X & ...\\
16 24 56.52 & $-$24 59 38.20 & K5 & X & ...\\
16 25 22.43 & $-$24 02 5.70 & K5 & X & ...\\
16 26 23.68 & $-$24 43 13.90 & K5 & X & ...\\
16 27 39.43 & $-$24 39 15.50 & K5 & X & o \\
16 27 40.29 & $-$24 22 4.00 & K5 & X & o \\
16 28 16.73 & $-$24 05 14.30 & K5 & X & ...\\
16 28 23.33 & $-$24 22 40.60 & K5 & X & ...\\
\hline
\end{tabular}
\end{minipage}
\end{table*}

\begin{figure*}
\begin{center}
\includegraphics[scale=0.5]{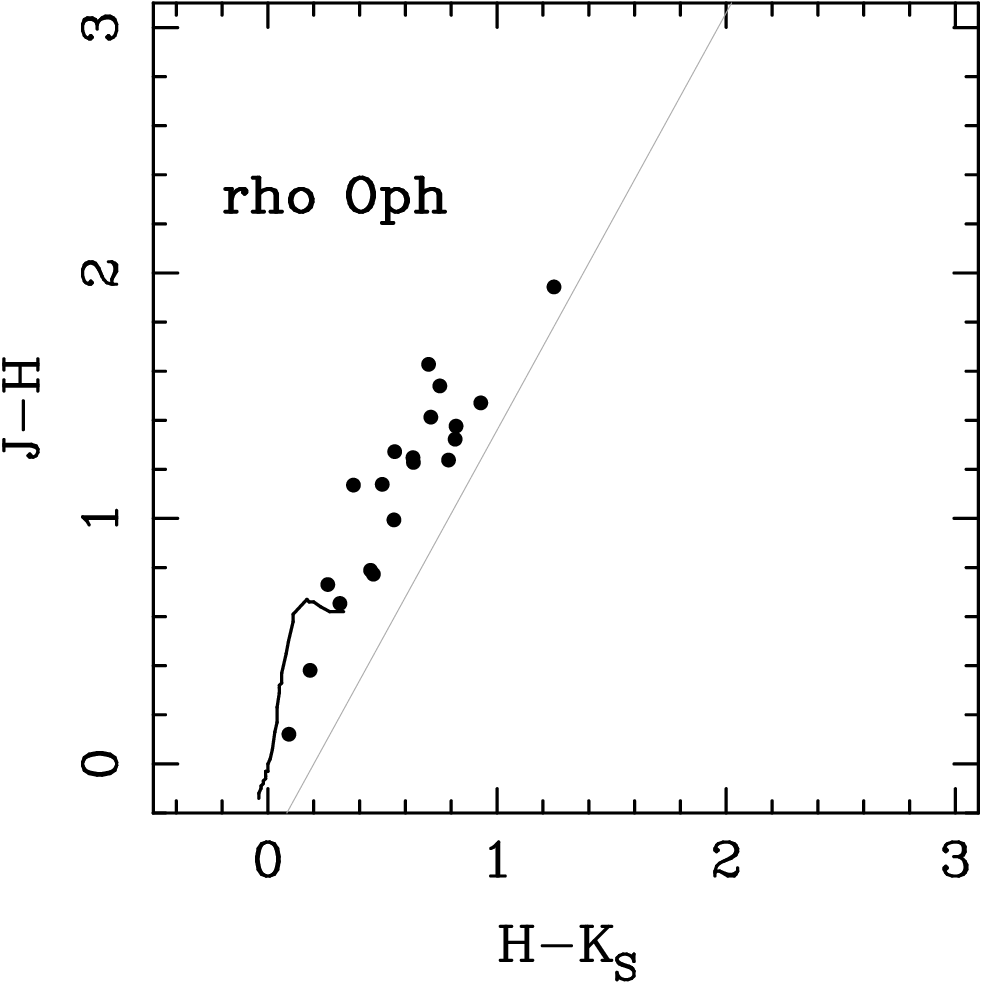}
\caption{$\rho$ Oph.}
\end{center}
\end{figure*}

\clearpage
\begin{table*}
 \centering
 \begin{minipage}{18cm}
\caption{Taurus.}
{The $\dag$ {and $\ddag$} marks show the members that are identified as
  w/disks or w/o disks from SEDs in \citet{Furlan2006} {and
  \citet{Furlan2011}, respectively}.}\\
\begin{tabular}{llllll}
\hline
 Name & SpT & $K$\,disk & MIR\,disk \\
\hline
 V892 Tau	& B9	& o	& o$\dag$ \\
 AB Aur 	& A0	& o	& o$\dag$\\
 HP Tau/G2	& G0	& \color{black} X & X$\dag$\\ 
 RY Tau 	& G1	& X	& o$\dag$\\
 SU Aur 	& G1	& X	& o$\dag$\\
 HD 283572 &  G5	& X	& X ($-$2.82) \\
\color{black}IRAS 04278+2253&  G8	& X	& \color{black} o$\ddag$\\
\color{black} LkCa 19& K0	& X	& \color{black} X$\ddag$\\
 T Tau	& K0	& o	& o$\dag$\\

 HBC 388	& K1	& X	& X$\dag$\\

\color{black} HQ Tau& K2	& X	& \color{black} o$\dag$ \\
 IT Tau	& K2	& X	& o ($-$1.47) \\

 CW Tau	& K3	& o	& o$\dag$\\
 HP Tau	& K3	& o	& o$\dag$\\
 RW Aur	& K3	& X	& o$\dag$\\
 V773 Tau	& K3	& X	& o$\dag$\\
 HBC 356 & K3	& X	& X$\dag$\\
 V410 Tau & K3	& X	& X ($-$2.79) \\
\color{black}2MASS J04390525+2337450&  K5 & X & \color{black} o$\ddag$\\
\color{black} DR Tau &  K5 & o & \color{black} o$\dag$ \\
 DS Tau &  K5 &  o&  o$\dag$\\
 FV Tau (A, B) &  K5 & X &  o ($-$0.90) \\
\color{black} FS Tau B (Haro 6-5B) &  K5 & o & \color{black} o$\ddag$\\
 HN Tau (A, B) &  K5 &  o&  o$\dag$\\
 LkCa 15 &  K5 &  X&  o $\dag$ \\
 UX Tau (A, Ba, Bb, C) & K5 & X & o$\dag$\\
\color{black} V807 Tau &  K5 & X & o$\dag$ \\
 HBC 392 &  K5 &  X&  X$\dag$ \\
HBC 427 &  K5 &  X&  X$\dag$ \\ 
\hline
\end{tabular}

\medskip
{\bf Notes:}\\
The sample includes two low-mass stars with measured dynamical masses,
Lk Ca 15 (\citealt{Hillenbrand2004}; 0.84\,$M_\odot$) and V807 Tau
(\citealt{Schaefer2006}; 1.15\,$M_\odot$).  By our criteria, both
objects have spectral type of K5 and are classified as IM stars. The
dynamical mass of V807 Tau is almost within the mass uncertainty of our
method as described in section 2.2. Lk Ca 15 has an estimated age
(3--5\,Myr; \citealt{Simon2000}) that is much higher than the average
Taurus cluster age (1.5\,Myr; see also Fig. 3 in Simon et al. 2000),
which we employed for our IM star selection (see \S\,2.2 and
\S\,8.1.2). However, these stars are included for consistency with the
other clusters. As discussed in section 2.2 up to 15\,\% of our sample
of IM stars could be low mass stars.
\end{minipage}
\end{table*}

\begin{figure*}
\begin{center}
\includegraphics[scale=0.5]{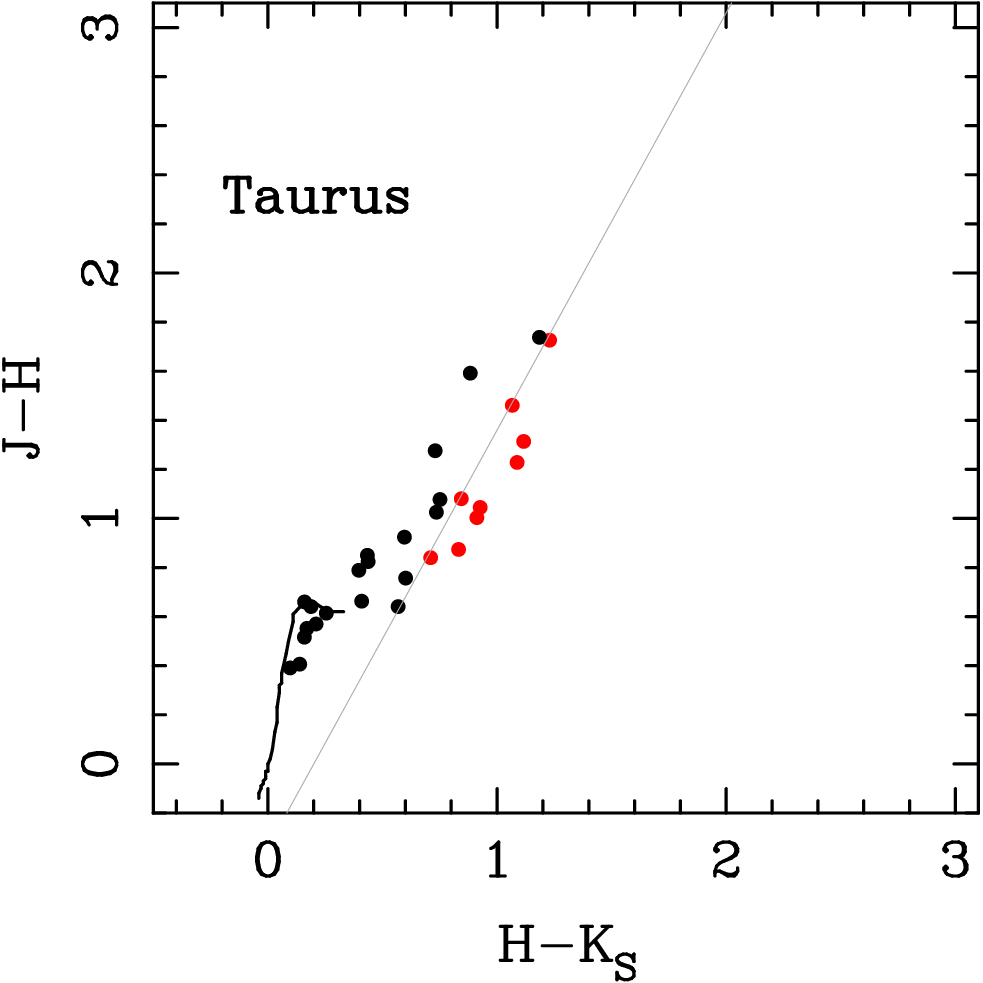}
\caption{Taurus.}
\end{center}
\end{figure*}

\clearpage

\begin{table*}
 \centering
 \begin{minipage}{18cm}
\caption{Cha I.}
  \begin{tabular}{lllll}
\hline
RAJ2000 & DEJ2000 & SpT & $K$\,disk & MIR\,disk \\
(h:m:s) & (d:m:s) &   & \\
\hline
11 05 57.81 & $-$76 07 48.9 & B6.5 & X & ... \\
11 08 03.30 & $-$77 39 17.4 & B9.5  & o & ... \\
%
11 09 50.03 &$-$76 36 47.7 & B9 & X & ... \\
%
10 46 37.95 & $-$77 36 03.6 & F0  & X & ... \\
%
11 06 15.41 & $-$77 21 56.8 & G5 & X & X (-2.67)\\
%
11 07 20.74 & $-$77 38 07.3 & G2 & o & ... \\
%
11 08 15.10 & $-$77 33 53.2 & G7 & o & ... \\
%
11 12 27.72 & $-$76 44 22.3 & G9 & X & o (-1.27)\\
%
11 12 42.69 & $-$77 22 23.1 & G8 & X & ... \\
%
10 58 16.77 &$-$77 17 17.1 & K0 & X & ... \\
%
10 59 06.99 &$-$77 01 40.4 & K2 & X & ...\\
%
11 10 38.02 &$-$77 32 39.9 & K3  & X & o (-1.24) \\
%
11 12 24.41 &$-$76 37 06.4 & K3.5& X & o (-0.57)\\
%
11 12 43.00 & $-$76 37 04.9 & K4.5  & X & X (-2.78)\\
%
11 04 09.09 & $-$76 27 19.4 & K5  & X & ... \\
%
11 09 53.41 & $-$76 34 25.5 & K5  & o & ... \\
%
11 10 00.11 & $-$76 34 57.9 & K5  & o & ... \\
\hline
\end{tabular}
\end{minipage}
\end{table*}

\begin{figure*}
\begin{center}
\includegraphics[scale=0.5]{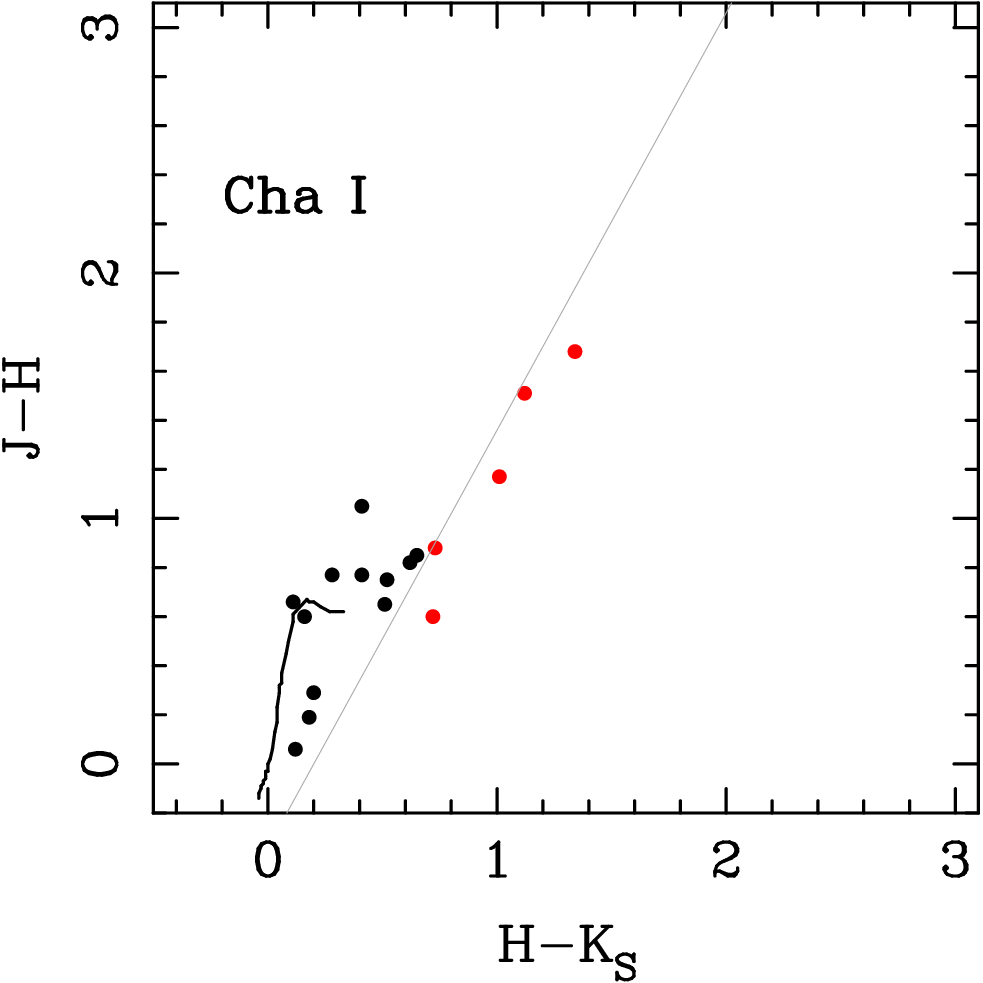}
\caption{Cha I.}
\end{center}
\end{figure*}


\clearpage
\begin{table*}
 \centering
 \begin{minipage}{18cm}
\caption{NGC 2068/71.}
  \begin{tabular}{llllll}
\hline
Name & RAJ2000 & DEJ2000 & SpT & $K$\,disk & MIR\,disk \\
(FM2008) & (h:m:s) & (d:m:s) &   & \\
\hline
1173 & 05 47 10.98 & $+$00 19 14.81 & G6  & X & o  ($-$1.08)  \\
1099 & 05 47 06.00 & $+$00 32 08.48 & K0  & o & o  ($-$1.04)  \\
 618 & 05 46 22.44 & $-$00 08 52.62 & K1  & X & o  ($-$1.64)  \\
 571 & 05 46 18.30 & $+$00 06 57.85 & K1  & X & o  ($-$0.73)  \\
 515 & 05 46 11.86 & $+$00 32 25.91 & K2  & X & X  ($-$2.24)  \\
 590 & 05 46 19.47 & $-$00 05 20.00 & K2.5 & X & o  ($-$0.65)  \\
 984 & 05 46 56.54 & $+$00 20 52.91 & K3  & X & X  ($-$2.46)  \\
 458 & 05 46 07.89 & $-$00 11 56.87 & K3  & X & o  ($-$1.92)  \\
 739 & 05 46 34.54 & $+$00 06 43.45 & K4  & \color{black} X  & o 
		       ($-$1.12) \\ 
 581 & 05 46 18.89 & $-$00 05 38.11 & K4  & \color{black} X  & o
		       ($-$1.27)  \\ 
 177 & 05 45 41.94 & $-$00 12 05.33 & K4  & X  & X  ($-$2.63)  \\
1116 & 05 47 06.96 & $+$00 00 47.74 & K4.5 & \color{black} X  & o
		       ($-$0.95)  \\ 
 584 & 05 46 19.06 & $+$00 03 29.59 & K5  & o  & o  ($-$1.00) \\
\hline
\end{tabular}
\end{minipage}
\end{table*}

\begin{figure*}
\begin{center}
\includegraphics[scale=0.5]{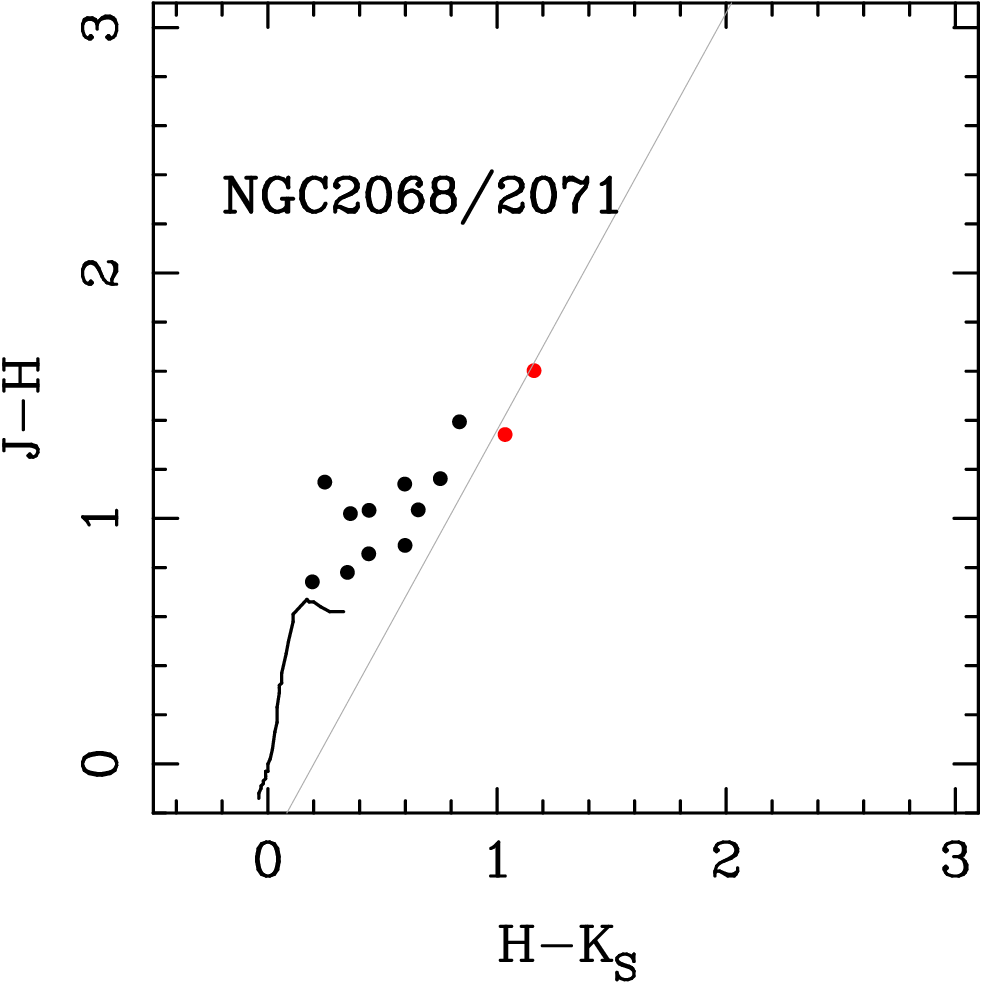}
\caption{NGC 2068/71.}
\end{center}
\end{figure*}


\clearpage
\begin{table*}
 \centering
 \begin{minipage}{18cm}
\caption{IC 348.}
  \begin{tabular}{lllll}
\hline
RAJ2000 & DEJ2000 & SpT & $K$\,disk & MIR\,disk \\
(h:m:s) & (d:m:s) &   & \\
\hline
03 44 34.20 & +32 09 46.3 & B5 & X & X ($-$2.58)\\
03 44 08.48 & +32 07 16.5 & A0 & X & X ($-$2.79) \\
03 44 50.65 & +32 19 06.8 & A0 & X & X ($-$2.82)\\
03 44 30.82 & +32 09 55.8 & A2 & X & o ($-$0.73)\\
03 44 09.15 & +32 07 09.3 & A2 & X & X ($-$2.53)\\
03 44 35.36 & +32 10 04.6 & A2 & X & o ($-$1.37) \\
03 44 32.06 & +32 11 44.0 & A3 & ... & ... \\
03 45 01.42 & +32 05 02.0 & A4 & X & X ($-$2.69) \\
03 44 47.72 & +32 19 11.9 & A4 & X & X ($-$2.70) \\
03 44 19.13 & +32 09 31.4 & F0  & X & X ($-$2.81) \\
03 44 31.19 & +32 06 22.1 & F0  & X & X ($-$2.79) \\
03 44 24.66 & +32 10 15.0 & F2  & X & X ($-$2.82) \\
03 44 23.99 & +32 11 00.0 & G0  & X & X ($-$2.77)\\
03 44 31.96 & +32 11 43.9 & G0  & X & o (1.01) \\
03 44 18.16 & +32 04 57.0 & G1  & X & o ($-$1.66) \\
03 45 07.61 & +32 10 28.1 & G1  & X & X ($-$2.67) \\
03 44 36.94 & +32 06 45.4 & G3  & X & o ($-$1.93) \\
03 45 07.96 & +32 04 02.1 & G4  & X & X ($-$2.68) \\
03 43 51.24 & +32 13 09.4 & G5  & X & X ($-$2.70) \\
03 44 32.74 & +32 08 37.5 & G6  & X & X ($-$2.75) \\
03 44 39.17 & +32 09 18.3 & G8  & X & X ($-$2.89) \\
03 44 26.03 & +32 04 30.4 & G8  & X & o ($-$1.31) \\
03 45 01.52 & +32 10 51.5 & K0  & X & X ($-$2.60) \\
03 44 16.43 & +32 09 55.2 & K0  & X & X ($-$2.77) \\
03 43 55.51 & +32 09 32.5 & K0  & X & X ($-$2.61) \\
03 44 08.86 & +32 16 10.7 & K0  & X & X ($-$2.72)\\
03 44 56.15 & +32 09 15.5 & K0  & X & o ($-$1.88)\\
03 44 40.13 & +32 11 34.3 & K2  & X & X ($-$2.66) \\
03 44 31.53 & +32 08 45.0 & K2  & X & X ($-$2.70) \\
03 44 39.25 & +32 07 35.5 & K3  & X & X ($-$2.60) \\
03 44 38.72 & +32 08 42.0 & K3  & X & X ($-$2.91) \\
03 44 05.00 & +32 09 53.8 & K3.5  & X & X ($-$2.77) \\
03 45 01.74 & 	+32 14 27.9 & 	K4  & X & X ($-$2.75) \\
03 44 55.63 & 	+32 09 20.2 & 	K4  & X & X ($-$2.74)\\
03 44 24.29 & 	+32 10 19.4 & 	K5  & X & X ($-$2.68) \\
\hline
\end{tabular}
\end{minipage}
\end{table*}

\begin{figure*}
\begin{center}
\includegraphics[scale=0.5]{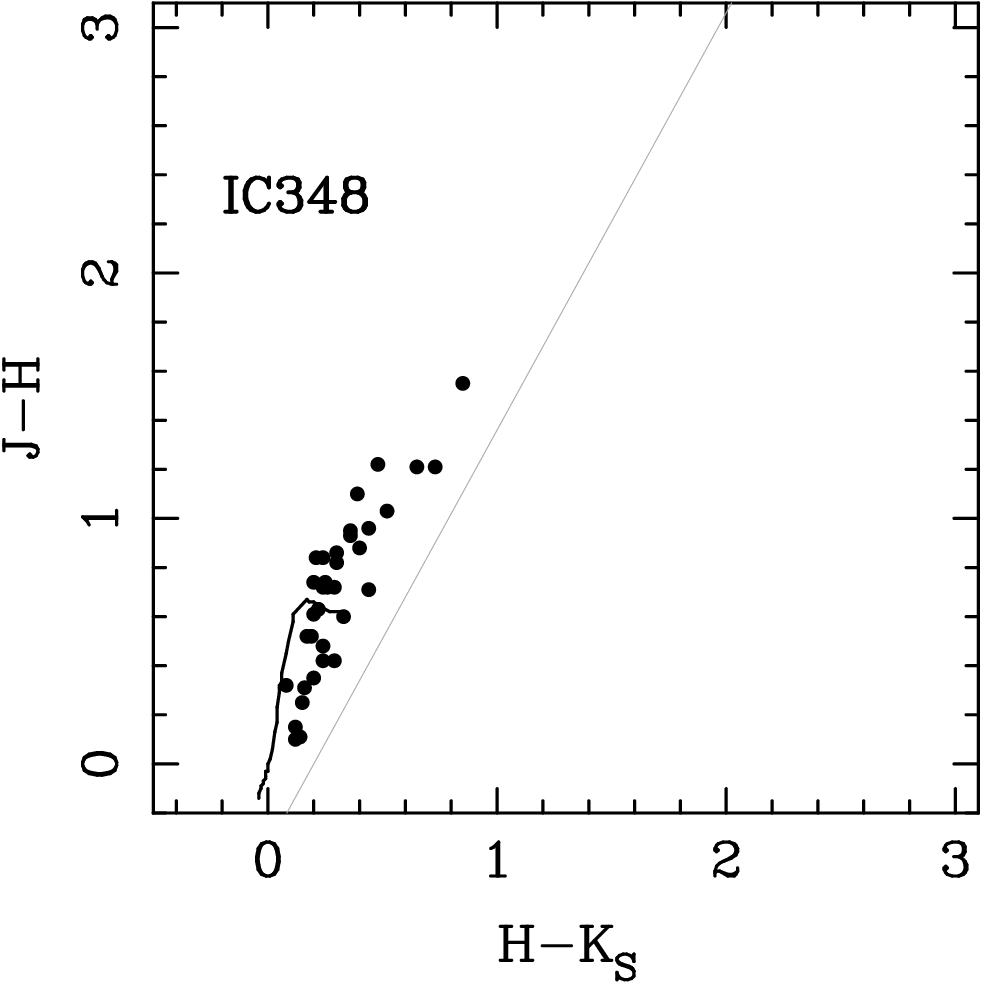}
\caption{IC 348.}
\end{center}
\end{figure*}


\clearpage
\begin{table*}
 \centering
 \begin{minipage}{18cm}
\caption{$\sigma$ Ori.}
The spectral type with $*$ mark is from SIMBAD database.\\

  \begin{tabular}{llllll}
\hline
Name & RAJ2000 & DEJ2000 & SpT & $K$\,disk & MIR\,disk \\
 & (h:m:s) & (d:m:s) &   & \\
\hline
HD 37525 & 05 39 01.49131	  & $-$02 38 56.3650	  & B5 &  X & X ($-$2.85) \\
HD 294271 & 05 38 36.5494	  & $-$02 33 12.740	  & B5V* &  X & X ($-$2.91) \\
2MASS J05383422$-$0234160 & 05 38 34.235	  & $-$02 34 16.08	  & B8V* &  X & X ($-$2.81) \\
HD 37545 & 05 39 09.2145	  & $-$02 56 34.732	  & B9* &  X & X ($-$2.79) \\
V1147 Ori & 05 39 46.1950	  & $-$02 40 32.054	  & B9 &  X & X ($-$2.81) \\
HD 294272 & 05 38 34.799	  & $-$02 34 15.78	  & B9.5III &  X & X ($-$2.91) \\
HD 37564 & 05 39 15.0594	  & $-$02 31 37.618	  & A0* &  X & X ($-$2.51) \\
HD 294275 & 05 37 31.8728	  & $-$02 45 18.473	  & A1V* &  X & X ($-$2.81) \\
HD 294279 & 05 38 31.3795	  & $-$02 55 03.075	  & A2* &  X & X ($-$2.76) \\
HD 294273 & 05 38 27.5241	  & $-$02 43 32.596	  & A3* &  X & X ($-$2.87) \\

HD 294299 & 05 39 40.572	  & $-$02 25 46.82	  & F2* &  X & X ($-$2.76) \\
HD 294268 & 05 38 14.1139	  & $-$02 15 59.741	  & F8* &  X & o  (0.17) \\
HD 294274 & 05 37 45.3662	  & $-$02 44 12.491	  & G0* &  X & X ($-$2.85) \\
HD 294298 & 05 39 59.318	  & $-$02 22 54.35	  & G0* &  X & X ($-$2.84) \\
2MASS J05375303$-$0233344 & 05 37 53.036	  & $-$02 33 34.41	  & K0 &  X & X ($-$2.82) \\
2MASS J05383848$-$0234550 & 05 38 38.486	  & $-$02 34 55.02	  & K0 &  X & X ($-$2.63) \\
2MASS J05393654$-$0242171 & 05 39 36.543	  & $-$02 42 17.16	  & K0* &  X & X ($-$2.79) \\
2MASS J05375440$-$0239298 & 05 37 54.405	  & $-$02 39 29.85	  & K0: &  X & X ($-$2.86) \\
TY Ori & 05 38 35.873 & 	 $-$02 43 51.22 & 	 K3* &  X & o  ($-$1.85) \\
2MASS J05384129$-$0237225 & 05 38 41.292 &  $-$02 37 22.57 &  K3 &  X & X ($-$2.78) \\
2MASS J05384803$-$0227141 & 05 38 48.036 &  $-$02 27 14.19 &  K3 &  X & o  ($-$1.55) \\
2MASS J05385410$-$0249297 & 05 38 54.107 &  $-$02 49 29.77 &  K3* &  X & X ($-$2.80) \\
TX Ori & 05 38 33.685 &  $-$02 44 14.15 &  K4* &  X & o  ($-$0.97) \\
\hline
\end{tabular}
\end{minipage}
\end{table*}

\begin{figure*}
\begin{center}
\includegraphics[scale=0.5]{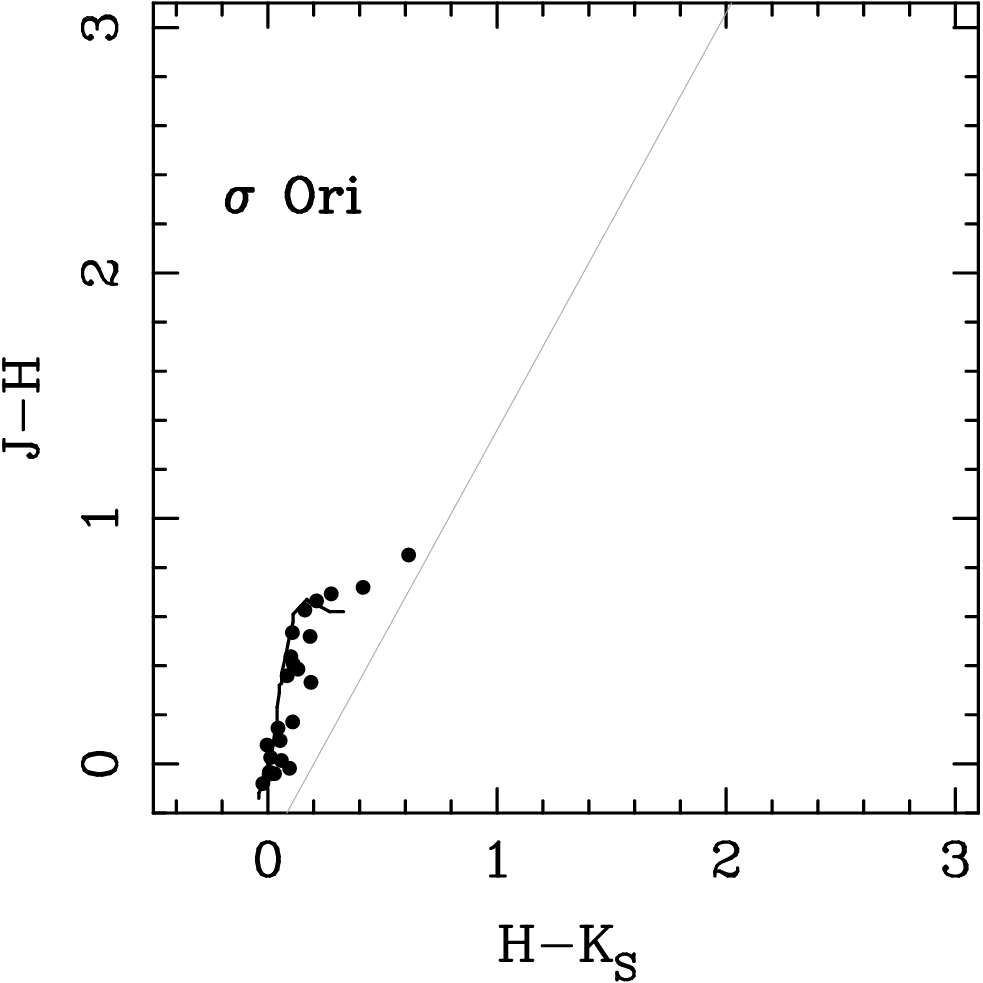}
\caption{$\sigma$ Ori.}
\end{center}
\end{figure*}


\clearpage
\begin{table*}
 \centering
 \begin{minipage}{18cm}
\caption{NGC 2264.}
  \begin{tabular}{llllll}
\hline
 RAJ2000 & DEJ2000 & SpT & $K$\,disk \\
(h:m:s) & (d:m:s) &   & \\
\hline
06 41 32.7 & +09 53 24 & A2 & X \\ 
06 41 27.7 & +09 55 13 & A2 & X \\ 
06 41 36.6 & +09 37 56 & A5 & X \\ 
06 41 24.3 & +09 46 10 & A5 & X \\ 
06 40 39.3 & +09 59 22 & A5 & X \\ 
06 40 47.0 & +09 55 03 & F0 & X \\ 
06 40 37.8 & +09 40 11 & F0 & X \\ 
06 41 26.2 & +09 47 22 & F1 & X \\ 
06 41 24.3 & +09 56 09 & F1 & X \\ 
06 40 50.4 & +09 54 16 & F1 & X \\ 
06 41 26.0 & +09 57 15 & F3 & X \\ 
06 40 33.5 & +09 42 55 & F5 & X \\ 
06 41 13.8 & +09 55 44 & G0 & X \\ 
06 40 37.3 & +09 42 15 & G0 & X \\ 
06 39 41.6 & +09 34 40 & G0 & X \\ 
06 40 41.4 & +09 54 13 & G1 & X \\ 
06 41 08.9 & +09 46 01 & G2.5 & X \\ 
06 41 03.5 & +09 31 19 & G3 & X \\ 
06 40 56.6 & +09 54 10 & G4 & X \\ 
06 41 29.2 & +09 39 36 & G5 & X \\ 
06 41 04.5 & +09 51 50 & G5 & X \\ 
06 39 46.8 & +09 40 54 & G5 & X \\ 
06 40 59.4 & +09 55 20 & G6 & X \\ 
06 40 59.4 & +09 55 20 & G6 & X \\ 
06 40 21.1 & +09 36 32 & G6 & X \\ 
06 41 06.8 & +09 34 46 & G6: & X \\ 
06 40 09.7 & +09 41 43 & G9 & X \\ 
06 41 02.6 & +09 34 56 & K0 & X \\ 
06 41 02.3 & +09 51 52 & K0 & X \\ 
06 39 43.6 & +09 36 04 & K0 & X \\ 
06 41 23.3 & +09 52 42 & K0: & X \\ 
06 41 15.4 & +09 46 40 & K1 & X \\ 
06 41 06.9 & +09 23 22 & K1.5 & X \\ 
06 41 00.3 & +09 58 49 & K1.5 & X \\ 
06 41 01.0 & +09 32 45 & K1: & X \\ 
06 41 36.8 & +09 58 20 & K2 & X \\ 
06 41 31.6 & +09 48 33 & K2 & X \\ 
06 41 27.2 & +09 35 07 & K2 & X \\ 
06 41 05.0 & +09 50 46 & K2 & X \\ 
06 40 58.8 & +09 30 57 & K2 & X \\ 
06 40 48.8 & +09 32 43 & K2 & X \\ 
06 40 47.6 & +09 49 29 & K2 & X \\ 
06 40 45.2 & +09 28 45 & K2 & X \\ 
06 40 30.0 & +09 50 10 & K2 & X \\ 
06 41 04.2 & +09 52 02 & K3 & X \\ 
06 41 00.5 & +09 45 03 & K3 & X \\ 
06 41 21.5 & +09 58 35 & K4 & X \\ 
06 41 18.3 & +09 33 54 & K4 & X \\ 
06 41 16.8 & +09 27 30 & K4 & X \\ 
06 41 09.4 & +09 59 38 & K4 & X \\ 
06 41 01.6 & +10 00 36 & K4 & X \\ 
06 40 51.6 & +09 43 24 & K4 & X \\ 
06 40 39.2 & +09 50 58 & K4 & X \\ 
06 40 30.7 & +09 46 11 & K4 & X \\ 
06 40 28.8 & +09 31 01 & K4 & X \\ 
06 40 16.1 & +09 57 37 & K4 & X \\ 
\hline
\end{tabular}
\end{minipage}
\end{table*}

\begin{figure*}
\begin{center}
\includegraphics[scale=0.5]{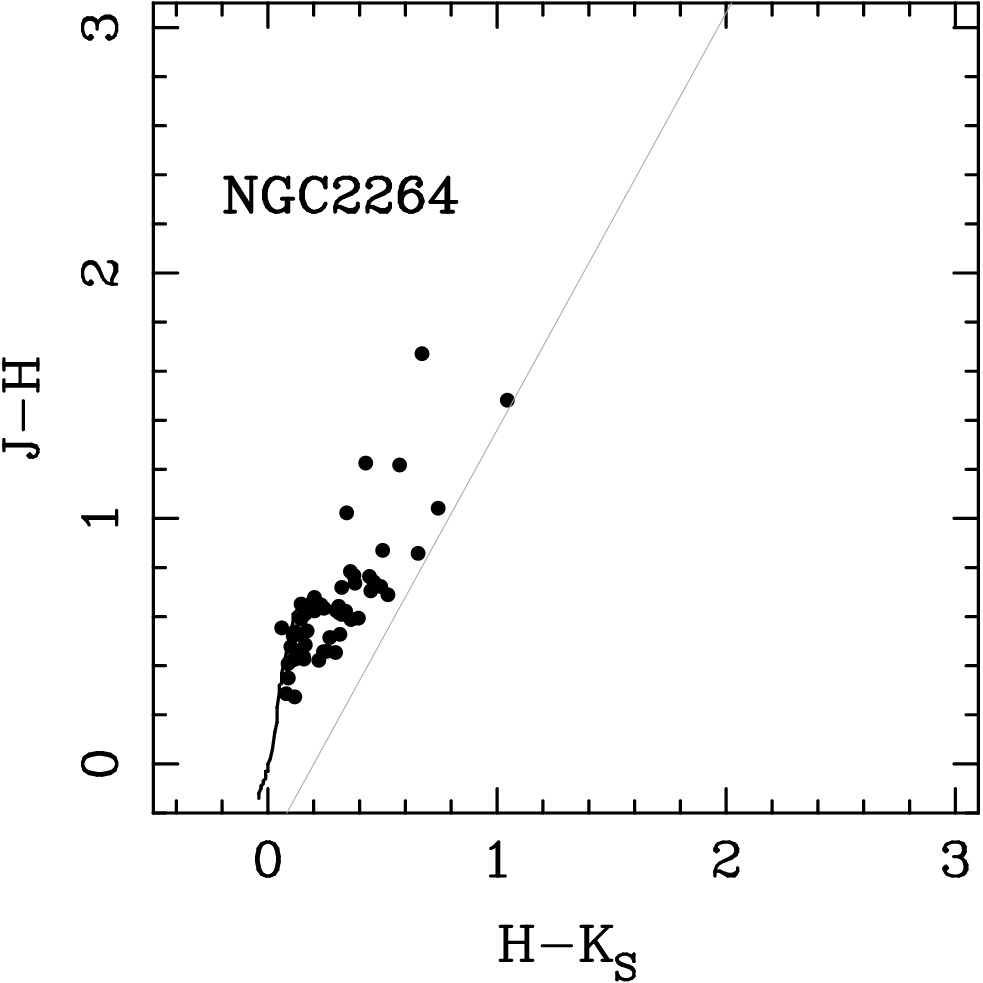}
\caption{NGC 2264.}
\end{center}
\end{figure*}


\begin{table*}
 \centering
 \begin{minipage}{18cm}
\caption{{Tr 37.}
The spectral type with $*$ mark is from SIMBAD database.}

  \begin{tabular}{llllll}
\hline
Name & SpT & $K$\,disk & MIR\,disk \\
\hline
 CCDM+5734Ae & B3 & X & X ($-$2.61)\\
 MVA-63 & B4 & X & ... \\
 MVA-1312 & B4 & X & ... \\
 CCDM+5734Aw & B5 & X & X ($-$2.61) \\
 MVA-805 & B6 & X & ...\\
 MVA-437 & B7 & X & X ($-$2.78) \\
 AG+561491 & B7 & X & \\
 MVA-426 & B7 & o & ...\\
 MVA-468 & B7 & X & X ($-$2.32) \\ 
 MVA-252 & B7 & X & ...\\
 MVA-182 & B8 & X & ...\\
 KUN-196 & B9 & X & o ($-$1.86) \\
 MVA-662 & B9 & X & X ($-$2.60) \\ 
 MVA-535 & B9 & X & X ($-$2.80) \\ 
 MVA-463 & A0 & X & ... \\
 SBZ-2-46 & A0 & X & ...\\
 MVA-81 & A0 & X & ...\\
 tr37-185 & A1 & X & ...\\
 BD+572362 & A1 & X & ... \\
 MVA-497 & A1 & X & ...\\
 MVA-566 & A1 & X & ...\\
 KUN-318 & A1 & X & X ($-$2.90) \\
 KUN-197 & A2 & X & X ($-$2.93) \\
 MVA-660 & A2 & X & ...\\
 MVA-258 & A2 & X & ...\\
 MVA-164 & A3 & X & ...\\
 BD+572355 & A4 & X & X ($-$2.67) \\
 MVA-169 & A4 & X & ...\\
 MVA-640 & A7 & X & ...\\
 BD+572356 & A7 & X & X ($-$2.74) \\
 MVA-545 & A7 & X & ...\\
 MVA-224e & A7 & X & X ($-$2.94) \\
 KUN-89 & A8 & X & X ($-$2.66) \\
 MVA-472 & A8 & X & ...\\
 MVA-564 & A9 & X & ...\\
 MVA-657 & F0 & X & ...\\
 KUN-87 & F0 & X & ...\\
 MVA-447 & F0 & X & ...\\
 KUN-93 & F1 & X & X ($-$2.79) \\
 KUN-327 & F1 & X & ...\\
 KUN-100 & F3 & X & ...\\
 KUN-198 & F3 & X & X ($-$2.58) \\
 KUN-191 & F5 & X & X ($-$2.83) \\
 KUN-97 & F6 & X & ...\\
 KUN-58 & F6 & X & ...\\
 KUN-85 & F7 & X & ...\\
 MVA-523 & F7 & X & ...\\
 MVA-232 & F7 & X & X ($-$2.77) \\ 
 KUN-92 & F9 & X & ...\\
 KUN-86 & F9 & X & o ($-$2.15) \\
 KUN-84 & F9 & X & ...\\
 KUN-83 & F9 & X & X ($-$3.04) \\
 MVA-234 & F9 & X & ...\\
 KUN-56 & F9 & X & ...\\
 KUN-314S & A* & o & ...\\
KUN-56 & F9.0 & X & ...\\
$[$SHB2004$]$ 11-581 & G & X & X ($-$2.79) \\
$[$SHB2004$]$ 11-1864 & G--K & ... & X ($-$2.35) \\
$[$SHB2004$]$ 93-361 & G1 & ... & ...\\
$[$SHB2004$]$ 13-277 & G1 & X & o ($-$0.76) \\
$[$SHB2004$]$ 73-537 & G1.5 & ... & o ($-$0.88) \\
$[$SHB2004$]$ 12-1091 & G2.5 & X & o ($-$1.17) \\
\hline
  \end{tabular}
\end{minipage}
\end{table*}

\begin{table*}
 \centering
 \begin{minipage}{18cm}
\contcaption{}
  \begin{tabular}{llllll}
\hline
Name & SpT & $K$\,disk & MIR disk \\
\hline
$[$SHB2004$]$ 22-404 &  G7 & X & ...\\
$[$SHB2004$]$ 21-1974 & G7.5 & X & ...\\
$[$SHB2004$]$ 82-272 & G9 & X & o ($-$1.20) \\
$[$SHB2004$]$ 13-669 & K1 & X & o ($-$1.01) \\
$[$SHB2004$]$ 13-236 & K2 & X & o ($-$1.35) \\
$[$SHB2004$]$ 11-2031 & K2 & X & o ($-$0.75) \\
$[$SHB2004$]$ 24-542 & K4 & X & X ($-$2.70) \\ 
$[$SHB2004$]$ 13-1087 & K4 & X &  X ($-$2.69) \\
$[$SHB2004$]$ 12-94 & K4 & X & X ($-$2.55) \\
\hline
\end{tabular}
\end{minipage}
\end{table*}

\clearpage
\begin{figure*}
\begin{center}
\includegraphics[scale=0.5]{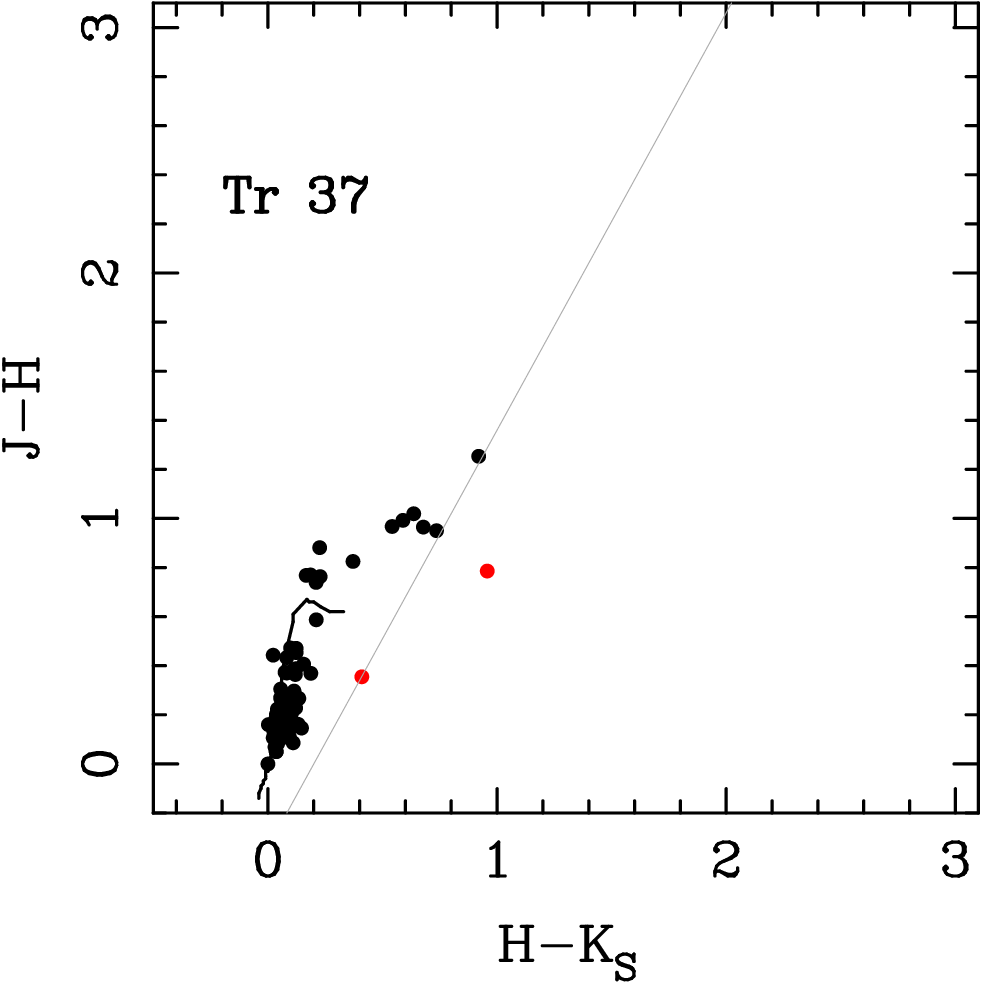}
\caption{Tr 37.}
\end{center}
\end{figure*}


\clearpage

\begin{table*}
 \centering
 \begin{minipage}{18cm}
\caption{Ori OB1bc.}
  \begin{tabular}{llllll}
\hline
 RAJ2000 & DEJ2000 & SpT & $K$\,disk \\
 (h:m:s) & (d:m:s) &   & \\
\hline
05 41 08.1 & $-$03 37 57 & B3 & X \\ 
05 35 35.9 & $-$03 15 10 & B3 & X \\ 
05 35 12.8 & $-$00 44 07 & B3 & X \\ 
05 33 07.3 & $-$01 43 02 & B3 & X \\ 
05 37 45.9 & $-$00 46 42 & B4 & X \\ 
05 35 22.3 & $-$04 25 28 & B4 & X \\ 
05 48 46.0 & +00 43 32 & B5 & X \\ 
05 39 02.4 & $-$05 11 40 & B5 & X \\ 
05 39 01.5 & $-$02 38 56 & B5 & X \\ 
05 37 34.8 & $-$01 25 20 & B5 & X \\ 
05 37 14.5 & $-$01 40 04 & B5 & X \\ 
05 36 17.8 & $-$01 38 07 & B6 & X \\ 
05 35 09.2 & $-$00 16 11 & B6 & X \\ 
05 59 37.7 & $-$01 26 39 & B7 & X \\ 
05 43 43.8 & $-$00 56 19 & B7 & X \\ 
05 40 25.3 & $-$04 25 16 & B7 & X \\ 
05 38 06.5 & $-$00 11 03 & B7 & X \\ 
05 34 56.5 & $-$00 07 22 & B7 & X \\ 
05 34 19.8 & +04 49 30 & B7 & X \\ 
05 20 07.8 & $-$05 50 46 & B7 & X \\ 
05 53 27.1 & +00 46 45 & B8 & X \\ 
05 49 32.7 & $-$00 40 55 & B8 & X \\ 
05 42 48.8 & +04 51 09 & B8 & X \\ 
05 38 31.3 & $-$00 08 52 & B8 & X \\ 
05 37 30.3 & $-$00 14 25 & B8 & X \\ 
05 33 07.5 & $-$05 20 26 & B8 & X \\ 
05 32 14.8 & $-$04 31 06 & B8 & X \\ 
05 30 48.7 & +00 01 43 & B8 & X \\ 
05 30 43.0 & $-$05 29 27 & B8 & X \\ 
05 28 52.6 & $-$00 36 11 & B8 & X \\ 
05 16 34.3 & $-$05 03 41 & B8 & X \\ 
05 14 52.8 & $-$04 37 36 & B8 & X \\ 
05 59 14.6 & $-$04 21 34 & B9 & X \\ 
05 58 36.1 & $-$02 05 57 & B9 & X \\ 
05 51 09.5 & $-$04 34 57 & B9 & X \\ 
05 49 13.1 & +01 27 30 & B9 & X \\ 
05 46 41.3 & +02 14 27 & B9 & X \\ 
05 42 17.6 & +02 22 02 & B9 & X \\ 
05 41 02.3 & $-$02 43 01 & B9 & o \\
05 39 55.4 & $-$03 19 50 & B9 & X \\ 
05 39 45.2 & +04 26 05 & B9 & X \\ 
05 38 50.2 & $-$04 16 18 & B9 & X \\ 
05 36 14.1 & $-$02 15 32 & B9 & X \\ 
05 35 39.9 & $-$03 18 58 & B9 & X \\ 
05 35 13.8 & $-$02 22 52 & B9 & X \\ 
05 33 45.5 & $-$00 01 44 & B9 & X \\ 
05 33 26.1 & +00 37 17 & B9 & X \\ 
05 33 05.6 & $-$01 43 16 & B9 & X \\ 
05 33 03.7 & $-$01 14 28 & B9 & X \\ 
05 30 10.4 & $-$05 12 06 & B9 & X \\ 
05 29 08.9 & $-$05 47 28 & B9 & X \\ 
05 28 40.4 & $-$02 44 01 & B9 & X \\ 
05 27 43.2 & $-$00 15 33 & B9 & X \\ 
05 20 32.9 & $-$05 17 17 & B9 & X \\ 
05 20 28.9 & $-$05 48 44 & B9 & X \\ 
05 17 54.8 & $-$04 29 24 & B9 & X \\ 
05 15 05.2 & $-$05 15 09 & B9 & X \\ 
05 10 47.6 & $-$05 10 11 & B9 & X \\ 
05 59 35.6 & $-$04 20 15 & A0 & X \\ 
05 55 57.3 & +00 50 10 & A0 & X \\ 
05 50 24.1 & +01 46 43 & A0 & X \\ 
05 49 53.7 & $-$00 11 01 & A0 & X \\ 
05 46 43.2 & +02 42 26 & A0 & X \\ 
\hline
\end{tabular}
\end{minipage}
\end{table*}

\begin{table*}
 \centering
 \begin{minipage}{18cm}
\contcaption{}
  \begin{tabular}{llllll}
\hline
 RAJ2000 & DEJ2000 & SpT & $K$\,disk \\
 (h:m:s) & (d:m:s) &   & \\
\hline
05 46 12.4 & $-$01 31 25 & A0 & X \\ 
05 44 48.6 & $-$00 03 43 & A0 & X \\ 
05 43 11.9 & $-$04 59 50 & A0 & o \\
05 42 58.8 & $-$04 49 58 & A0 & X \\ 
05 40 40.6 & $-$03 55 11 & A0 & X \\ 
05 35 37.5 & $-$03 34 42 & A0 & X \\ 
05 32 49.8 & $-$02 11 49 & A0 & X \\ 
05 56 26.6 & $-$01 49 27 & A1 & X \\ 
05 34 23.7 & +05 25 11 & A1 & X \\ 
05 31 21.2 & $-$02 05 57 & A1 & X \\ 
05 50 28.6 & $-$04 58 37 & A2 & X \\ 
05 50 23.9 & +04 57 24 & A2 & X \\ 
05 38 09.2 & $-$00 10 56 & A2 & o \\
05 16 06.6 & $-$04 27 51 & A2 & X \\ 
05 45 15.1 & $-$05 06 41 & A3 & X \\ 
05 26 16.3 & $-$03 04 34 & A3 & X \\ 
05 52 22.6 & $-$00 55 03 & A4 & X \\ 
05 37 40.5 & $-$02 26 37 & A4 & X \\ 
05 32 16.7 & $-$03 33 51 & A4 & X \\ 
05 56 49.4 & $-$03 04 17 & A5 & X \\ 
05 50 13.1 & +02 24 53 & A5 & X \\ 
05 44 18.8 & +00 08 40 & A9 & o \\
05 26 41.1 & $-$05 09 24 & F0 & X \\ 
05 31 18.4 & $-$05 42 14 & F1 & X \\ 
05 02 44.0 & $-$05 42 22 & F1 & X \\ 
05 44 16.9 & $-$02 20 36 & F2 & X \\ 
05 57 01.0 & $-$02 10 00 & F4 & X \\ 
05 31 04.7 & $-$03 56 00 & F5 & X \\ 
05 18 26.7 & $-$04 37 16 & F6 & X \\ 
05 40 24.4 & +02 04 20 & G3 & X \\ 
\hline
\end{tabular}
\end{minipage}
\end{table*}

\clearpage
\begin{figure*}
\begin{center}
\includegraphics[scale=0.5]{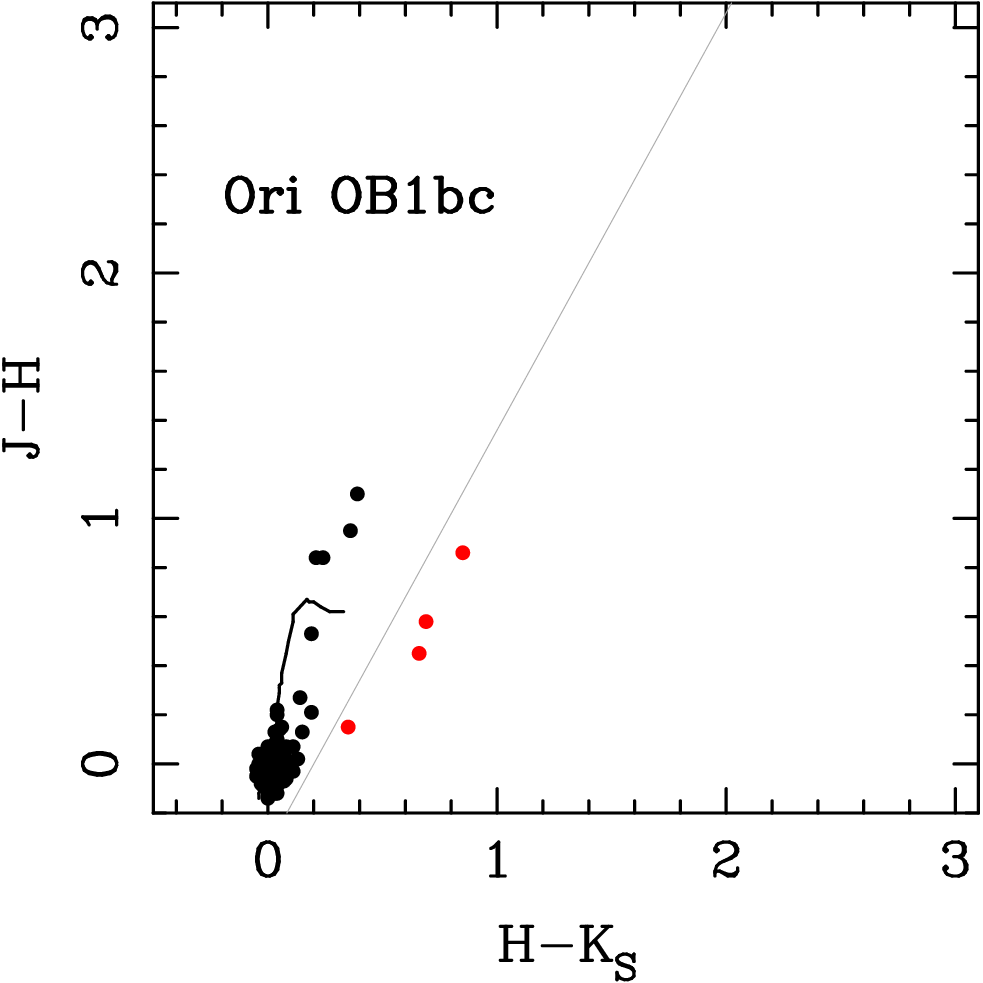}
\caption{Ori OB1bc.}
\end{center}
\end{figure*}


\clearpage
\begin{table*}
 \centering
 \begin{minipage}{18cm}
\caption{Upper Sco.} 
  \begin{tabular}{llllll}
\hline
Name & SpT & $K$\,disk & MIR disk \\
\hline
HIP 78168 & B3V & X & X ($-$2.94) \\
HIP 78246 & B5V & X & X ($-$2.93) \\
HIP 77858 & B5V & X & X ($-$2.90) \\
HIP 79530 & B6IV & X & X ($-$2.92) \\
HIP 77900 & B7V & X & X ($-$2.94) \\
HIP 78207 & B8Ia/Iab & X & o ($-$2.02) \\
HIP 80338 & B8II & X & X ($-$2.86) \\
HIP 77909 & B8III/IV & X & X ($-$2.92) \\
HIP 79739 & B8V & X & X ($-$2.92) \\
HIP 78877 & B8V & X & X ($-$2.90) \\
HIP 78956 & B9.5V & X & X ($-$2.89) \\
HIP 78549 & B9.5V & X & X ($-$2.93) \\
HIP 80024 & B9II/III & X & X ($-$2.91) \\
HIP 80493 & B9V & X & X ($-$2.91) \\
HIP 79897 & B9V & X & X ($-$2.93) \\
HIP 79785 & B9V & X & X ($-$2.89) \\
HIP 79771 & B9V & X & X ($-$2.89) \\
HIP 79439 & B9V & X & X ($-$2.83) \\
HIP 79410 & B9V & X & X ($-$2.75) \\
HIP 78968 & B9V & X & X (-3.01) \\
HIP 78809 & B9V & X & X ($-$2.92) \\
HIP 78702 & B9V & X & X ($-$2.90) \\
HIP 78530 & B9V & X & X ($-$2.92) \\
HIP 77911 & B9V & X & X ($-$2.88) \\
HIP 76633 & B9V & X & X ($-$2.86) \\
HIP 76071 & B9V & X & X ($-$2.94) \\
HIP 80311 & A0V & X & X ($-$2.87) \\
HIP 79878 & A0V & X & X ($-$2.88) \\
HIP 79860 & A0V & X & X ($-$2.87) \\
HIP 79156 & A0V & X & X ($-$2.77) \\
HIP 79124 & A0V & X & X ($-$2.90) \\
HIP 78847 & A0V & X & X ($-$2.88) \\
HIP 78196 & A0V & X & X ($-$2.93) \\
HIP 78099 & A0V & X & X ($-$2.92) \\
HIP 76310 & A0V & X & X ($-$2.83) \\
HIP 80324 & A0V+A0V & X & X ($-$2.88) \\
HIP 79733 & A1mA9-F2 & X & X ($-$2.88) \\
HIP 77545 & A2/3V & X & X ($-$2.82) \\
HIP 79392 & A2IV & X & X ($-$2.86) \\
HIP 78494 & A2mA7-F2 & X & X ($-$2.90) \\
HIP 79250 & A3III/IV & X & X ($-$2.91) \\
HIP 82397 & A3V & X & X ($-$2.90) \\
HIP 79366 & A3V & X & X ($-$2.95) \\
HIP 77960 & A4IV/V & X & X ($-$2.95) \\
HIP 77815 & A5V & X & X ($-$2.88) \\
HIP 80059 & A7III/IV & X & X ($-$2.77) \\
HIP 77457 & A7IV & X & X ($-$2.91) \\
HIP 80130 & A9V & X & X ($-$2.98) \\
HIP 80088 & A9V & X & X ($-$2.77) \\
HIP 78996 & A9V & X & X ($-$2.73) \\
HIP 78963 & A9V & X & X ($-$2.96) \\
HIP 79643 & F2 & X & X ($-$2.84) \\
HIP 78233 & F2/3IV/V & X & X ($-$2.84) \\
HIP 82319 & F3V & X & X ($-$2.87) \\
HIP 80896 & F3V & X & X ($-$2.93) \\
HIP 79097 & F3V & X & X ($-$2.86) \\
HIP 79083 & F3V & X & X ($-$2.89) \\
HIP 79644 & F5 & X & X ($-$2.85) \\
HIP 79606 & F6 & X & X ($-$2.84) \\
RX J1550.9-2534 & F9 & X & X ($-$2.83) \\
\hline
\end{tabular}
\end{minipage}
\end{table*}

\begin{table*}
 \centering
 \begin{minipage}{18cm}
\contcaption{}
  \begin{tabular}{llllll}
\hline
$[$PZ99$]$ J160000.7-250941 & G0 & X & X ($-$2.83) \\
HD 149598 & G0 & X & X ($-$2.79) \\
HD 146516 & G0IV & X & X ($-$2.81) \\
HIP 78483 & G0V & X & X ($-$2.80) \\
HD 147810 & G1 & X & X ($-$2.83) \\
$[$PZ99$]$ J155812.7-232835 & G2 & X & X ($-$2.79) \\
HIP 79462 & G2V & X & X ($-$2.80) \\
PPM 747978 & G3 & X & X ($-$2.81) \\
PPM 747651 & G3 & X & X ($-$2.78) \\
HD 142361 & G3V & X & X ($-$2.78) \\
$[$PZ99$]$  J161402.1-230101 & G4 & X & X ($-$2.75) \\
HD 142987 & G4 & X & X ($-$2.75) \\
$[$PZ99$]$ J161459.2-275023 & G5 & X & X ($-$2.74) \\
PPM 732705 & G6 & X & X ($-$2.79) \\
RX J1541.1-2656 & G7 & X & X ($-$2.76) \\
$[$PZ99$]$ J161618.0-233947 & G7 & X & X ($-$2.74) \\
SAO 183706 & G8e & X & X ($-$2.81) \\
RX J1600.6-2159 & G9 & X & X ($-$2.78) \\
$[$PZ99$]$ J161318.6-221248 & G9 & X & X ($-$2.75) \\
RX J1603.6-2245 & G9 & X & X ($-$2.77) \\
RX J1548.0-2908 & G9 & X & X ($-$2.78) \\
$[$PZ99$]$ J161411.0-230536 & K0 & X & o ($-$1.18) \\
RX J1602.8-2401A & K0 & X & X ($-$2.70) \\
$[$PZ99$]$ J161933.9-222828 & K0 & X & X ($-$2.68) \\
$[$PZ99$]$ J161329.3-231106 & K1 & X & X ($-$2.72) \\
ScoPMS 21 & K1IV & X & X ($-$2.74) \\
$[$PZ99$]$ J160814.7-190833 & K2 & X & X ($-$2.75) \\
$[$PZ99$]$ J160421.7-213028 & K2 & X & X ($-$2.55) \\
ScoPMS 27 & K2IV & X & X ($-$2.75) \\
$[$PZ99$]$ J155847.8-175800 & K3 & X & X ($-$2.73) \\
$[$PZ99$]$ J153557.8-232405 & K3: & X & X ($-$2.75) \\
RX J1558.1-2405A & K4 & X & X ($-$2.75) \\
$[$PZ99$]$ J161302.7-225744 & K4 & X & X ($-$2.76) \\
$[$PZ99$]$ J160251.2-240156 & K4 & X & X ($-$2.73) \\
\hline
\end{tabular}
\end{minipage}
\end{table*}

\clearpage
\begin{figure*}
\begin{center}
\includegraphics[scale=0.5]{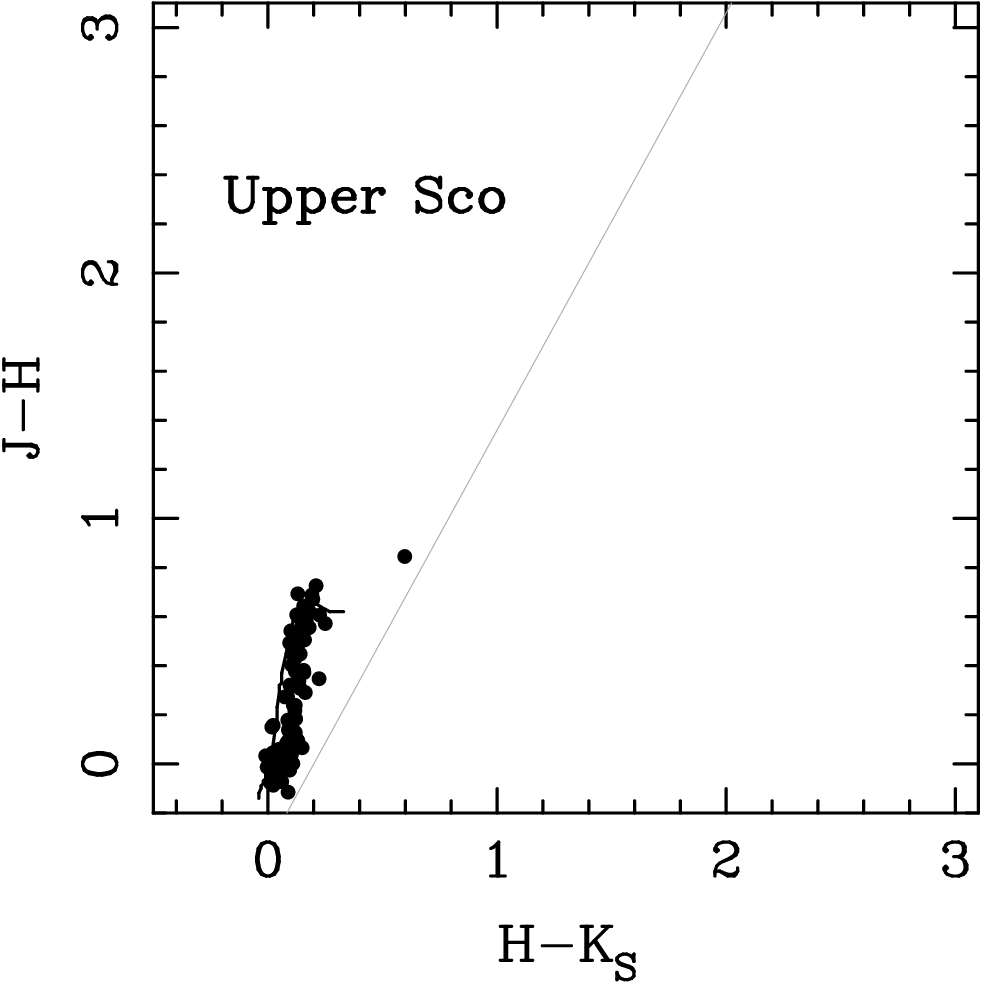}
\caption{Upper Sco.}
\end{center}
\end{figure*}


\clearpage
\begin{table*}
 \centering
 \begin{minipage}{18cm}
\caption{NGC 2362.}
  \begin{tabular}{llllll}
\hline
RAJ2000 & DEJ2000 & SpT & $K$\,disk & MIR disk \\
(h:m:s) & (d:m:s) &   & \\
\hline
07 18 58.40 & $-$24 57 41.2 & B3 & X & X ($-$2.88) \\
07 18 36.85 & $-$24 56 05.7 & B3 & X & X ($-$2.95) \\
07 18 40.81 & $-$24 58 27.5 & B5 & X & X ($-$2.91) \\
07 18 45.74 & $-$24 59 35.6 & B7 & X & X ($-$2.93) \\
07 18 38.10 & $-$24 59 01.6 & B9 & X & X ($-$2.88) \\
07 18 54.54 & $-$24 57 29.2 & A0 & X & X ($-$2.72) \\
07 18 35.48 & $-$24 58 59.5 & F2 & X & X ($-$2.91) \\
07 18 34.01 & $-$24 58 04.6 & F2 & X & X ($-$2.79) \\
07 18 48.54 & $-$25 01 48.6 & G2 & X & X ($-$2.73) \\
07 18 46.89 & $-$24 57 01.6 & G6 & X & X ($-$2.81) \\
07 18 59.61 & $-$24 58 51.3 & G8 & X & X ($-$2.87) \\
07 18 32.46 & $-$24 58 09.3 & K1 & X & X ($-$2.73) \\
07 18 24.51 & $-$24 54 32.3 & K1 & X & X ($-$2.78) \\
07 18 43.36 & $-$24 56 17.9 & K2 & X & X ($-$2.87) \\
07 18 40.20 & $-$24 55 13.1 & K2 & X & X ($-$2.91) \\
07 18 46.46 & $-$24 57 09.6 & K3 & X & X ($-$3.02) \\
07 18 31.63 & $-$25 01 47.5 & K3 & X & X ($-$2.89) \\
07 18 50.90 & $-$24 57 03.5 & K4 & X & X ($-$2.78) \\
07 18 35.31 & $-$25 00 35.3 & K4 & X & X ($-$2.87) \\
\hline
\end{tabular}
\end{minipage}
\end{table*}

\begin{figure*}
\begin{center}
\includegraphics[scale=0.5]{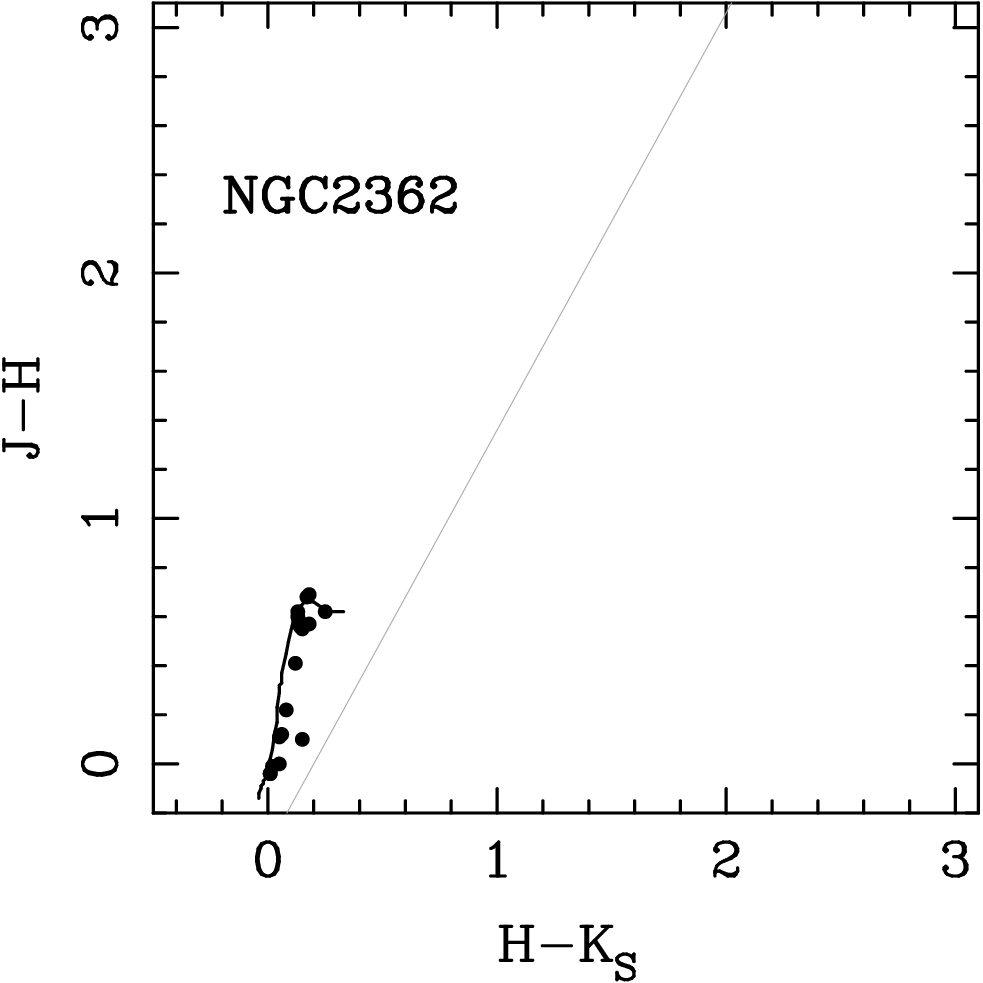}
\caption{NGC 2362.}
\end{center}
\end{figure*}


\clearpage
\begin{table*}
 \centering
 \begin{minipage}{18cm}
\caption{$\gamma$ Vel.}
The spectral type with $*$ mark is from SIMBAD database.\\

  \begin{tabular}{llllll}
\hline
RAJ2000 & DEJ2000 & SpT & $K$\,disk & MIR disk \\
 (h:m:s) & (d:m:s) &   & \\
\hline
08 11 3929 & $-$47 21 06.5 & B2/B3III/IV & X & X ($-$2.89) \\
08 08 5123 & $-$47 10 27.7 & B3V & X & X ($-$2.87) \\
08 07 4074 & $-$47 15 17.5 & B8IV & X & X ($-$2.85) \\
08 08 2188 & $-$47 09 28.6 & B8Vne* & X & X ($-$2.72) \\
08 09 1107 & $-$46 59 53.4 & B9V & X & X ($-$2.80) \\
08 09 0430 & $-$47 41 02.4 & A0/A1V & X & X ($-$2.78) \\
08 08 2593 & $-$47 36 06.9 & A0V & X & X ($-$2.85) \\
08 09 0738 & $-$47 38 13.6 & A0V & X & X ($-$2.79) \\
08 11 1618 & $-$47 13 18.8 & A0V & X & X ($-$2.82) \\
08 09 1637 & $-$47 13 37.4 & A1/A2V & X & X ($-$2.80) \\
08 08 0690 & $-$47 15 07.4 & A1V & X & X ($-$2.85) \\
08 10 3253 & $-$47 12 40.9 & A2* & X & X ($-$2.86) \\
08 10 5813 & $-$47 29 13.6 & A2V & X & X ($-$2.85) \\
08 11 2187 & $-$47 11 28.1 & A5* & X & X ($-$2.82) \\
08 09 3482 & $-$47 21 06.9 & F0* & X & X ($-$2.86) \\
08 09 3763 & $-$47 21 25.6 & F0* & X & X ($-$2.78) \\
08 10 4836 & $-$47 34 55.9 & F5* & X & X ($-$2.79) \\
\hline
\end{tabular}
\end{minipage}
\end{table*}

\begin{figure*}
\begin{center}
\includegraphics[scale=0.5]{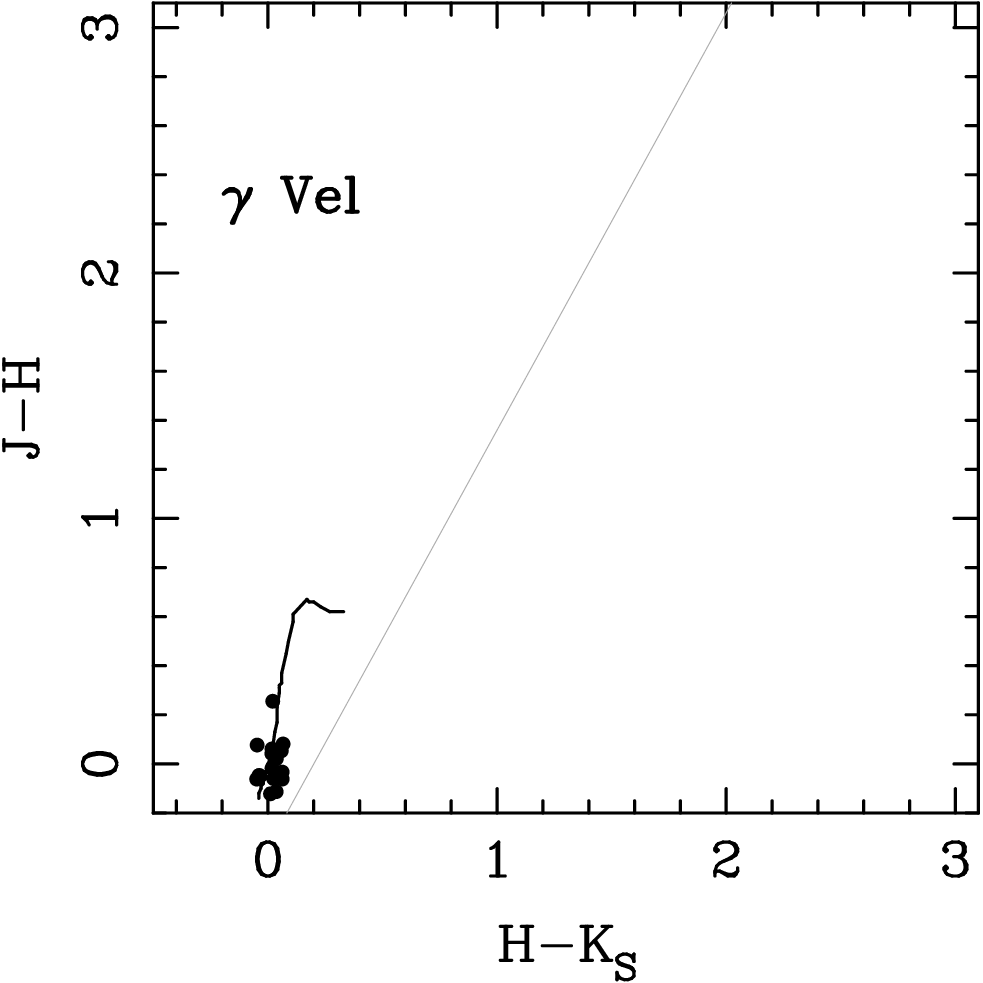}
\caption{$\gamma$ Vel.}
\end{center}
\end{figure*}


\clearpage
\begin{table*}
 \centering
 \begin{minipage}{18cm}
\caption{$\lambda$ Ori. }
  \begin{tabular}{llllll}
\hline
Name & RAJ2000 & DEJ2000 & SpT & $K$\,disk & MIR\,disk\\
 & (h:m:s) & (d:m:s) &   & \\
\hline
HD36895 & 05 35 1280 & +09 36 47.8 & B3 & X & X ($-$2.91) \\ 
HD245203 & 05 35 1380 & +09 41 49.4 & B8 & X & X ($-$2.89) \\ 
HD37035 & 05 35 5825 & +09 31 54.1 & B9 & X & X ($-$2.84) \\ 
HD37110 & 05 36 2962 & +09 37 54.2 & B8 & X & X ($-$2.75) \\ 
HD37051 & 05 36 0418 & +09 49 55.0 & B9 & X & X ($-$2.77) \\ 
HD245140 & 05 34 5817 & +09 56 26.7 & B9 & X & X ($-$2.81) \\ 
HD245168 & 05 35 02968 & +09  56 04.1 &  B9 & ... & X ($-$2.78) \\ 
HD37034 & 05 35 5938 & +09 42 48.0 & A0 & X & X ($-$2.73) \\ 
HD245185 & 05 35 0960 & +10 01 51.5 & A0 & o & o (0.29) \\  
HD245385 & 05 36 1338 & +09 59 24.4 & A0 & ... & X ($-$2.83) \\ 
HD244908 & 05 33 4712 & +09 40 26.1 & A2 & ... & X ($-$2.85) \\ 
HD245386 & 05 36 2132 & +09 50 41.4 & A2 & ... & X ($-$2.89) \\ 
HD37159 & 05 36 5811 & +10 16 58.6 & A3 & X & X ($-$2.80) \\ 
... & 05 34 4857 & +09 30 57.1 & A4 & ... & X ($-$2.73) \\ 
HD245275 & 05 35 4485 & +09 55 24.3 & A5 & ... & X ($-$2.83) \\ 
HD244927 & 05 33 5042 & +10 04 21.1 & A7 & X & X ($-$2.87) \\ 
... & 05 34 5914 & +09 33 50.8 & F3 & ... & X ($-$2.88) \\ 
299-3 & 05 36 0529 & +10 21 27.1 & F3 & ... & X ($-$2.87) \\ 
HD245370 & 05 36 0940 & +10 01 25.4 & F4 & X & X ($-$2.58) \\ 
... & 05 33 4028 & +09 48 01.3 & F6 & ... & X ($-$2.86) \\ 
... & 05 33 5032 & +09 58 18.5 & F7 & X & X ($-$2.86) \\ 
... & 05 35 2468 & +10 11 45.2 & F7 & ... & X ($-$2.86) \\ 
HD244907 & 05 33 5115 & +09 46 42.1 & F8 & X & X ($-$2.84) \\ 
h-star & 05 35 0920 & +10 02 51.8 & F8 & ... & X ($-$2.87) \\ 
... & 05 36 5226 & +09 29 58.4 & F9 & ... & X ($-$2.88) \\ 
... & 05 35 4220 & +10 13 44.7 & G0 & ... & X ($-$2.83) \\ 
\hline
\end{tabular}
\end{minipage}
\end{table*}

\begin{figure*}
\begin{center}
\includegraphics[scale=0.5]{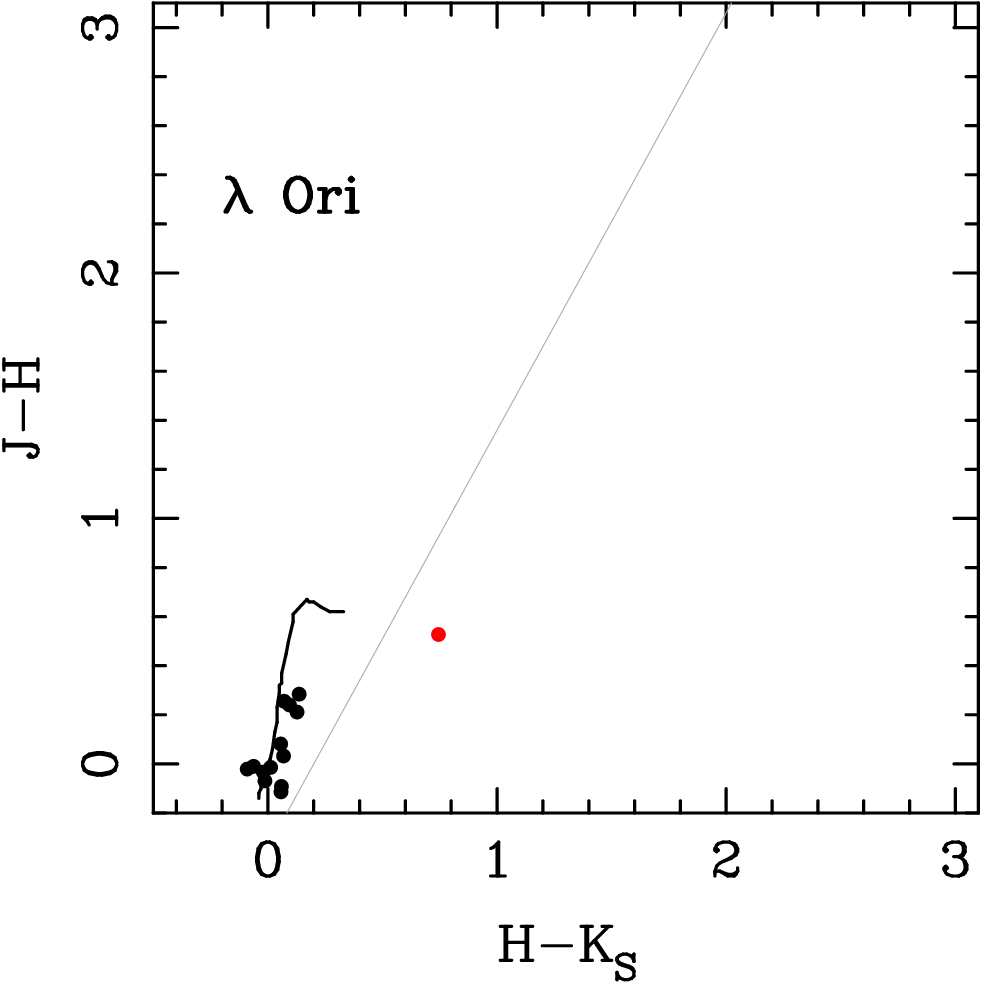}
\caption{$\lambda$ Ori.}
\end{center}
\end{figure*}


\clearpage

\begin{table*}
 \centering
 \begin{minipage}{18cm}
\caption{Per OB2.}
  \begin{tabular}{llllll}
\hline
 RAJ2000 & DEJ2000 & SpT & $K$\,disk \\
 (h:m:s) & (d:m:s) &   & \\
\hline
04 06 39.0 & +32 23 06 & B3 & X \\ 
03 47 52.7 & +33 36 00 & B3 & X \\ 
03 47 25.7 & +29 52 33 & B3 & X \\ 
04 06 55.8 & +33 26 47 & B4 & X \\ 
03 49 07.3 & +32 15 51 & B6 & X \\ 
03 25 50.1 & +30 55 54 & B6 & X \\ 
03 50 51.3 & +35 05 59 & B7 & X \\ 
03 44 40.7 & +29 49 21 & B7 & X \\ 
03 55 58.7 & +32 09 48 & B8 & X \\ 
04 12 45.2 & +31 47 41 & B9 & X \\ 
04 07 24.5 & +33 05 17 & B9 & X \\ 
04 02 56.6 & +31 55 54 & B9 & X \\ 
03 58 35.5 & +31 24 30 & B9 & X \\ 
03 55 54.9 & +32 09 18 & B9 & X \\ 
03 54 20.7 & +30 59 55 & B9 & X \\ 
03 44 51.3 & +30 08 09 & B9 & X \\ 
03 28 17.4 & +29 52 07 & B9 & X \\ 
03 20 53.5 & +38 53 07 & B9 & X \\ 
03 07 51.0 & +33 03 18 & B9 & X \\ 
03 06 35.1 & +38 36 07 & B9 & X \\ 
03 58 55.5 & +32 45 23 & A0 & X \\ 
03 34 57.9 & +29 18 48 & A0 & X \\ 
03 11 57.6 & +38 32 17 & A0 & X \\ 
03 10 06.3 & +38 20 44 & A0 & X \\ 
03 03 11.3 & +41 20 07 & A1 & X \\ 
03 40 40.3 & +29 27 17 & A5 & X \\ 
03 02 23.6 & +43 11 02 & F8 & X \\ 
03 36 00.0 & +23 54 51 & G4 & X \\ 
03 55 06.9 & +27 03 52 & G5 & X \\ 
03 22 11.9 & +27 36 27 & G6 & X \\ 
03 46 09.5 & +28 51 33 & G8 & X \\ 
\hline
\end{tabular}
\end{minipage}
\end{table*}

\begin{figure*}
\begin{center}
\includegraphics[scale=0.5]{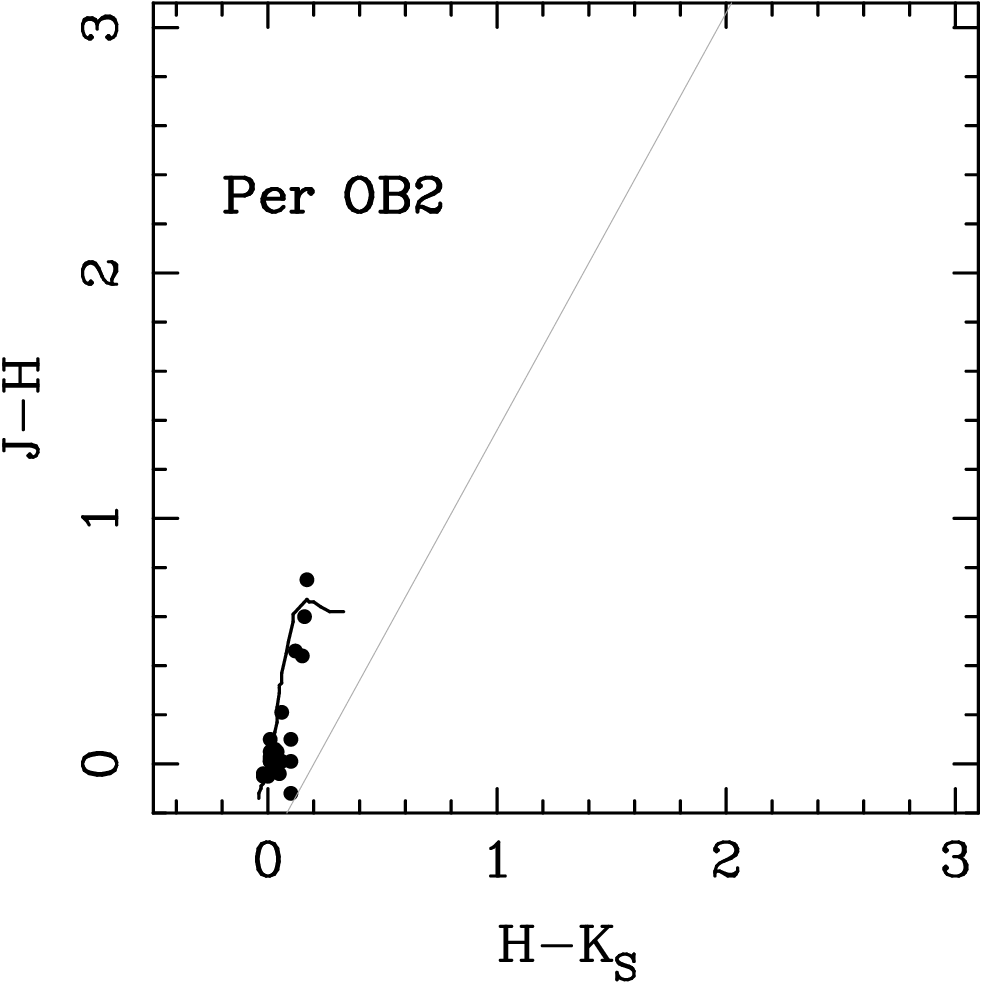}
\caption{Per OB2.}
\end{center}
\end{figure*}


\clearpage
\begin{table*}
 \centering
 \begin{minipage}{18cm}
\caption{$\eta$ Cha.}
  \begin{tabular}{llllll}
\hline
Name & RAJ2000 & DEJ2000 & SpT & $K$\,disk \\
 & (h:m:s) & (d:m:s) &   & \\
\hline
$\eta$ Cha & 08 41 19.51 & $-$78 57 48.1 & B8V & X \\
HD 75505 & 08 41 44.71 & $-$79 02 53.3 &A1V & X \\
RS Cha & 08 43 12.22 & $-$79 04 12.3 & A7V & X \\
\hline
\end{tabular}
\end{minipage}
\end{table*}

\begin{figure*}
\begin{center}
\includegraphics[scale=0.5]{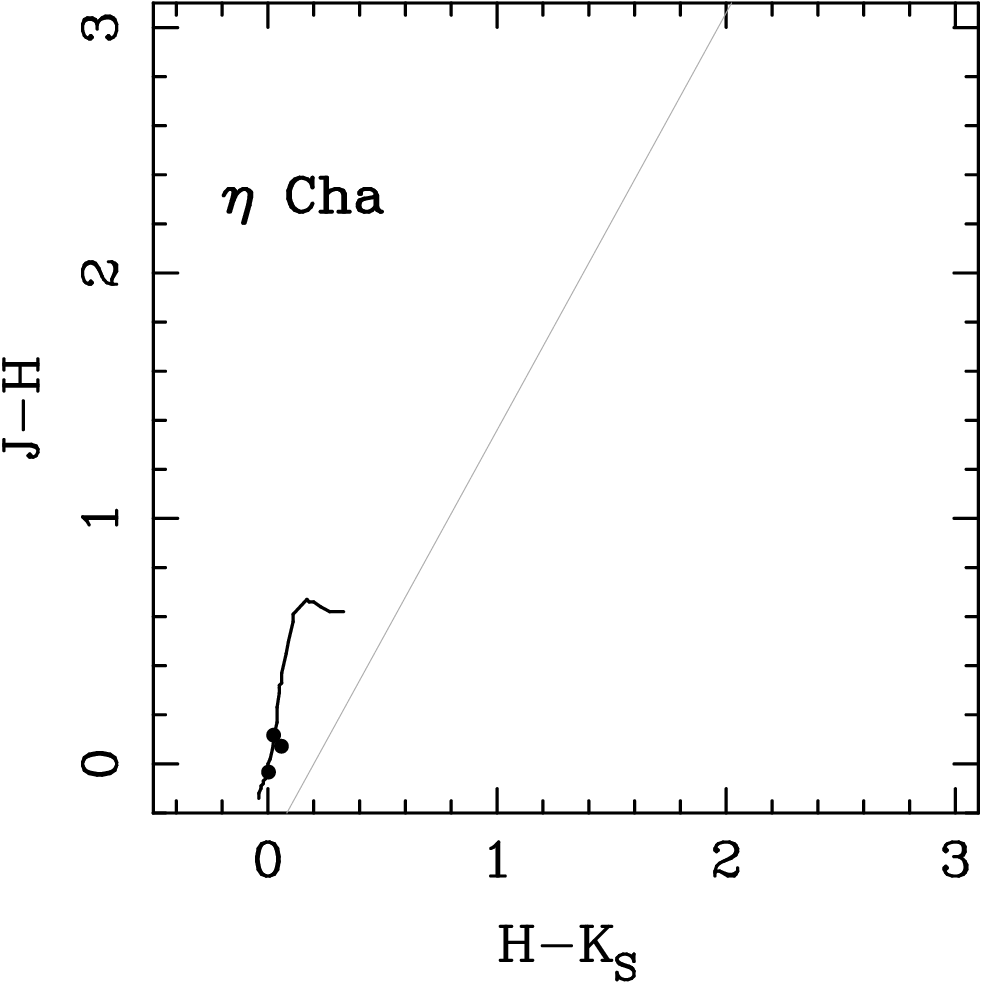}
\caption{$\eta$ Cha.}
\end{center}
\end{figure*}


\clearpage
\begin{table*}
 \centering
 \begin{minipage}{18cm}
\caption{Ori OB1a.}

  \begin{tabular}{llllll}
\hline
 RAJ2000 & DEJ2000 & SpT & $K$\,disk \\
 (h:m:s) & (d:m:s) &   & \\
\hline
05 31 29.9 & +01 41 24 & B3 & X \\ 
05 27 09.4 & $-$01 22 02 & B3 & X \\ 
05 21 31.8 & $-$00 24 59 & B3 & X \\ 
05 28 45.3 & +01 38 38 & B4 & X \\ 
05 27 45.8 & $-$02 08 44 & B4 & X \\ 
05 26 54.3 & +03 36 53 & B4 & X \\ 
05 24 36.1 & +02 21 11 & B4 & X \\ 
05 22 51.0 & +03 33 08 & B4 & X \\ 
05 13 39.1 & $-$03 37 19 & B4 & X \\ 
05 04 54.5 & $-$03 02 23 & B5 & X \\ 
05 37 56.3 & +00 59 15 & B6 & X \\ 
05 37 53.5 & +00 58 07 & B6 & X \\ 
05 33 08.9 & +03 07 52 & B6 & X \\ 
05 31 41.2 & +02 49 58 & B6 & X \\ 
05 28 48.5 & +02 09 53 & B6 & X \\ 
05 27 44.7 & $-$01 48 47 & B6 & X \\ 
05 25 01.2 & $-$02 48 56 & B6 & X \\ 
05 18 01.0 & $-$00 02 16 & B6 & X \\ 
05 27 54.2 & +01 06 18 & B7 & X \\ 
05 23 10.1 & +01 08 23 & B7 & X \\ 
05 02 44.6 & +03 27 28 & B7 & X \\ 
05 30 04.4 & $-$01 44 59 & B8 & X \\ 
05 29 55.6 & +02 08 32 & B8 & X \\ 
05 29 36.4 & +05 13 38 & B8 & X \\ 
05 28 12.6 & $-$01 56 29 & B8 & X \\ 
05 28 10.1 & +00 47 14 & B8 & X \\ 
05 27 36.9 & +01 06 27 & B8 & X \\ 
05 25 11.4 & +01 55 24 & B8 & X \\ 
05 23 51.4 & +00 51 46 & B8 & X \\ 
05 23 01.9 & +01 41 49 & B8 & X \\ 
05 21 03.3 & +04 28 41 & B8 & X \\ 
05 08 21.4 & $-$02 17 23 & B8 & X \\ 
05 06 22.9 & +02 40 24 & B8 & X \\ 
05 34 26.0 & +01 21 37 & B9 & X \\ 
05 33 21.9 & +02 22 36 & B9 & X \\ 
05 32 39.5 & +02 05 32 & B9 & X \\ 
05 27 20.6 & +02 12 57 & B9 & X \\ 
05 26 48.1 & +02 04 06 & B9 & X \\ 
05 26 06.0 & +00 50 02 & B9 & X \\ 
05 25 55.9 & $-$02 20 08 & B9 & X \\ 
05 23 50.4 & +02 04 56 & B9 & X \\ 
05 23 28.1 & $-$01 00 09 & B9 & X \\ 
05 23 22.9 & $-$01 26 27 & B9 & X \\ 
05 22 43.1 & +00 08 21 & B9 & X \\ 
05 21 28.4 & $-$01 32 46 & B9 & X \\ 
05 20 24.7 & $-$03 30 35 & B9 & X \\ 
05 19 38.8 & $-$01 06 31 & B9 & X \\ 
05 19 07.5 & $-$01 05 56 & B9 & X \\ 
05 18 30.0 & $-$01 08 18 & B9 & X \\ 
05 17 09.8 & $-$02 34 48 & B9 & X \\ 
05 16 43.8 & $-$00 53 20 & B9 & X \\ 
05 13 37.9 & +04 12 40 & B9 & X \\ 
05 12 50.0 & $-$01 33 49 & B9 & X \\ 
05 10 57.3 & $-$01 45 50 & B9 & X \\ 
05 10 04.9 & +02 56 09 & B9 & X \\ 
05 07 35.9 & +04 32 30 & B9 & X \\ 
05 07 29.4 & $-$03 18 41 & B9 & X \\ 
05 03 21.6 & $-$02 58 57 & B9 & X \\ 
05 00 48.8 & $-$00 30 03 & B9 & X \\ 
05 00 39.8 & +03 15 55 & B9 & X \\ 
05 40 17.1 & +00 58 21 & A0 & X \\ 
05 39 43.2 & +00 54 27 & A0 & X \\ 
05 33 27.4 & +02 39 06 & A0 & X \\ 
\hline
\end{tabular}
\end{minipage}
\end{table*}

\begin{table*}
 \centering
 \begin{minipage}{18cm}
\contcaption{}
  \begin{tabular}{llllll}
\hline
 RAJ2000 & DEJ2000 & SpT & $K$\,disk \\
 (h:m:s) & (d:m:s) &   & \\
\hline
05 27 15.8 & +05 01 09 & A0 & X \\ 
05 22 57.8 & $-$02 08 59 & A0 & X \\ 
05 19 40.1 & $-$01 21 22 & A0 & X \\ 
05 19 03.5 & +00 26 19 & A0 & X \\ 
05 05 54.6 & +01 51 08 & A0 & X \\ 
05 05 04.2 & $-$01 18 41 & A0 & X \\ 
05 28 10.5 & $-$01 46 18 & A1 & X \\ 
05 22 38.8 & $-$01 02 34 & A1 & X \\ 
05 14 37.0 & +02 33 49 & A1 & X \\ 
05 05 01.0 & +02 38 44 & A1 & X \\ 
05 03 21.7 & $-$00 00 16 & A1 & X \\ 
05 30 52.6 & +01 36 41 & A2 & X \\ 
05 18 29.9 & +02 05 29 & A2 & X \\ 
05 13 26.8 & $-$02 37 37 & A2 & X \\ 
05 24 08.0 & +02 27 47 & A3 & o \\ 
05 23 53.3 & $-$03 04 59 & A3 & X \\ 
05 15 57.6 & +01 19 39 & A3 & X \\ 
05 08 06.4 & +03 44 55 & A3 & X \\ 
05 20 52.9 & +01 01 00 & A4 & X \\ 
05 18 24.3 & $-$02 32 07 & A5 & X \\ 
05 36 29.4 & +03 18 30 & A6 & X \\ 
05 28 09.6 & +03 37 23 & A6 & X \\ 
05 02 43.5 & +05 49 50 & A7 & X \\ 
05 30 53.3 & +05 41 34 & A8 & X \\ 
05 24 42.8 & +01 43 48 & A8 & o \\ 
05 15 46.4 & $-$01 16 40 & A8 & X \\ 
05 28 10.7 & +02 22 35 & F1 & X \\ 
05 02 19.0 & $-$01 11 55 & F1 & X \\ 
05 00 34.4 & +00 00 20 & F1 & X \\ 
05 37 48.2 & +02 44 47 & F3 & X \\ 
05 36 41.8 & +02 41 11 & F4 & X \\ 
05 19 36.8 & +01 33 02 & F7 & X \\ 
05 08 12.6 & +01 08 36 & G1 & X \\ 
05 05 38.6 & +01 27 31 & G1 & X \\ 
05 02 15.1 & $-$01 43 08 & G6 & X \\ 
\hline
\end{tabular}
\end{minipage}
\end{table*}

\clearpage
\begin{figure*}
\begin{center}
\includegraphics[scale=0.5]{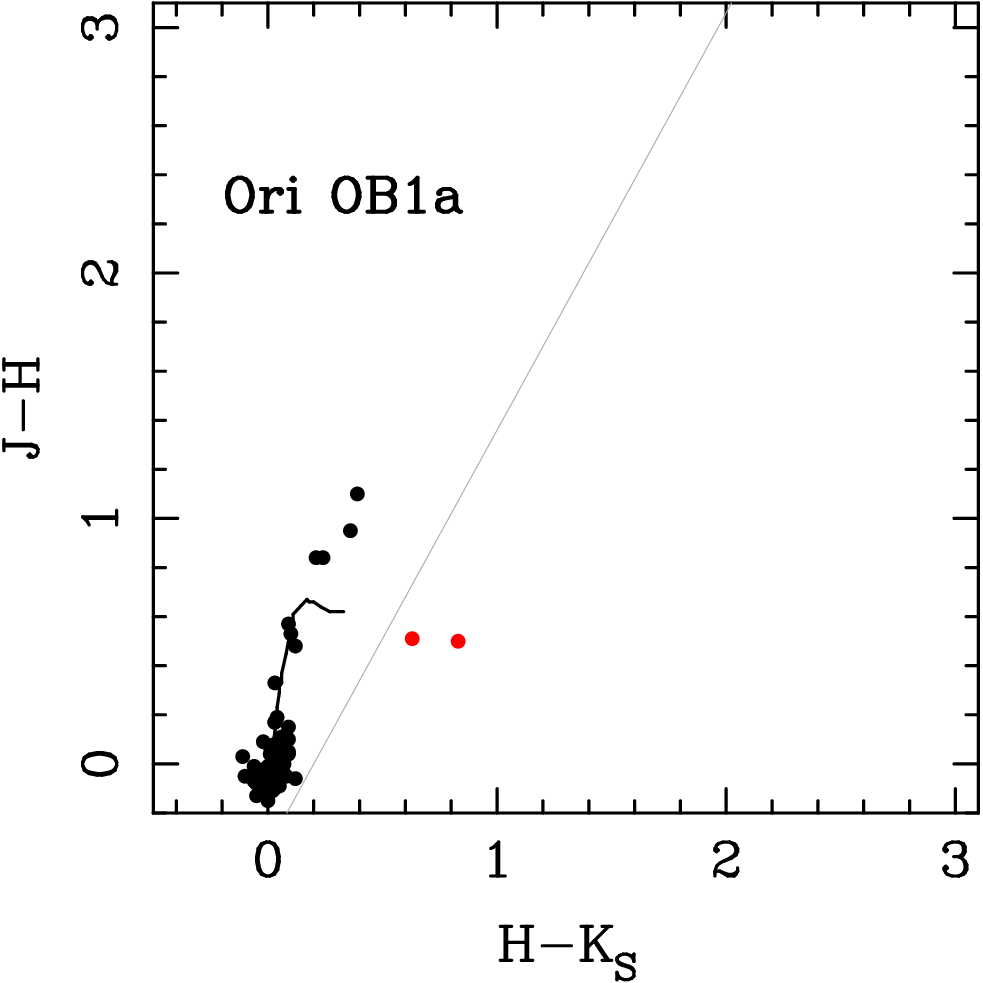}
\caption{Ori OB1a.}
\end{center}
\end{figure*}

\clearpage
\begin{table*}
 \centering
 \begin{minipage}{18cm}
\caption{NGC 7160.}
  \begin{tabular}{llllll}
\hline
Name & RAJ2000 & DEJ2000 & SpT & $K$\,disk & MIR\,disk\\
 & (h:m:s) & (d:m:s) &   & \\
\hline
DG-513 & 21 52 32.89 & +62 23 56.9 & B5.0 & X & X ($-$2.42) \\
DG-32 & 21 54 22.86 & +62 27 55.3 & B5.5 & X & X ($-$2.80) \\
DG-940 & 21 56 07.66 & +62 34 06.9 & B8.5 & X & X ($-$2.52) \\
DG-37 & 21 54 19.40 & +62 28 06.7 & B9.0 & X & X ($-$2.91) \\
DG-424 & 21 51 55.66 & +62 27 13.8 & B9.0 & X & X ($-$2.96) \\
DG-36 & 21 54 14.24 & +62 45 57.7 & B9.5 & X & X ($-$2.99) \\
DG-720 & 21 54 02.85 & +62 26 34.8 & A0.0 & X & X ($-$2.86) \\
DG-39 & 21 53 27.80 & +62 35 18.7 & A0.0 & X & X ($-$2.68) \\
DG-460 & 21 52 11.46 & +62 38 45.6 & A0.0 & X & X ($-$2.96) \\
DG-682 & 21 53 45.12 & +62 36 54.8 & A2.0 & X & X ($-$2.57) \\
DG-529 & 21 52 39.25 & +62 44 49.5 & A2.0 & X & X ($-$2.91) \\
DG-934 & 21 56 03.59 & +62 38 54.8 & A2.5 & X & X ($-$2.95) \\
DG-853 & 21 55 07.13 & +62 43 33.7 & A2.5 & X & X ($-$2.65) \\
DG-45 & 21 53 45.51 & +62 40 57.4 & A3.0 & X & X ($-$2.98) \\
DG-67 & 21 52 59.85 & +62 42 06.4 & A4.0 & X & X ($-$2.76) \\
DG-920 & 21 55 54.91 & +62 44 33.4 & A4.5 & X & X ($-$2.87) \\
DG-954 & 21 56 15.84 & +62 45 44.1 & A5.0 & X & ... \\
DG-47 & 21 53 55.61 & +62 36 18.0 & A5.0 & X & X ($-$2.82) \\
DG-687 & 21 53 46.15 & +62 46 35.4 & A5.0 & X & X ($-$2.90) \\
DG-946 & 21 56 10.81 & +62 34 54.9 & A5.5 & X & X ($-$2.43) \\
DG-409 & 21 51 45.66 & +62 42 58.2 & A5.5 & X & X ($-$2.99) \\
DG-42 & 21 53 36.84 & +62 32 48.5 & A6.0 & X & X ($-$2.99) \\
DG-398 & 21 51 42.28 & +62 33 14.5 & A6.0 & X & X ($-$2.92) \\
DG-685 & 21 53 45.42 & +62 45 25.0 & A6.5 & X & X ($-$3.04) \\
DG-382 & 21 51 31.43 & +62 28 46.2 & A6.5 & X & X ($-$2.86) \\
DG-907 & 21 55 43.05 & +62 42 28.9 & A7.0 & X & X ($-$2.93) \\
DG-65 & 21 54 36.79 & +62 33 59.5 & A7.0 & X & X ($-$2.87) \\
DG-526 & 21 52 38.57 & +62 45 52.3 & A7.0 & X & X ($-$2.96) \\
DG-481 & 21 52 21.13 & +62 45 03.4 & A7.0 & X & o ($-$1.75) \\
DG-794 & 21 54 33.51 & +62 47 53.1 & A8.0 & X & X ($-$2.83) \\
DG-49 & 21 53 51.91 & +62 33 24.5 & A8.0 & X & X ($-$2.92) \\
DG-531 & 21 52 39.31 & +62 46 58.1 & A8.0 & X & X ($-$2.50) \\
DG-725 & 21 54 05.40 & +62 43 42.7 & A8.5 & X & X ($-$2.95) \\
DG-899 & 21 55 38.35 & +62 45 53.0 & A9.0 & X & X ($-$2.85) \\
DG-399 & 21 51 41.82 & +62 47 13.8 & F0.0 & X & X ($-$2.90) \\
DG-408 & 21 51 45.44 & +62 47 05.3 & F0.5 & X & X ($-$2.95) \\
DG-48 & 21 54 13.27 & +62 43 09.2 & F1.0 & X & X ($-$2.86) \\
DG-952 & 21 56 14.29 & +62 41 41.7 & F1.5 & X & X ($-$2.93) \\
DG-60 & 21 54 33.59 & +62 28 52.9 & F1.5 & X & X ($-$2.89) \\
DG-41 & 21 53 19.42 & +62 37 38.7 & F2.0 & X & X ($-$2.90) \\
DG-936 & 21 56 05.45 & +62 26 53.5 & F2.5 & X & X ($-$2.86) \\
DG-52 & 21 55 19.88 & +62 39 15.0 & F3.0 & X & X ($-$2.93) \\
DG-55 & 21 53 33.08 & +62 37 03.1 & F3.0 & X & X ($-$2.94) \\
DG-58 & 21 54 15.89 & +62 36 04.5 & F3.5 & X & X ($-$2.94) \\
DG-472 & 21 52 20.24 & +62 27 58.8 & F4.5 & X & X ($-$2.71) \\
DG-423 & 21 51 54.64 & +62 44 06.7 & F4.5 & X & X ($-$2.83) \\
DG-949 & 21 56 11.16 & +62 47 04.6 & F5.0 & X & ... \\
DG-59 & 21 54 39.40 & +62 36 21.9 & F5.0 & X & X ($-$2.97) \\
DG-62 & 21 54 30.34 & +62 31 15.7 & F5.0 & X & X ($-$2.62) \\
DG-603 & 21 53 07.41 & +62 27 19.6 & F5.0 & X & X ($-$2.87) \\
DG-394 & 21 51 38.59 & +62 35 50.6 & F5.0 & X & X ($-$2.98) \\
DG-392 & 21 51 37.31 & +62 38 06.5 & F5.0 & X & X ($-$2.82) \\
DG-912 & 21 55 47.60 & +62 35 43.2 & F5.5 & X & o ($-$2.19) \\
DG-533 & 21 52 40.42 & +62 46 06.6 & F5.5 & X & X ($-$2.93) \\
DG-825 & 21 54 52.61 & +62 45 28.2 & F6.0 & X & X ($-$2.87) \\
DG-921 & 21 55 55.07 & +62 43 55.5 & F6.5 & X & X ($-$2.88) \\
DG-422 & 21 51 54.87 & +62 38 34.6 & F6.5 & X & X ($-$2.84) \\
DG-64 & 21 53 22.52 & +62 34 24.9 & F7.0 & X & X ($-$2.77) \\
DG-40 & 21 52 49.71 & +62 31 30.9 & F7.0 & X & X ($-$2.91) \\
DG-61 & 21 53 30.14 & +62 30 09.7 & F8.0 & X & X ($-$2.96) \\
DG-414 & 21 51 46.81 & +62 46 11.6 & F8.0 & X & X ($-$2.91) \\
DG-644 & 21 53 27.03 & +62 44 50.4 & F8.5 & X & X ($-$2.97) \\
\hline
\end{tabular}
\end{minipage}
\end{table*}

\begin{table*}
 \centering
 \begin{minipage}{18cm}
\contcaption{}
  \begin{tabular}{llllll}
\hline
Name & RAJ2000 & DEJ2000 & SpT & $K$\,disk & MIR disk \\
 & (h:m:s) & (d:m:s) &   & \\
\hline
DG-462 & 21 52 12.89 & +62 44 08.6 & F8.5 & X & X ($-$2.84) \\
DG-50 & 21 53 35.48 & +62 30 03.2 & F9.0 & X & X ($-$2.88) \\
DG-349 & 21 51 17.53 & +62 43 41.1 & F9.5 & X & X ($-$2.94) \\
DG-56 & 21 54 07.28 & +62 44 26.0 & G0.5 & X & X ($-$2.92) \\
DG-455 & 21 52 09.98 & +62 25 31.2 & G2.0 & X & X ($-$2.96) \\
DG-895 & 21 55 36.36 & +62 43 53.8 & G2.5 & X & X ($-$2.86) \\
DG-371 & 21 51 26.25 & +62 29 16.1 & G3.5 & X & X ($-$2.88) \\
$[$SHB2004$]$ 03-180 & 21 53 54.11 & +62 38 10.2 & F9 & X & X ($-$2.78) \\
$[$SHB2004$]$ 03-479 & 21 54 17.22 & +62 41 33.8 & G0 & X & X ($-$2.88) \\
$[$SHB2004$]$ 03-872 & 21 53 52.96 & +62 45 24.8 & G0 & X & ... \\
$[$SHB2004$]$ 03-791 & 21 54 56.24 & +62 44 42.2 & G2 & X & X ($-$2.68) \\
$[$SHB2004$]$ 03-228 & 21 53 59.59 & +62 38 43.3 & G2 & X & X ($-$2.88) \\
$[$SHB2004$]$ 03-654 & 21 53 58.11 & +62 43 21.3 & G2 & X & X ($-$2.89) \\
$[$SHB2004$]$ 04-1027 & 21 53 07.62 & +62 46 14.4 & G2 & X & X ($-$2.93) \\
$[$SHB2004$]$ 01-615 & 21 52 35.60 & +62 29 08.2 & G2 & X & X ($-$2.77) \\
$[$SHB2004$]$ 03-835 & 21 54 04.98 & +62 45 04.8 & G3.5 & X & X ($-$2.91) \\
$[$SHB2004$]$ 02-592 & 21 54 08.60 & +62 30 14.0 & G4 & X & ... \\
$[$SHB2004$]$ 04-521 & 21 52 58.91 & +62 41 34.7 & G4 & X & X ($-$2.84) \\
$[$SHB2004$]$ 03-500 & 21 55 04.86 & +62 41 47.9 & G5 & X & X ($-$2.84) \\
$[$SHB2004$]$ 01-1164 & 21 53 32.07 & +62 34 05.3 & G5.5 & X & X ($-$2.91) \\
\hline
\end{tabular}
\end{minipage}
\end{table*}

\clearpage
\begin{figure*}
\begin{center}
\includegraphics[scale=0.5]{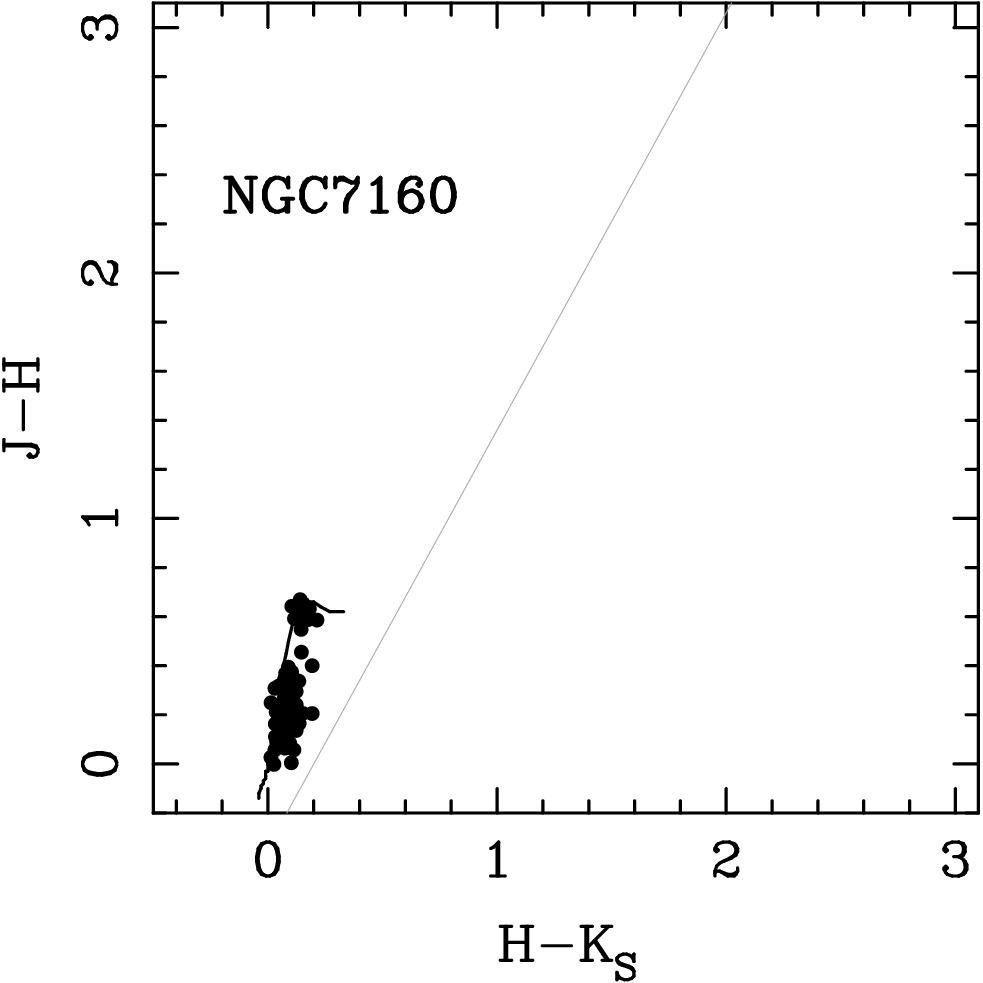}
\caption{NGC 7160.}
\end{center}
\end{figure*}

\bsp

\label{lastpage}

\end{document}